\documentclass{article}
\usepackage{amscd,amsmath,amssymb,amsfonts,emlines2}
\usepackage[dvips]{graphicx}
\makeatletter \@addtoreset{equation}{section} \makeatother
\def\Z{\mathbb{Z}} \def\R{\mathbb{R}}
\def\C{\mathbb{C}} \def\H{\mathcal{H}}

\title{The hyperbolic field theory on the plane of double variable}
\author{Dmitriy G. Pavlov\thanks{geom2004@mail.ru}\ \  {\em and}\ \  Sergey S. Kokarev\thanks{logos-center@mail.ru}}
\date{Research Institute of Hypercomplex system in Geometry and Physics (Fryazino), Regional Scientific-Educational Center
\, "Logos"\,, (Yaroslavl) }

\begin{document}\maketitle
\begin{abstract}
By analogy to the theory of harmonic fields on the complex plane, we build the theory of
    wave-like fields on the plane of double variable.
We construct the hyperbolic analogues of point vortices, sources,
vortice-sources and their higher-order
    multipole generalizations.
We examine the physical aspects and the possibility of extension to the space of
    polynumbers of higher dimensions.
\end{abstract}
{\bf Keywords:} double numbers, $h$-holomorphic functions, hypercharges, multipoles,
    hyperbolic field theory, basis.
%
%
\section{Introduction}

In paper \cite{7} were stated considerations on the theory of holomorphic functions of double variable.
    As applications to Physics there were presented 2-dimensional problems which lead to the wave equations,
    for which instead of initial-boundary conditions is provided the form of the Space-Time equipotential surface.

The reason for passing from complex numbers to double ones shows
that
    using the theory of $h$-holomorphic functions  the plane of double variable
    we can build the theory of wave fields, which, by analogy to harmonic fields on the complex plane
    can be understood as a reduced (to dimension two) field theory, which satisfy the Maxwell
    equations or their variants, built on the basis of the polynumbers algebra and (or)
    on the basis of the related to this algebra  Berwald-Moor metric.

In this paper we develop such a theory of $h$-holomorphic fields
and discuss their main properties
    and characteristic features, analogous to the harmonic fields from the complex
    plane.
    We study several examples, analyze their possible meaning for Physics, and
    show the way towards the extension of the developed construction to the space $\mathcal{H}_n$.
%
%
\section{$\C$-holomorphic field theory on $\R^2$}\label{complll}

The whole present section is a reminder of basic facts of the standard theory of functions
    of complex variable in vide, which are convenient in view of their rephrasing on the
    double plane.
All the proofs of used facts can be found in well known textbooks (\cite{6,shab}).
%
%
\subsection{Analytical interpretation}

We remind that an arbitrary smooth mapping $f:\R^2\to\R^2$ of the plane into itself
    can be represented by a couple of components
\begin{equation}\label{mapc}(x,y)\mapsto(x',y'):\ \  x'=f_1(x,y);\quad y'=f_2(x,y),\end{equation}
where $f_1,f_2:\R^2\to\R$ are smooth functions. Using the formulas:
\begin{equation}\label{form1}x=\frac{z+\bar z}{2};\quad y=\frac{z-\bar z}{2i},\end{equation}
where $z=x+iy$, and the bar denotes the complex conjugation, these mappings may be
    always written in vide
\begin{equation}\label{mapcc}
(z,\bar z)\mapsto(z',\bar z'):\ \  z'=F_1(z,\bar z);\quad \bar z'=F_2(z,\bar
z).\end{equation}
Among all the smooth mappings of this form, remarkable properties are exhibited by the mappings,
    so called {\it holomorphic}, which satisfy the condition:
\begin{equation}\label{anc}F_{,\bar z}=0,\end{equation}
and the mappings, called {\it anti-holomorphic}, which satisfy the condition:
\begin{equation}\label{aanc}F_{,z}=0.\end{equation}
The holomorphicity or the anti-holomorphicity may be satisfied at some point, or in
    some domain of the complex plane%
\footnote{Usually, we say that a function is holomorphic at some point, if the function
    is differentiable which satisfies the condition (\ref{anc}), which is satisfied in
    some neighborhood of the point}.

We remind several properties of holomorphic functions. By splitting a function $F$
    into its real and imaginary parts: $F(z)=u(x,y)+iv(x,y)$, using the relation between
    the differentiation operators relative to complex and real variables:
\begin{equation}\label{diffc}\frac{\partial}{\partial z}=\frac{1}{2}
    \left(\frac{\partial}{\partial x}-i\frac{\partial}{\partial y}\right);\quad
    \frac{\partial}{\partial \bar z}=\frac{1}{2}\left(\frac{\partial}{\partial x}+
    i\frac{\partial}{\partial y}\right),\end{equation}
which follow from (\ref{form1}) and re-writing the condition (\ref{anc}) in Cartesian coordinates,
    we get the relation
\begin{equation}\label{dbarz}\frac{dF}{d\bar z}=\frac{1}{2}[u_{,x}+iv_{,x}+i(u_{,y}+iv_{,y})]=0,\end{equation}
whose real and imaginary parts provide the well known {\it Cauchy-Riemann holomorphicity conditions
    (complex analiticity)}:
\begin{equation}\label{crc}u_{,x}=v_{,y};\quad u_{,y}=-v_{,x}.\end{equation}
Similarly, for anti-holomorphic functions we get the conditions:%
\footnote{Practically, these emerge from the condition (\ref{crc}) by changing
    the sign of $v$ to the opposite one}. %
\begin{equation}\label{acrc}u_{,x}=-v_{,y};\quad u_{,y}=v_{,x}.\end{equation}

By acting on a holomorphic or anti-holomorphic function $F$ by the real Laplace operator of second order:
\begin{equation}\label{lapl}\Delta\equiv 4\partial_z\partial_{\bar z}=\partial^2_x+\partial^2_y,\end{equation}
we infer the identities:
\begin{equation}\label{harmc}\Delta u=0;\quad \Delta v=0,\end{equation}
which express the harmonicity of both the real and the imaginary parts
    of holomorphic and anti-holomorphic functions.

We note that due to our definition, the derivative of a function (if this exists -- the existence
    of derivatives of arbitrary order is proved in any standard textbook of Theory of Functions
    of Complex Variable) is a holomorphic function as well:
\begin{equation}\label{diff3}F_{,\bar z}=0\Rightarrow (F_{,\bar z})_{,\bar z}=0\Rightarrow\dots
    \Rightarrow (F^{(n)}_{,\bar z})_{,\bar z}=0\end{equation}
(similarly, for anti-holomorphic functions, by changing $z\leftrightarrow \bar z$.)
To this context it is related the representation of a holomorphic function in the neighborhood
    of some point $z_0$, by the Taylor power series:
\begin{equation}\label{teylc}F(z)=\sum\limits_{k=0}^\infty c_k(z-z_0)^k,\end{equation}
where the complex coefficients are given by
\begin{equation}\label{teylcoeff}c_k=\frac{1}{k!}\left.\frac{d^kF}{dz^k}\right|_{z=z_0}.\end{equation}
In the neighborhood of the point $z_0$, at which the holomorphicity condition is broken,
    the function $F$ can be represented by the more general {\it Lorant  series}:
\begin{equation}\label{loran}F(z)=\sum\limits_{k=-\infty}^\infty c_k(z-z_0)^k.\end{equation}
The part of the Lorant  series which contains negative powers of
$(z-z_0)$ is called
    {\it the main part}, and the part which contains non-negative powers is called {\it the right part}.
If the main part contains only a finite number of terms and $n$ is
the maximal absolute value of
    the power, which corresponds to the nonzero coefficient $c_{-n}$, then they say, that the function $F$
    has at the point $z_0$ a pole of order $n$. In the case when $n=\infty$, the point
    $z_0$ is called {\it essential singularity}.
Another type of singular points are the branching points, for which the value of the function
    is changed, while following a path which surrounds this point.
In the neighborhood of such points, the Lorant series has to be
replaced by the
    generalized {\it Pisot series}, which contains non-integer powers of $z-z_0$ or
    $\ln(z-z_0)$.\par\bigskip

{\bf Example 1}. {\small
The function $F(z)=\ln z$ is everywhere holomorphic on the complex plane, excluding the points $z=0$
    and $z=\infty$. Its coordinate representation is given by the expression:
\begin{equation}\label{ln}\ln z=\ln\rho+2\pi ik\varphi=\ln{\sqrt{x^2+y^2}}+2\pi ik
    \text{arg}\,z,\quad k\in \mathbb{Z}.\end{equation}
The singular points $z=0$ and $z=\infty$ are the branching points of the logarithm}.\par\bigskip

{\bf Example 2}. {\small
We examine the function $Z=\frac{1}{2}\left(z+z^{-1}\right)$, which is called the {\it Zhukowskiy
    function}. This function is everywhere holomorphic, excluding the points $z=0$ and $z=\infty$,
    which are poles of first order. The expansion of this function into its real and imaginary parts
    has the form:
\begin{equation}\label{zhcc}Z=\frac{x(x^2+y^2+1)}{2(x^2+y^2)}+i\frac{y(x^2+y^2-1)}{2(x^2+y^2)}.\end{equation} }
%
%
\subsection{Topological interpretation}

The smooth functions $u$ and $v$, which are related by the conditions (\ref{crc}),
    are not independent and define each other up to an additive constant.
    Indeed, assigning to the function $v$ the arbitrary value $v_0$ at the point $(x_0,y_0)$,
    its value at an arbitrary point $(x,y)$ -- due to the Cauchy-Riemann conditions
    can be determined using the function $u$, by means of the following integral:
\begin{equation}\label{harmc1}v(x,y)=v_0+\int\limits_{(x_0,y_0)}^{(x,y)}u_{,x}\, dy-u_{,y}\,dx,\end{equation}
which can be computed over any path which connects the points $(x_0,y_0)$ and $(x,y)$ and lie inside the
    holomorphicity domain of the function $F=u+iv$.

In fact, the result of this integration in (\ref{harmc1}) may depend on the choice of the path,
    which joins the initial and the final point. In such a case, we deal with {\it a
    multi-valued
    analytic function}, which may have a finite or countable number of branches. A more detailed answer
    on the uniqueness of the definition of conjugate functions and in general, on the monovaluedness of
    integrals of holomorphic functions is provided by the fundamental {\em Cauchy theorem},
    according to which, a function which is holomorphic in the domain $D$ and continuous on the
    piecewise-smooth boundary $\partial D$ functions, is expressed by the equality
\begin{equation}\label{coshc1}\oint\limits_{\partial D}F(z)\, dz=0\end{equation}
(for anti-holomorphic functions one should perform in this relation $z$ into $\bar z$).
    This theorem follows from the holomorphicity condition (\ref{anc}) and the equality:
\begin{equation}\label{darbu}\oint\limits_{\partial D}F(z)\, dz=\int\limits_{D}F_{,\bar z}\,
    d\bar z\wedge dz,\end{equation}
which represents the complex form of the Poincar\'e-Darboux theorem regarding the integration of
    differential forms.

We examine now a pair of points $z_1$ and $z_2$ on the complex plane and we consider some
    function, which is holomorphic everywhere on the complex plane, excluding a countable
    number of singular points.

\begin{figure}[htb]\centering \unitlength=0.50mm \special{em:linewidth 0.4pt}
    \linethickness{0.4pt} \footnotesize \unitlength=0.70mm
    \special{em:linewidth 0.4pt} \linethickness{0.4pt}\input{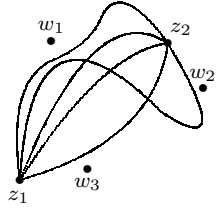}
    \caption{\small Homotopic paths relative to the singular points $w_1,w_2,w_3$ of a holomorphic
    function $F$.}\label{psi}
\end{figure}

\noindent The whole family of paths $\{\gamma_{z_1,z_2}\}$ with fixed initial and final points
    splits into classes of paths relative to their relation to the singular points of the function $F$.
    Namely, two paths $\gamma_1$ and $\gamma_2$ are mutually homotopic,
    and we write: $\gamma_1\stackrel{F}{\sim}\gamma_2$, if there exists a continuous deformation
    of one path into the other, such that no path of the deformation process contains any of
    the singular points of the function $F$.
We shall denote the class of paths which are homotopic to $\gamma$, as $[\gamma]$.
    Then from the Cauchy theorem we immediately infer the following equality:
\begin{equation}\label{hom}\int\limits_{\gamma} F(z)\, dz=\text{const}_{[\gamma]},\end{equation}
which expresses the independence of the integral on the path which belongs to the class $[\gamma]$
    of paths, which are homotopic to $\gamma$.

From the equality
\begin{equation}\label{form4}\oint\limits_{\mathcal{C}}(z-z_0)^m\, dz =2\pi i\delta_{m,-1},\end{equation}
which holds true for any closed contour $\mathcal{C}$ which surrounds the point $z_0$,
    and is easy to check for a circle with center at $z_0$. Then, using the power expansion
    of the holomorphic function we yield the integral Cauchy formula
\begin{equation}\label{coshc2}F(z_0)=\frac{1}{2\pi i}\oint
    \limits_{\mathcal{C}}\frac{F(z)}{z-z_0}\,dz,\end{equation}
where $\mathcal{C}$ is a piecewise-smooth closed contour, which surrounds the domain
    $D\subset \C$, $F(z)$ is an arbitrary holomorphic function in the domain $D$ the function,
    and $z_0\in D$.
The formula (\ref{coshc1}) admits a generalization to multi-connected domains, and the
    formula (\ref{coshc2}) admits a generalization to contours, which contain the point from infinity
    and contours, which pass through the point $z_0$ and even exhibiting non-smoothness at this point.

From formula (\ref{form4}) and from the expansion of the function $F(z)$ in the Maclaurin series (\ref{loran}),
    it follows the well known rule for computing integrals of holomorphic functions which have singular points:
\begin{equation}\label{intcc}\oint\limits_{\mathcal{C}}F(z)\, dz=2\pi i\sum\limits_kc_{-1}^{(k)},\end{equation}
where the sum in the right hand side extends over all the singular points, which fall inside the contour
    $\mathcal{C}$. Since the main contribution in the integral is given only by the coefficients $c_{-1}$
    (these ones are called {\it residues} of the function $F$ at the corresponding singular points),
    then the definition of classes of homotopic paths on which the integrals are constant, can be
    significantly relaxed: we call homotopic all the paths, which are obtained one from another
    by a continuous deformation, which allows traversing singular points whose residues are zero.

The multi-valued analytic functions can be easily described by means of their graphs
    in the 4-dimensional space $\C\times \C$. These are called {\it Riemannian surfaces}
    of the corresponding functions. On its Riemannian surface, an analytic function is uniquely defined
    (is one-valued).
%
%
\subsection{Geometric interpretation}

We examine the quadratic form
\begin{equation}\label{quad}\eta=\text{Re}(dz\otimes d\bar z)=dx\otimes dx+dy\otimes dy.\end{equation}
We can say that the complex structure algebraically induces on the plane $\R^2$ an
    Euclidean metric.
Relative to the transformations provided by holomorphic functions $F(z)$, the form $\eta$ behaves as follows:
\begin{equation}\label{Bc}\eta\mapsto \eta'=|F'(z)|^2\eta,\end{equation}
where $F'(z)=dF/dz$. The formula (\ref{Bc}) means that the function $F(z)$ in its holomorphicity domain,
    assuming that $F'(z)\neq0$, provides a conformal mapping of the complex plane into itself,
    i.e., it preserves the angles. We note that\  $|F'|^2=|\nabla u|^2=|\nabla v|^2=\Delta_F$,
    where $\nabla$ is the gradient operator relative to the Euclidean metric, and $\Delta_F$ is
    the Jacobian of the mapping $F$.
As it follows from condition (\ref{crc}) or from the conformality, the lines $u=\text{const}$ and
    the lines $v=\text{const}$, for any holomorphic function $F(z)$, determine on the plane $\C$
    an orthogonal curvilinear coordinate system, since at each point we have the equality:
\begin{equation}\label{ortogo}\nabla u\cdot\nabla v=u_{,x}v_{,x}+u_{,y}v_{,y}
    =-u_{,x}u_{,y}+u_{,y}u_{,x}=0.\end{equation}
This clarifies the geometric meaning of the conjugation relation which exists between the functions
    $u$ and $v$: {\it the conjugate functions have mutually-orthogonal surfaces of constant level
    and have equal norms of their gradients at each point}.\par\bigskip

{\bf Example 3}. {\small
Due to the formulas $z^n=\rho^ne^{in\varphi}$ and (\ref{ln}), it is obvious that
    the power function maps the orthogonal net of polar coordinates into another orthogonal net
    of polar coordinates on the image-plane, and the logarithm maps the polar net into the Cartesian
    net on the image-plane. In the figure, we show the image of the polar net under the mapping
    given by the Zhukowskiy function.

{\centering\small\refstepcounter{figure}\label{riszhc}
    \includegraphics[width=.4\textwidth]{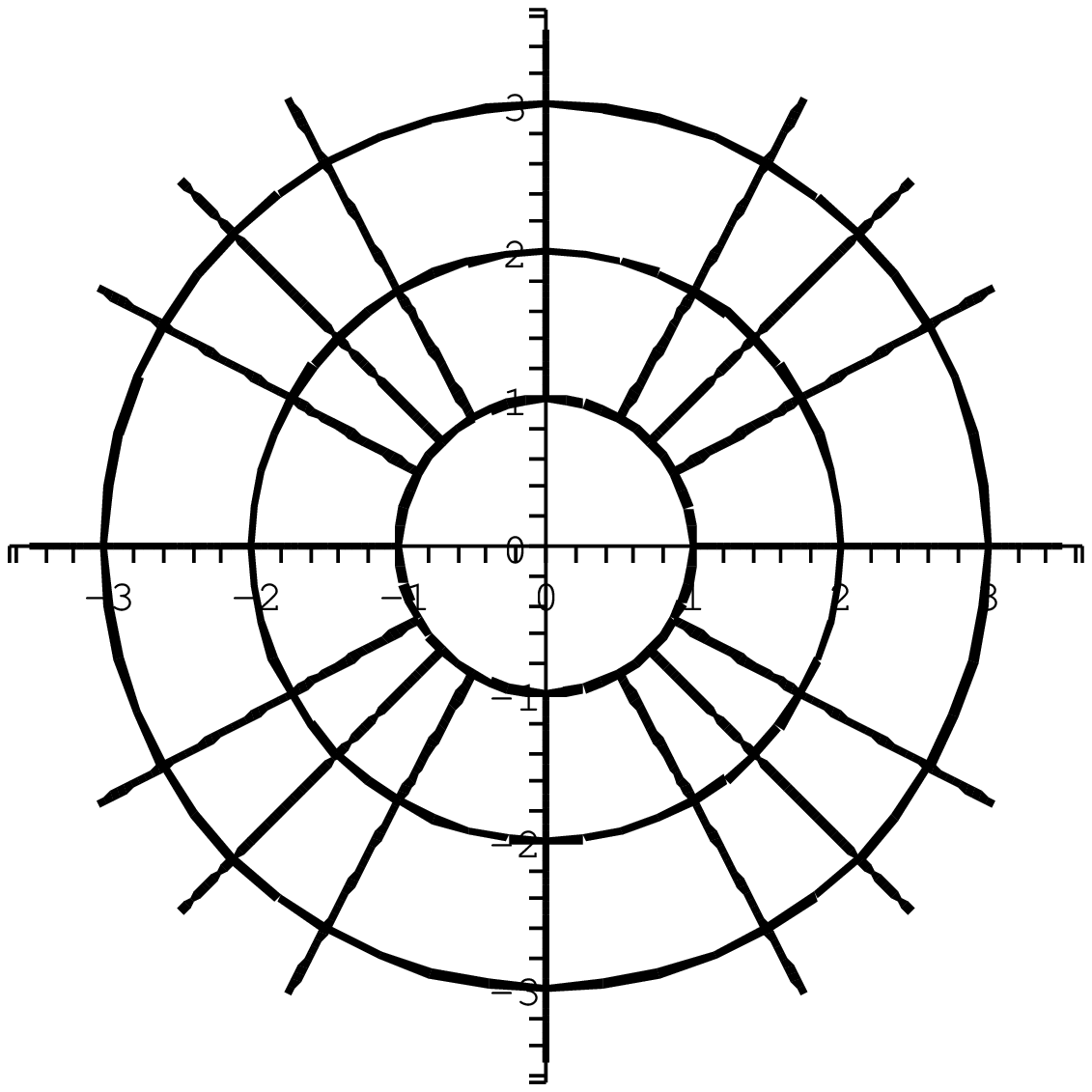}\includegraphics[width=.4\textwidth]{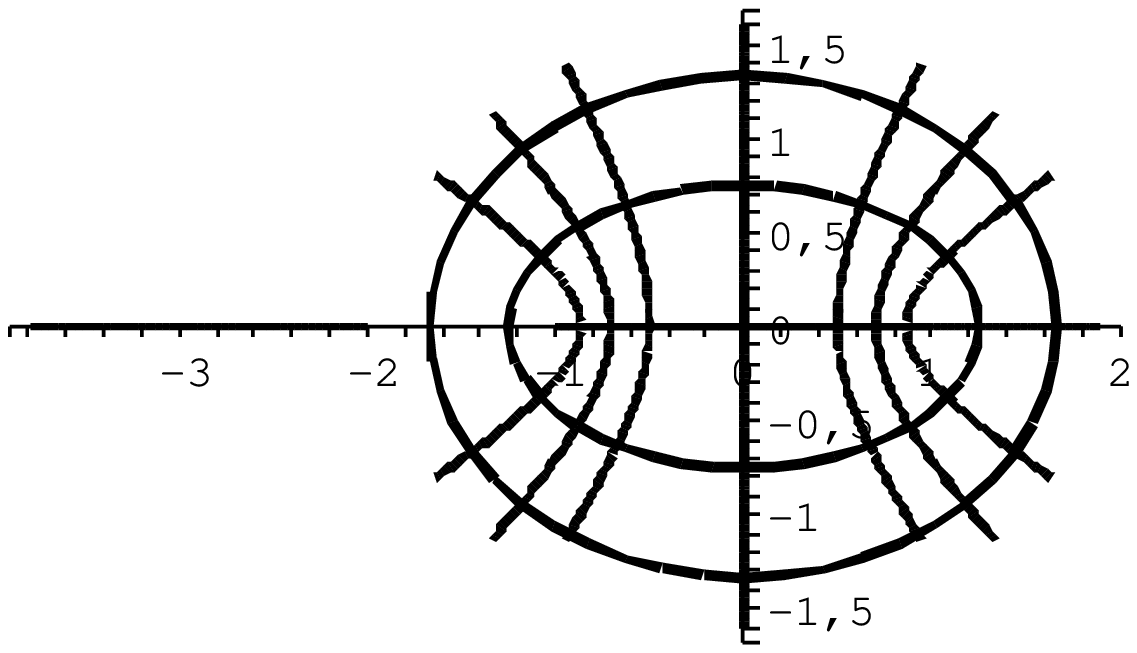}
    \medskip\nopagebreak\par Fig.~\thefigure. In the left figure are represented the lines of the polar
    coordinate system; on the right -- their images under the mapping $z\mapsto (z+z^{-1})/2$.}}\par\bigskip

We note that a conformal mapping can be regarded both in active (deformation of the plane),
    and in passive (change of coordinates) meaning.
Independently of interpretation, the metric $\eta$ remains Euclidean, unlike the conformal mappings
    of the more general form: $\eta\to \eta'=e^{2\phi}\eta$, where $\phi$ is an arbitrary smooth
    function depending on the coordinates $x$ and $y$.
It can be shown that the condition of vanishing of the curvature tensor for the metric $e^{2\phi}\eta$
    (where $\eta$ is the Euclidean metric) is equivalent to the equality $e^{2\phi}=|F'|^2$
    for some holomorphic function $F(z)$.
%
%
\subsection{Physical interpretation}

In physical applications, to each holomorphic function $F(z)$ corresponds a 2-dimensional problem
    of electrostatics, magnetostatics, problem of potential flows for an incompressible fluid and
    many others.
What is common for all such problems is the Laplace equation, which is satisfied by the
    potential function (accordingly, the electrostatic potential, magnetostatic potential and
    potential of velocities) in a source-free domain.

We remind the fundamental facts of the applications of holomorphic functions using as
    an example a problem from electrostatics.
To the holomorphicity domain, we can associate to the function $F(z)=u+iv$ an electrostatic field
    in the free of charges space. The real part $u$ of this function is the potential
    of the electrostatic field ({\it potential function}), and the imaginary part $v$
    is {\it the force function} of this field.
In other words, lines $u=\text{const}$ are equi-potential lines of the electrostatic
    field, and the lines $v=\text{const}$ coincide with the strength lines of this field.
As confirmed by physical considerations, these families of lines are mutually orthogonal
    (cf. (\ref{ortogo})), and each of the functions $u$ and $v$ satisfies the Laplace equation
    in the considered domain. The main idea, on which rely the applications of the Theory of Physical
    Complex Fields for solving problems of plane electrostatics, consists on finding such a
    holomorphic function $F(z)$, which maps the borders $\Gamma_i$ of all the sources of
    the studied problem into straight lines $\text{Re}\, F(z)|_{\Gamma_i}=\text{const}_i$.
Indeed, in such a case the borders of all the circuits become equipotential, and due to
    the theorem of existence and uniqueness of solutions of the Laplace equation with
    with given boundary conditions, the obtained solution will be unique (up to a physical
    negligible renormalization of the potential).
If the needed holomorphic function is known (for its finding there exist no universal recipes),
    then the strength of the electrostatic field can be found using the formula:
\begin{equation}\label{voltage}E=E_x+iE_y=-\overline{\frac{dF}{dz}}
    =-\frac{d\bar F}{d\bar z}=-u_{,x}-iu_{,y},\end{equation}
Which should be understood as a complex form of the vector field given by the gradient
    of the function $u$. We note that formula (\ref{voltage}) is obtained by considering
    the relations (\ref{diffc}) and the Cauchy-Riemann conditions (\ref{crc}).

Due to the relation $E=E(\bar z)$ (anti-holomorphicity of the strength), which follows from (\ref{voltage})
    by taking into consideration (\ref{diffc}), we obtain the identity:
\begin{equation}\label{potsolc1}\frac{\partial E}{\partial z}
    =\frac{1}{2}[E_{x,x}+E_{y,y}-i(E_{x,y}-E_{y,x})]=0,\end{equation}
which is equivalent to two identities:
\begin{equation}\label{potsol2}\text{div}\, E\equiv E_{x,x}+E_{y,y}=0;\quad \text{rot}\,
    E\equiv E_{y,x}-E_{x,y}=0,\end{equation}
which correspondingly express the properties of {\it solenoidality} and {\it potentiality}
    of the electrostatic field. We note that these conditions are automatically satisfied,
    if the potential $u$ is the real part of some holomorphic function.

We examine now the integral
\begin{equation}\label{potcirc1}\Phi[E,\gamma]=\int\limits_{\gamma}E\, d\bar z
    =\int\limits_{\gamma}E_{x}\,dx+E_{y}\, dy+i\int\limits_{\gamma}E_{y}\,dx-E_{x}\,dy
    =\Gamma[E,\gamma]-i\Pi[E,\gamma]\end{equation}
along some path $\gamma$.
Its real part $\Gamma[E,\gamma]$ is called {\it the circulation of the field $E$ along the path
    $\gamma$}, and the quantity $\Pi[E,\gamma]$, opposed to the imaginary part, is called
    {\it the flow of the field $E$ through the line $\gamma$}.
By considering the definition (\ref{voltage}) and the Cauchy-Riemann conditions for these
    quantities, we get the following expressions in terms of the variation of the components
    of the complex potential:
\begin{equation}\label{cpot}\Gamma[E,\gamma]=-\delta_\gamma u;\quad \Pi[E,\gamma]
    =-\delta_\gamma v,\end{equation}
which can be simultaneously examined as an illustrative physical meaning of the components
    of the complex potential $F(z)$. We used here the notation $\delta_\gamma Q\equiv Q(z_2)-Q(z_1)$
    for the variation of the function $Q$ along the path $\gamma$ with initial point $z_1$ and
    final point $z_2$.

We examine further several important examples.\par\bigskip

{\bf Example 6: the field of the point source.} {\small
The complex potential
\begin{equation}\label{qul2}F(z)=-q\ln z\end{equation}
describes the field of a point charge $q$ on the plane (in 3-dimensional space it corresponds
    to an infinite straight charged thread, with the linear density of the charge $q$).
From the expression (\ref{ln}) and formula (\ref{voltage}) it follows the formula for the
    strength, which can be written as
\begin{equation}\label{Elnc}E=q\frac{z}{|z|^2},\end{equation}
of the 2-dimensional Coulomb law. From the relations (\ref{cpot}), we get for any circle
    with center at the origin of the coordinate system, where is the charge located (and hence for
    any closed contour, which surrounds once the point $z=0$):
\begin{equation}\label{formuc}\Gamma[E]=0,\quad  \Pi[E]=2\pi q,\end{equation}
which expresses the potentiality of the electrostatic field and the 2-dimensional
    Gauss theorem, accordingly}.\par\bigskip

{\bf Example 7: conducting neutral cylinder posed in a constant homogeneous field}.
{\small
The complex potential $F(z)=2iE_0RZ(z/R)=iE_0(z+R^2z^{-1})$ describes the electrostatic field
    around the conducting pole neutral cylinder of radius $R$, which is displaced in a
    homogeneous electric field $E_0$, orthogonal to its symmetry axis. From the expression
    (\ref{zhcc}) and formula (\ref{voltage}), follows the expression of the strength
\begin{equation}\label{Ezhc}E=iE_0-iE_0R^2\frac{z^2}{|z|^4}.\end{equation}
The force lines of this field can be easier obtained from the form of the force functions from (\ref{zhcc})
    (by considering the multiplier $i$, this is exactly the function $u$).
    These are represented in Fig.\ref{silazhc}}.

{\centering\small\refstepcounter{figure}\label{silazhc}
    \includegraphics[width=.5\textwidth,clip]{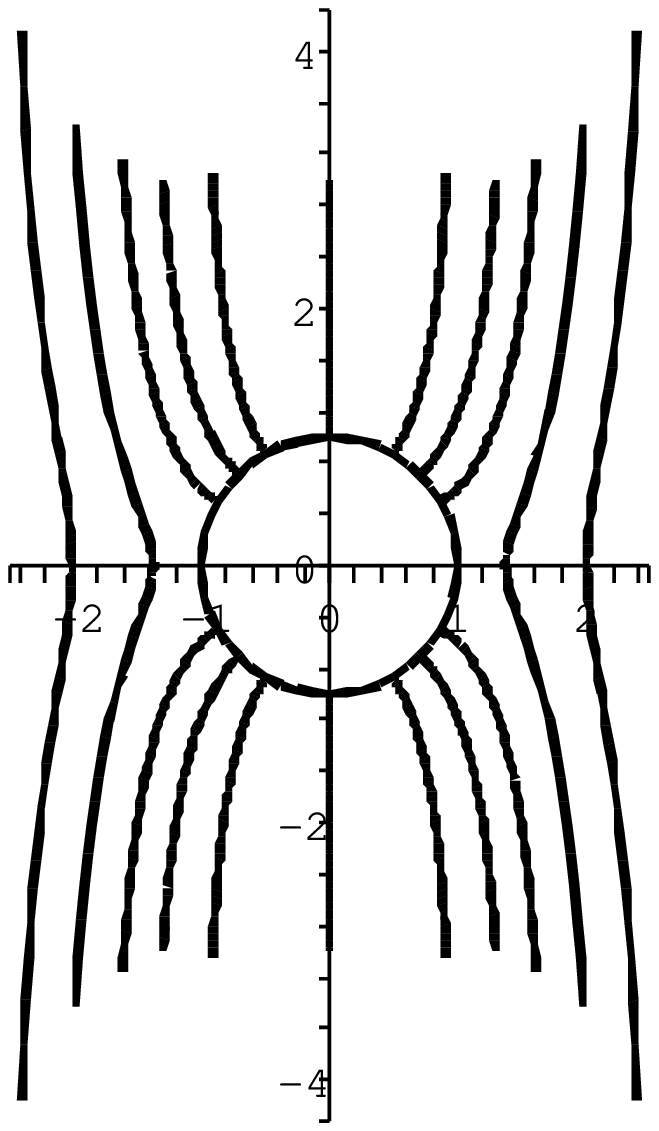}\par
    \nopagebreak\par Fig.~\thefigure. Force lines of the electric field around the cylinder of radius
    $R=1$. The outer electric field is oriented along the imaginary axis}.\par
%
%
\subsubsection{Magnetic interpretation}

The examples presented above, and generally, any solution of the plane electrostatic problem,
    written in terms of the complex potential, satisfy a remarkable {\it dual symmetry}.
For clarifying the idea of dual symmetry we examine the complex potential $F(z)$ of some
    electrostatic problem and study the complex potential $iF$.
For the strength of the new field $B$, due to (\ref{voltage}), we get the expression
\begin{equation}\label{Bcc}B=iE,\end{equation}
which geometrically represents a rotation of the electric vector $E$ at each point with the angle $\pi/2$.
    Physically, such a transformation can be regarded as {\it passing from the problem of electrostatics
    to the conjugate (dual) magnetostatic one}.
Applying such a transition to the situation of a point charge (charged thread) on the plane,
    we get the magnetic field of the linear current with the potential:
\begin{equation}\label{potmc}B=-q\arctan(y/x)+iq\ln\sqrt{x^2+y^2}.\end{equation}
Here, the force lines of the field $B$ are concentric circles, and the equi-potential surfaces are radial lines,
    which emerge from the source (the force and the potential functions have interchanged their places).
The formulas (\ref{cpot}) lead now to $\Gamma[B]=2\pi q$, $\Pi[B]=0$, which represents the
    law of the total current in magnitostatics, and the solenoidality of the magnetic field, accordingly.

The transition to the potential $iF$ in the previous example with the cylinder leads to the
    problem of the diamagnetic cylinder, posed in an exterior homogeneous magnetic field,
    orthogonal to its symmetry axis. The force lines of the magnetic field result now by using
    the imaginary part of the formula (\ref{zhcc}).

{\centering\small\refstepcounter{figure}\label{silazhc1}
    \includegraphics[width=.5\textwidth,clip]{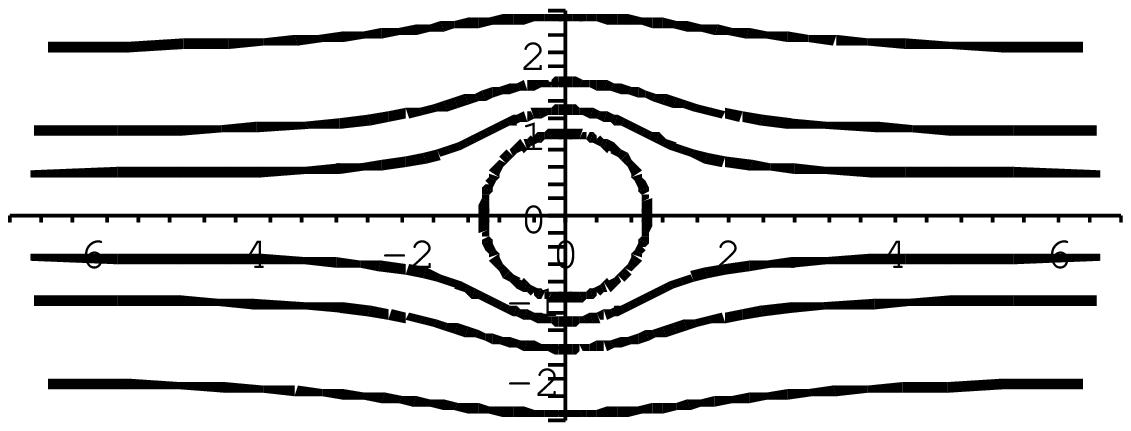}
    \nopagebreak\par Fig.~\thefigure. Force lines of the magnetic field around the diamagnetic
    cylinder of radius $R=1$. The outer homogeneous magnetic field is oriented along the real axis.\par}
%
%
\subsubsection{The vortex source}

There exists an interesting possibility, of unifying the two dual-conjugate pictures which were examined
    above, into a single one. We address first to solving the case with point charge
    and then examine the case, when this charge is complex: $Q=q-im$.
It is obvious that the logarithmic potential for such a charge:
\begin{equation}\label{vic1}F(z)=-Q\ln z
    =-q\ln\sqrt{x^2+y^2}-m\arctan(y/x)-i(q\arctan(y/x)-m\ln\sqrt{x^2+y^2}))\end{equation}
includes in itself both the particular cases of electric, and magnetic portraits
    which were examined above, and describes their superposition.
For avoiding the physically wrong addition of the electric with the magnetic fields,
    we can examine, as an example, the hypothetical electrostatics, in which
    the electrostatic field may be non-potential.
The source of this non-potentiality is exactly the imaginary part $m$ of the charge $Q$.
    In terms of complete electrodynamics this point of view means the passing to
    the extended dual-symmetric electrodynamics, in which there exist magnetic charges
    (monopoles) and magnetic fields.
The latter are the ones to be interpreted in plane problems as point or distributed
    vorticity sources of the electrostatic field.
In magnetic interpretation of the magnetic field there occur sources -- which are
    the magnetic charges themselves.

On Fig.\ref{vic2} it is shown the shape of force lines of the electrostatic field in superposition, for $q=m$.

{\centering\small\refstepcounter{figure}\label{vic2}
    \includegraphics[width=.45\textwidth,clip]{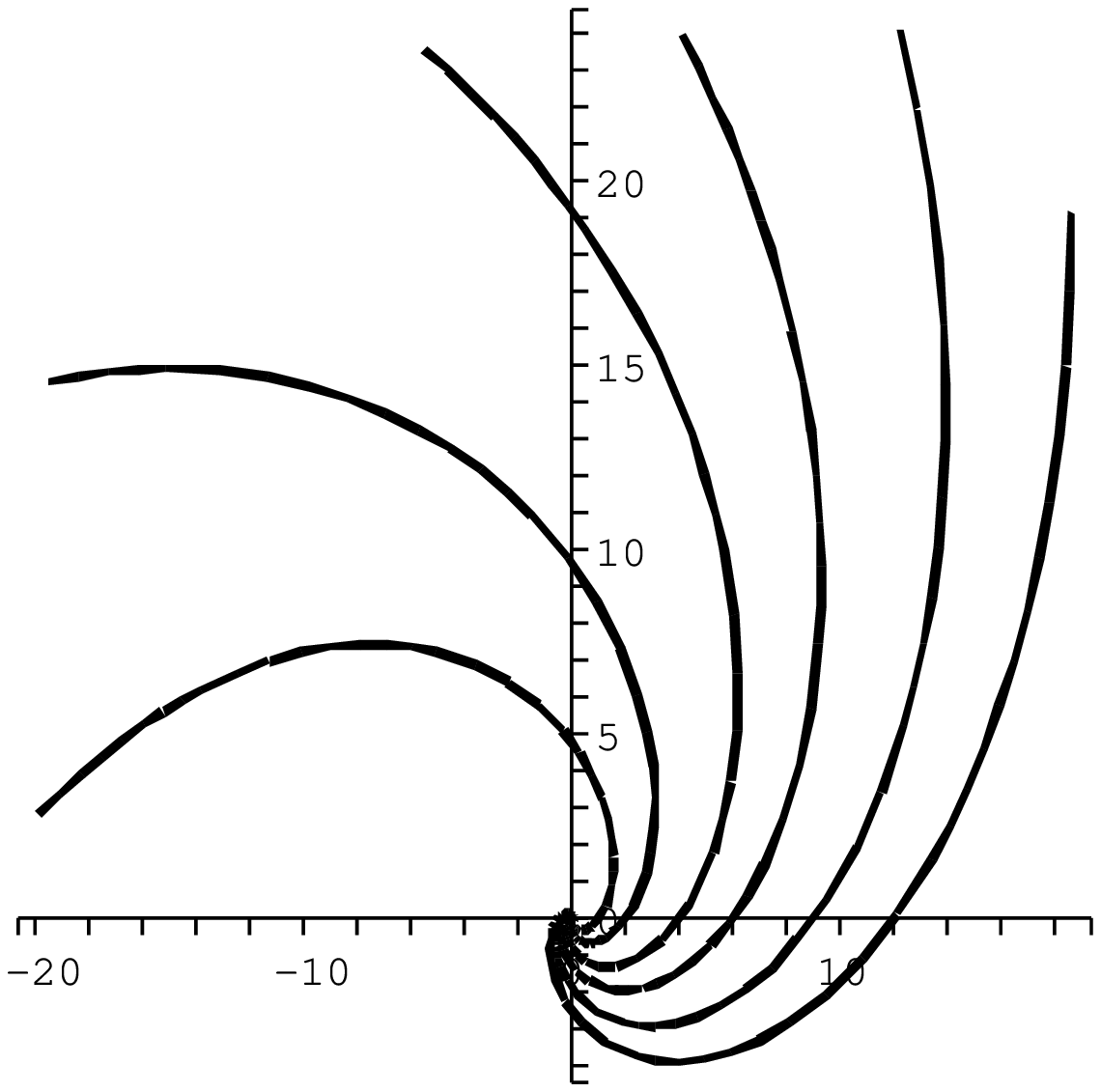}
    \nopagebreak\par Fig.~\thefigure. Force lines of the electrostatic field around
    the vorticity source with vortex and source having equal powers}.\par
%
%
\subsubsection{Multipole generalizations}

We examine two opposed point charges: the negative charge is displaced at the origin of coordinates,
    and the positive one -- at distance $\epsilon$ from the previous one, on the positive real semiaxis.
According to the superposition principle, the complex potential of such a system of charges will
    be defined by the equation:
\begin{equation}\label{dip1}F(z)=F_++F_-=-q\ln z+q\ln(z+\epsilon)=q(-\ln z+\ln z+\ln(1+\epsilon/z))
    =\frac{q\epsilon}{z}+o(\epsilon/z).\end{equation}
We examine the limit of this expression for $\epsilon\to 0$, $q\to\infty$, and for finite value of
    $p=q\epsilon$. As a result, we get the potential
\begin{equation}\label{dip2}F_1(z)=\frac{p}{z}=\frac{p\bar z}{|z|^2}\end{equation}
{\it of the point-like dipole with power (or with dipole moment) $p$}.
    The point-like dipole is the simplest system with with zero complete electric charge.
    In coordinates, the expression (\ref{dip2}) has the form:
\begin{equation}\label{dip3}F_1(x,y)=\frac{px}{x^2+y^2}-i\frac{py}{x^2+y^2}
    =\frac{pe^{-i\varphi}}{\rho}.\end{equation}
The strength lines of the dipole field are defined by the force function (by the
    imaginary part of the expression (\ref{dip3})) and in polar system of coordinates
    these are described by the family of equations:
\begin{equation}\label{dip4}\rho=C\sin\varphi, \quad \R\ni C>0,\end{equation}
Some of the lines are shown in Fig.\ref{dipp4}

{\centering\small\refstepcounter{figure}\label{dipp4}
    \includegraphics[width=.5\textwidth,clip]{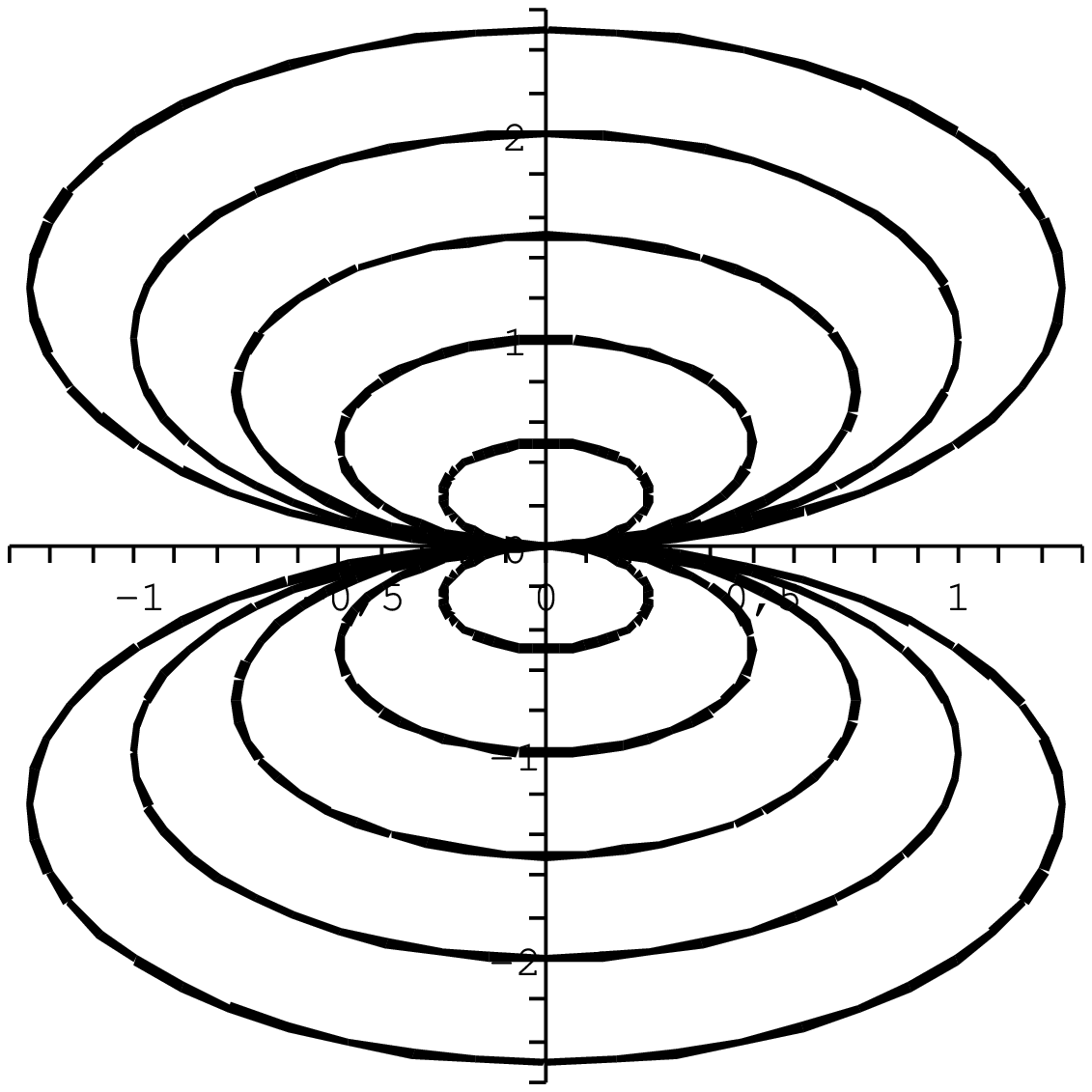}
    \nopagebreak\par Fig.~\thefigure. The force lines of the electrostatic field
    around the point-like dipole with unitary power}.\par\bigskip

The examination of the dipole, which is generated by the charge of vortex-source type
    is obtained from the previous one by replacing $p\mapsto p_e-ip_m$.
    The the formulas (\ref{dip3}) and (\ref{dip4}) get a more general form:
\begin{equation}\label{dip5}F_1(x,y)=\frac{|p|e^{-i(\varphi-\delta)}}{\rho};\quad \rho
    =C\sin(\varphi+\delta),  \quad \R\ni C>0,\end{equation}
where $\delta=\arctan (p_m/p_e)$, $|p|=\sqrt{p_e^2+p_m^2}$.
    It is easy to remark that the magnetic part of the dipole moment is responsible for
    the rotation of the picture \ref{dipp4} with the angle $-\delta$ and for the increase
    of the dipole power.

Analogously, by examining the pair of dipoles which have opposed orientation and same power $p$,
    relatively shifted by $\epsilon$ (this quantity can be regarded as a complex number) and
    passing to the limit $\epsilon\to0$, $p\to\infty$, $p\epsilon=Q^{(2)}<\infty$,
    we get {\it the potential point-quadrupole whose power is $Q^{(2)}$}:
\begin{equation}\label{dip6}F_2(z)=-\frac{Q^{(2)}}{z^2}
    =-\frac{|Q^{(2)}|e^{-i(2\varphi-\delta_2)}}{\rho^2},\end{equation}
where $|Q^{(2)}|=\sqrt{(Q^{(2)}_e)^2+(Q^{(2)}_m)^2}$, $\delta_2=\arctan{Q^{(2)}_m/Q^{(2)}_e}$.
    The strength lines are described by the family of equations
\begin{equation}\label{dip7}\rho=\sqrt{C\sin(2\varphi-\delta_2)}\end{equation}
and are shown in Fig.\ref{dipp7}

{\centering\small\refstepcounter{figure}\label{dipp7}
    \includegraphics[width=.45\textwidth,clip]{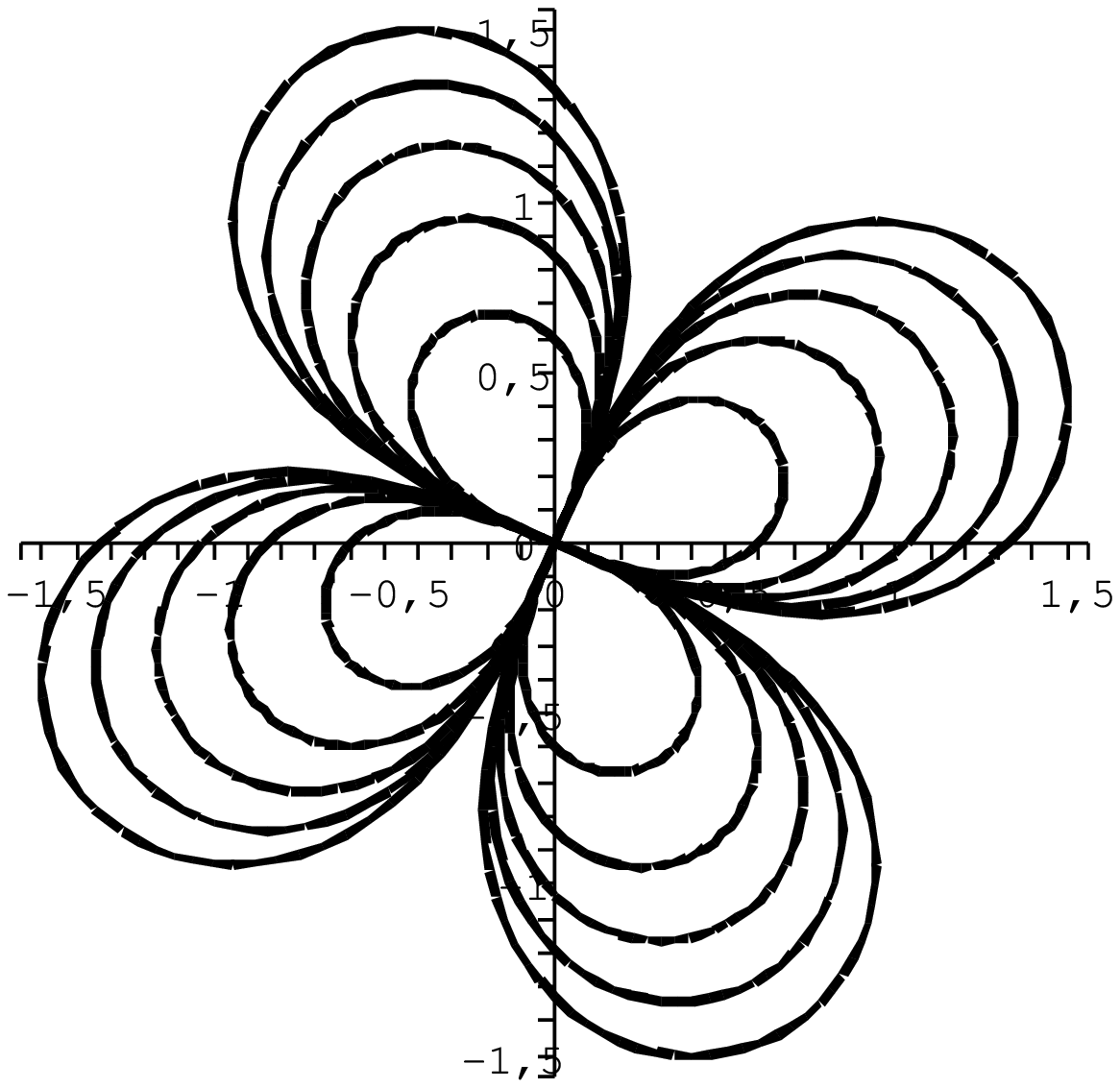}
    \nopagebreak\par Fig.~\thefigure. Force lines of the electrostatic field in
    the neighborhood of the point-quadrupole with $Q^{(2)}_e=Q^{(2)}_m=1/\sqrt2$.}\par\bigskip

Further, by induction, if is defined the potential $F_{n-1}$ of the point-like
    $2(n-1)$-multipole with power $Q^{(n-1)}$, then the potential $F_{n}$ of the point-like
    $2n$-multipole with power $Q^{(n)}$ is defined by the formula:
\begin{equation}\label{dip8}F_n(z)=\frac{Q^{(n)}}{Q^{(n-1)}}\frac{dF_{n-1}}{dz},\end{equation}
which leads to the following explicit formula for the potential of the $2n$-multipole:
\begin{equation}\label{dip9}F_n(z)=(-1)^{n+1}\frac{Q^{(n)}}{z^n}
    =(-1)^{n+1}\frac{|Q^{(n)}|e^{-i(n\varphi-\delta_n)}}{\rho^n},\end{equation}
where $|Q^{(n)}|=\sqrt{(Q^{(n)}_e)^2+(Q^{(n)}_m)^2}$, $\delta_n=\arctan{Q^{(n)}_m/Q^{(n)}_e}$.
    The equation for the force lines has in polar coordinates the following form:
\begin{equation}\label{dip10}\rho=\sqrt[n]{\sin(n\varphi-\delta_n)}\end{equation}
The shape of force lines for $n=5$ is shown in Fig.\ref{dipp10}.

{\centering\small\refstepcounter{figure}\label{dipp10}
    \includegraphics[width=.45\textwidth,clip]{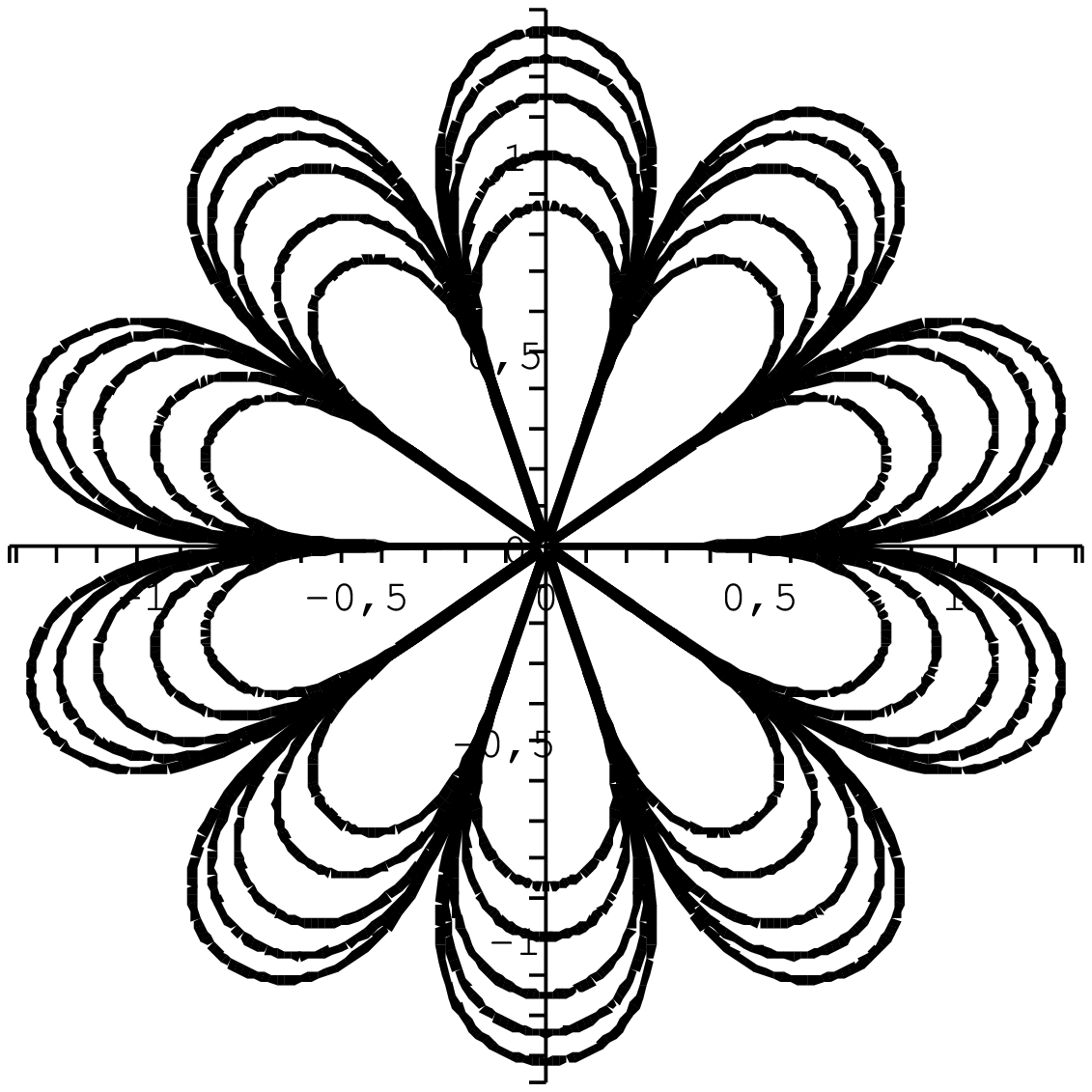}
    \nopagebreak\par Fig.~\thefigure. Force lines of the electrostatic field in
    the neighborhood of the point 10-pole for $Q^{(5)}_e=1$, $Q^{(5)}_m=0$.}\par
%
%
\section{$h$-analytic field theory on the plane $\mathcal{H}$}

In this section we develop the hyperbolic version of the previous constructions from the
    complex plane. First, we do not relate the obtained $h$-analytic fields with
    the electromagnetic field. Moreover, we choose a manner of reasoning which is reverse
    to the one used in the previous Section, where all the constructions emerged from physics
    (electromagnetostatics) towards formulations and definitions; we shall consider
    the hyperbolic analogues of these formulations as primary definitions, and only
    consequently we shall talk about their physical interpretations.

Before directly passing to field theory, we shall briefly remind some basic knowledge regarding
    the double plane and the defined on it $h$-holomorphic functions.
%
%
\subsection{Double numbers}

The algebra of double numbers $\mathcal{H}_2$ can be defined by means of a pair of generators
    $\{1,j\}$ with the multiplication table:
\begin{equation}\label{alg2}\begin{array}{c|c|c} & 1& j\\ \hline1& 1 & j\\ \hline j& j& 1\end{array}.\end{equation}
The elements of $\mathcal{H}_2$ will be written in the form: $\mathcal{H}_2\ni h=1\cdot t+jx$,
    where $t,x\in\R$, having in view further applications applications of this algebra
    for describing the 2-dimensional Space-Time%
\footnote{In the following, the speed of light will be assumed to be equal to 1}..

The particularity of the algebra of double numbers endowed with the multiplication table (\ref{alg2})
    relies on the fact, that this does not determine a numeric field, since it contains zero-divisors,
 i.e., the equation $h_1h_2=0$ may be satisfied by nonzero elements $h_1$ and $h_2$.
This is one of the reasons for which double numbers were not widely used in applications, as the
    complex numbers did. But exactly this feature shows in algebraic terms a very important
    circumstance of the 2-dimensional Space-Time -- the occurrence of light-cones.
The involutive operation {\it of complex conjugation} for double numbers is defined by analogy to
    the complex case: $h=t+jx\mapsto \bar h=t-jx$.
Geometrically, this operation describes
    the reflection of the hyperbolic plane with respect to the axis $\text{Im}\, h=0$.
    Like in the complex case, the couple $(h,\bar h)$ can be regarded as independent double
    coordinates on the hyperbolic plane, which relate to the Cartesian coordinates by means
    of the formulas (\ref{form1}) in which we replace $z,\bar z\to h,\bar h$.

The transition to hyperbolic polar coordinates and to the exponential form of representing
    double numbers exhibit a series of special features, which are absent in the case of
    complex numbers. The pair of lines $t\pm x=0$ which contain the subset of double numbers
    with zero squared norm\footnote{Strictly speaking, the occurrence of zero-divisors and
    the possibility of having negative values for the expression $h\bar h$ does not
    allow us to speak about the norm of a double number in its rigorous meaning.
For the brevity of terms and for preserving the partial analogy to the complex numbers,
    we shall call the quantity $\sqrt{|h\bar h|}$ {\em the norm} or {\em the module}
    of the double number (see further formula \ref{regih}).
The quantity $h\bar h$ will be called quadratic scalar of second order}. with vanishing
    quadratic scalar, split the whole hyperbolic plane into four quadrant-type domains,
    which are represented on the drawing by the numbers I, II, III and IV (Fig.\ref{psii}).
\begin{figure}[htb]\centering \unitlength=0.50mm \special{em:linewidth 0.4pt}
    \linethickness{0.4pt} \footnotesize \unitlength=0.70mm
    \special{em:linewidth 0.4pt} \linethickness{0.4pt}
    \input{pugol.pic}
    \caption{\small The domain $\R\sqcup-\R\sqcup\R\sqcup-\R$ of the change of the angle
    $\psi$ on the plane $\H$. The orientation is synchronized in opposed quadrants
    and is opposed in the neighboring ones. For different angles in different quadrants
    one can enumerate the angle $\psi$ by using the index $k$: $\psi_k$, $(k=1,2,3,4)$. }
    \label{psii}
\end{figure}

One can immediately verify that in each of the mentioned areas, the double numbers allow
    a hyperbolic trigonometric representation of the form:
\begin{equation}\label{trigh}h=t+jx=\epsilon\varrho(\cosh\psi+j\sinh\psi),\end{equation}
where for each quadrant take place the following definitions of quantities:
\begin{equation}\label{regih}\begin{array}{rlcc}
    \text{I}:&\epsilon=1,& \varrho=\sqrt{t^2-x^2},\quad \psi=\text{Arth}(x/t);\\
    \text{II}:&\epsilon=j,& \varrho=\sqrt{x^2-t^2},\quad \psi=\text{Arth}(t/x);\\
    \text{III}:&\epsilon=-1,& \varrho=\sqrt{t^2-x^2},\quad \psi=\text{Arth}(x/t);\\
    \text{IV}:&\epsilon=-j,& \varrho=\sqrt{x^2-t^2},\quad \psi=\text{Arcth}(t/x).
    \end{array}\end{equation}
The quantities $\varrho$ and $\psi$, defined in each of the quadrants by the formulas
    (\ref{trigh}), will be called {\em the module} and respectively {\em the argument}
    of the double number $h$.
In this way, in each of the quadrants we have $0\le\varrho<\infty$, and the
    quadrants themselves are parametrized by different copies of real lines,
    which together determine {\em the manifold $\Psi$ of angular variables}
    as a oriented disjoint sum $\R\sqcup -\R\sqcup \R\sqcup -\R$.
    Moreover, the manifold $\Psi$ can be suggestively represented by compactifying
    each copy of $\R$ into an open interval and further by gluing the intervals at their ends
    to obtain a circle with four pinched points.

We note, that the set of double numbers of zero norm is not described in any of the
    coordinate charts introduced above of the hyperbolic polar coordinate system.
    In the following, we shall call the subset of double numbers of the form
\begin{equation}\label{con}h_0+h_1(1\pm j),\end{equation}
(where $h_0,h_1$ are real double numbers) as {\em the cone of the number $h_0$}
    and will be denoted as $\text{Con}(h_0)$. All the points which lie in $\text{Con}(h_0)$,
    have their hyperbolic distance to the point $h_0$ equal to zero.
The hyperbolic Euler formula: $\cosh\psi+j\sinh\psi=e^{j\psi}$ can be verified by
    expanding the left and right sides into formal Maclaurin series, and comparing
    their real and imaginary parts. The hyperbolic Euler formula leads to the exponential
    representation of double numbers:
\begin{equation}\label{exph}h=t+jx=\epsilon\varrho e^{j\psi}=\epsilon e^{\Theta},\end{equation}
where in the last equality we passed to the "hyperbolic complex angle"
\begin{equation}\label{lnh}\Theta=\ln\varrho+j\psi\equiv\ln h.\end{equation}
Here, the product of a pair of double numbers reduces to adding their hyperbolic (complex) angles
    and the product of the sign factors $\epsilon$.
%
%
\subsection{$h$-holomorphic functions of double variable: analytic interpretation}\label{holooo}

The function $\ln h$, defined by formula (\ref{lnh}), is a simple
and important sample of the class of so-called {\em $h$-holomorphic
mappings of double variable}, whose definition emerges from
considerations similar to those which yield the definition of
analytic functions of complex variable. Any smooth mapping
$f:\R^2\to\R^2$ can be represented by a pair of real components
(\ref{mapc}), and we can pass to its representation by means of a
pair of double variables $\{h,\bar h\}$, as follows:
\begin{equation}\label{mapch}(h,\bar h)\mapsto(h',\bar h'):\ \
    h'=F_1(h,\bar h);\quad \bar h'=F_2(h,\bar h).\end{equation}
If $\R^2$ is regarded now as the plane $\H$ of two variables, then
    we can naturally limit ourselves to the mappings which preserve
    the hyperbolic complex structure of the plane, i.e., such
    mappings of the form $h\in\H \to s=F(h)\in \H$.
The differentiable functions $\R^2\to\R^2$, which satisfy
    the condition:%
\footnote{The notion of derivative of a function
    $F(h,\bar h)$ relative to its arguments is similar to the
    one used in real analysis. Namely, we can define the
    differentiability of a function $F$ at the point $(h,\bar h)$
    as the following property of its variation:
    $\Delta F=A(h,\bar h)\,\Delta h+B(h,\bar h)\,
    \Delta\bar h+o(\|\Delta h\|_{\mathcal{H}})$, where $\|\Delta h\|_{\mathcal{H}}
    \equiv[\Delta t^2-\Delta x^2]^{1/2}$ is the pseudo-Euclidean
    norm of the variation of the variable.
Passing to different limits for $\|\Delta h\|_{\mathcal{H}}\to 0$, we get the definition
    of partial derivatives or "directional derivatives". The problem, for such a definition,
    emerges only in the case of derivatives along the components of the cone:
    $\partial_{\text{Con}}=\partial_{\lambda(1\pm j)}$. We shall not examine in detail this question
    in the present paper}.
We shall call {\em $h$-holomorphic mappings of double variable $h$}
    the functions $\R^2\to\R^2$, which satisfy the condition:
%
%
\begin{equation}\label{anh}F_{,\bar h}=0.\end{equation}
We shall call {\em anti-holomorphic} mappings of double variable, the functions
    which satisfy the condition:
\begin{equation}\label{aanh}F_{,h}=0.\end{equation}
By analogy to holomorphic functions of complex variable, the
    holomorphic functions of double variable can be defined by formal
    power series, whose convergence often follows from the convergence
    of the corresponding real series.
The following statement holds true: {\it every $h$-holomorphic or $h$-antiholomorphic
    function of double variable maps zero-divisors into zero-divisors}.
Formally, this property is expressed by the equality:
\[F(\text{Con}_h)=\text{Con}_{F(h)},\]
for all the points $h$ from the holomorphicity domain of the function $F$.
    The proof is given in \cite{7}.

Considering the hyperbolic analogues of the formulas (\ref{diffc}):
\begin{equation}\label{diffh}\frac{\partial}{\partial h}=\frac{1}{2}\left(\frac{\partial}{\partial t}+
    j\frac{\partial}{\partial x}\right);\quad \frac{\partial}{\partial \bar h}
    =\frac{1}{2}\left(\frac{\partial}{\partial t}-j\frac{\partial}{\partial x}\right),\end{equation}
the condition (\ref{anh}) for the $h$-holomorphic function $F=U+jV$, written in Cartesian
    coordinates, gets the form:
\[F_{,\bar h}=(U+jV)_{,\bar h}=\frac{1}{2}[(U+jV)_{,t}-j(U+jV)_{,x}]=U_{,t}-V_{,x}+j(V_{,t}-U_{,x})]=0,\]
whence there follow {\it the Cauchy-Riemann analyticity conditions:}
\begin{equation}\label{crh}U_{,t}=V_{,x};\quad U_{,x}=V_{,t}.\end{equation}
%
It is easy to check that from the conditions (\ref{crh}), it follows
    the hyperbolic harmonicity of the real and of the imaginary parts of
    the analytic function $F$, which is expressed by the equations:
\begin{equation}\label{harmh}\Box U=\Box V=0,\end{equation}
where
\begin{equation}\label{hlap}\Box\equiv 4\partial_h\partial_{\bar h}=\partial^2_{t}-\partial^2_x\end{equation}
is the wave operator of order two -- the d'Alambertian ("the hyperbolic Laplacian").
%
%
\subsection{$h$-holomorphic functions of double variable: topological interpretation}

The curvilinear integrals of functions defined on the plane of double variable reduce to
    a pair of standard integrals of 1-forms on the Cartesian plane, and hence we do not
    thoroughly insist on their definition.
The formal proof of the hyperbolic Cauchy theorem is almost identical to the corresponding proof
    of the formula (\ref{coshc1}), and like in the complex case, it looks more compact in complex terms%
\footnote{Indeed, it remains the problem regarding the definition of the derivative $F_{,\bar h}$
    along the direction of the cone. If the contour is convex and curved, then the number of points for which
    the vector which is tangent to the contour lies on the cone is be finite or at most countable.
By excluding ("pinching") such points from the contour and by computing the integrals over the remaining
    non-connected pieces in the sense of their principal values, we are led to the definition
    of the integral in some improper sense. We shall not insist in this paper on the details of such definitions}.:
\[\oint\limits_{\Gamma}F(h)\, dh=\int\limits_{\Sigma}F_{\bar h}\,d\bar h\wedge dh=0\]
due to (\ref{anh}). From purely topological observation, analogous to the observations
    from the complex plane case, the integral of a holomorphic function will vanish
    as well on the border of a multi-connected domain.

\begin{figure}[htb]\centering\input{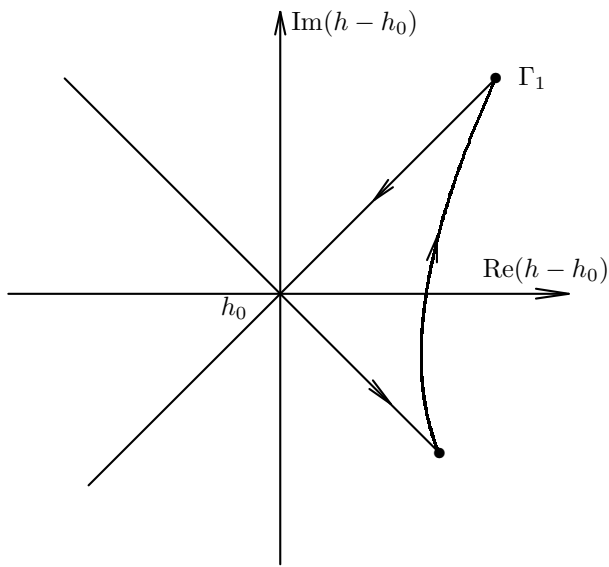}
    \caption{\small Towards deducing the integral Cauchy theorem on the plane of
    double variable}.\label{cont3}\end{figure}

The integral Cauchy formula, which relies on the following hyperbolic analogue of
    formula (\ref{form4}) is (\cite{7}):
\begin{equation}\label{rezt}\oint\limits_{\Gamma}(h-h_0)^\alpha\, dh= \left\{\begin{array}{lr}
    0,&\alpha\neq-1;\\j\ell_H,&\alpha=-1,\end{array}\right.\end{equation}
where $\ell_H=\int\limits_{-\infty}^{+\infty}d\psi$ is the dimension of the space of hyperbolic angles
    ("the fundamental constant"\, of hyperbolic geometry, analogous to the constant $2\pi$ on the
    Euclidean plane), has the form
\begin{equation}\label{coshh1}\oint\limits_{\Gamma}\frac{F(h)}{h-h_0}\, dh=0.\end{equation}
for the contour of the type shown in Fig.\ref{contt}.
\begin{figure}[htb]\centering\input{pcontt.pic}
    \caption{\small Towards deducing the integral Cauchy theorem on the plane of
    double variable}.\label{contt}\end{figure}
and the form
\begin{equation}\label{coshh4}F(h_0)=\frac{1}{\ell_Hj}\oint\limits_{\Gamma_0}\frac{F(h)}{h-h_0}\,dh,\end{equation}
for the contour of the type shown in Fig.\ref{cont3}.
%
%
\subsection{$h$-holomorphic functions of double variable: geometric interpretation}

We examine the quadratic form
\begin{equation}\label{quadh}\Theta=\text{Re}(dh\otimes d\bar h)=dt\otimes dt-dx\otimes dx.\end{equation}
By analogy to the complex case, we ca say that the algebra of double numbers on the plane $\R^2$
    algebraically induces a pseudo-Euclidean metric.
Relative to the mapping given by the holomorphic function $F(h)$, the form $\Theta$
    behaves as follows:
\begin{equation}\label{Bch}\Theta\mapsto \Theta'=|F'(h)|^2\Theta,\end{equation}
where $F'(h)=dF/dh$. The formula (\ref{Bch}) means that the function $F(h)$ in its holomorphicity domain
    and at the points, where $|F'(h)|^2\neq0$ provides a conformal mapping of the complex plane
    into itself, i.e., it preserves the hyperbolic angles.
This is tightly related to the previously fact that the cones $\text{Con}$ are invariant relative to the
    $h$-holomorphic mappings.
We note that $|F'|^2=|\triangledown U|^2=|\triangledown V|^2=\Delta_F$, where
    $\triangledown$ is the gradient operator relative to the pseudoEuclidean metric,
    and $\Delta_F$ is the Jacobian of the mapping $F$, regarded as a mapping $\R^2\to\R^2$.
As it follows from condition (\ref{crh}) or from the conformality property, the lines $U=\text{const}$
    and the lines $V=\text{const}$, for any holomorphic function $F(h)$, determine on the plane
    $\mathcal{H}$ an orthogonal (in hyperbolic sense) curvilinear system of coordinates, since
    at each point one has the equality:
\begin{equation}\label{ortoh}\triangledown U\cdot\triangledown V=U_{,t}V_{,t}-U_{,x}V_{,x}
    =U_{,t}U_{,x}-U_{,x}U_{,t}=0.\end{equation}
By analogy to the complex case, this clarifies the geometric meaning of the relation of
    {\it hyperbolic conjugation} of a pair of functions $U$ and $V$, which are the real and the
    imaginary parts of some $h$-holomorphic function $F(h)$: {\it the conjugation of functions
    provides mutually-orthogonal level-lines and equal norms of the gradients, at each point.}

We note that the set of conformal mappings of the pseudo-Euclidean metric, unlike
    the Euclidean case, is not limited to $h$-holomorphic functions alone.
    To this question we come back in Section \ref{confo}.
%
%
\subsection{Properties of elementary functions of double variable}\label{elemmm}

Since the properties of the hyperbolic analogues of elementary functions are
    quite poorly reflected in literature, it makes sense to insist on several such functions.
    A more complete account on elementary functions of double variable will be provided in \cite{7}.
%
%
\subsection{Power functions $F(h)=h^n$}

Unlike the power function of complex variable, the cases of even powers $n$ and odd powers $n$
    are essentially different.
Indeed, when passing to the exponential representation (\ref{exph}), we get:
\begin{equation}\label{step1}h=\epsilon\varrho e^{j\psi}\mapsto \epsilon^n\varrho e^{jn\psi}\end{equation}
Since for any even $n$ we have $\epsilon^n=1$, we conclude that {\em the power mapping
    $h\mapsto h^n$, for $n=2k$, $k\in\Z$ bijectively maps each of the quadrants I, II, III, IV
    onto quadrant I while mapping the cones $\text{Con}_\pm\to\text{Con}_{\pm}$.}
On the contrary, {\it for odd $n$, each of the coordinate quadrants is bijectively
    mapped into itself via the mapping $h\mapsto h^n$ $n=2k+1$, $k\in\mathbb{Z}$}.
As it is easy to see from (\ref{step1}), the coordinate net of lines $\varrho=\text{const}$,
    $\psi=\text{const}$ is mapped to the coordinate net of lines
    $\varrho'=\varrho^n=\text{const}$, $\psi'=n\psi=\text{const}$ for each integer value $n$.
    In the case of positive integers $n$, the radial lines expand for $\varrho>1$ and
    shrink for $\varrho<1$. Moreover, they rotate from the value $\psi=0$ in the sense related
    to the corresponding sign of the cone components.
For integer negative $n$ there takes place a supplementary inversion relative to the unit
    spheres $\varrho=1$ and an inversion of the space of angles $\Psi\to-\Psi$.
As an example of function with even $n$ we examine the function
    $w=h^2=x^2+y^2+2jxy=\varrho^2e^{2\psi}$.

\begin{figure}[htb]\centering\input{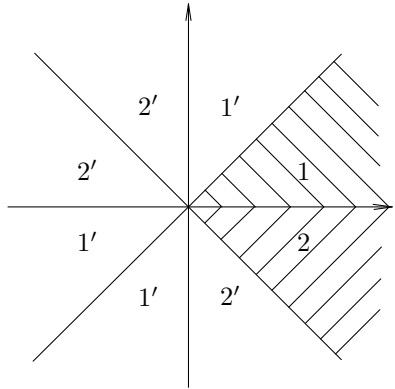}
    \caption{\small The global structure of the mapping $h\mapsto h^2$.}\label{quaddm}\end{figure}

On Fig.\ref{quaddm} there is represented the global structure of the mapping $h\mapsto h^2$:
    the quadrant 1-2 is mapped into itself (its borders map to the corresponding ones),
    and the mapping of the other quadrants into quadrant 1-2 is shown by the corresponding
    figures (accented figures which identify a quadrant, show how the corresponding quadrant
    maps to the quadrant 1-2). In this way, the mapping $h\mapsto h^2$ is 4-fold.
A similar situation occurs with the mapping: $h\to h^{2k}$ $k\in\Z$.

We present in Figs.\ref{mapindex1}-\ref{mapindex2} an illustrative representation of several
    simple power mappings.

{\centering\small\refstepcounter{figure}\label{mapindex1}
    \includegraphics[width=.5\textwidth]{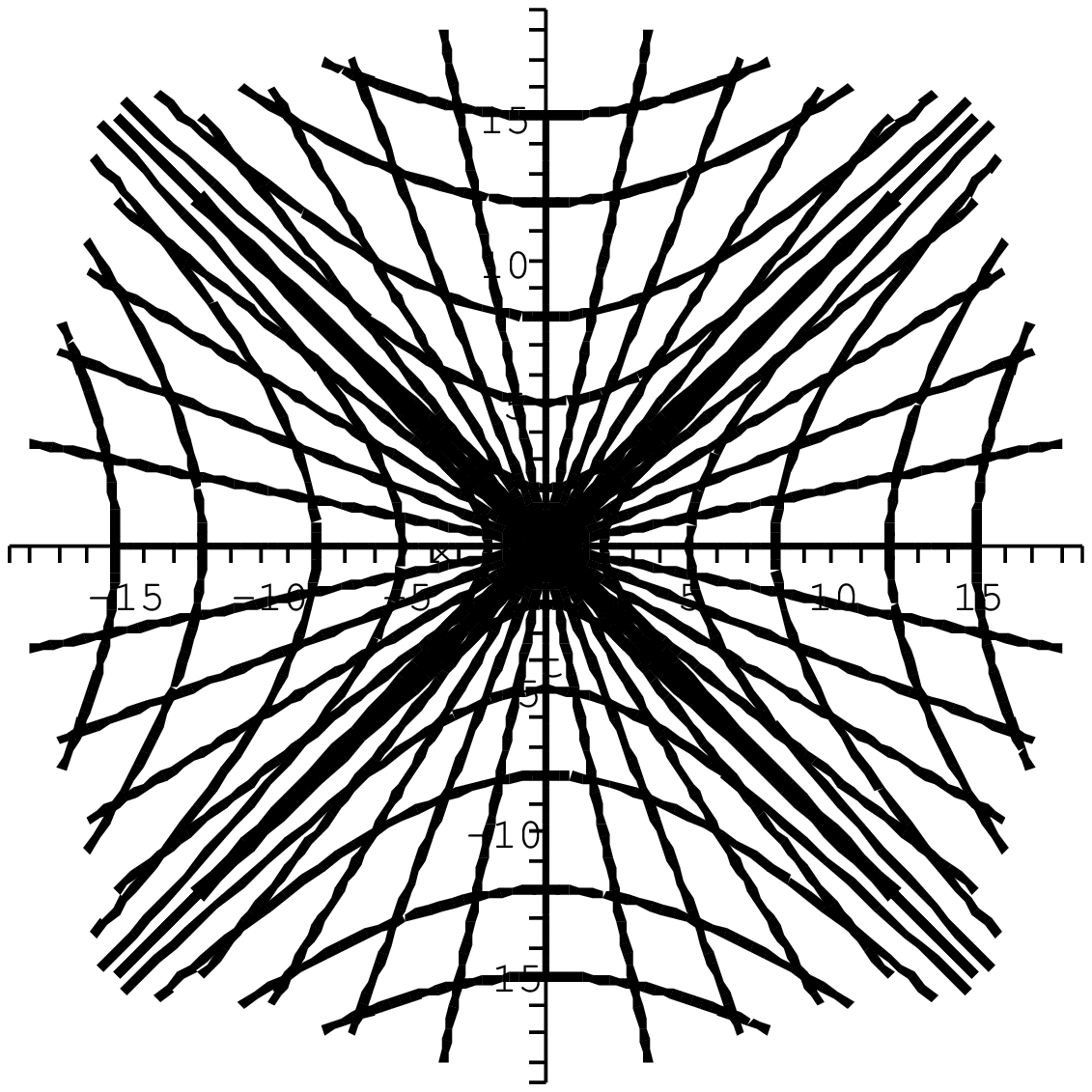}\includegraphics[width=.5\textwidth]{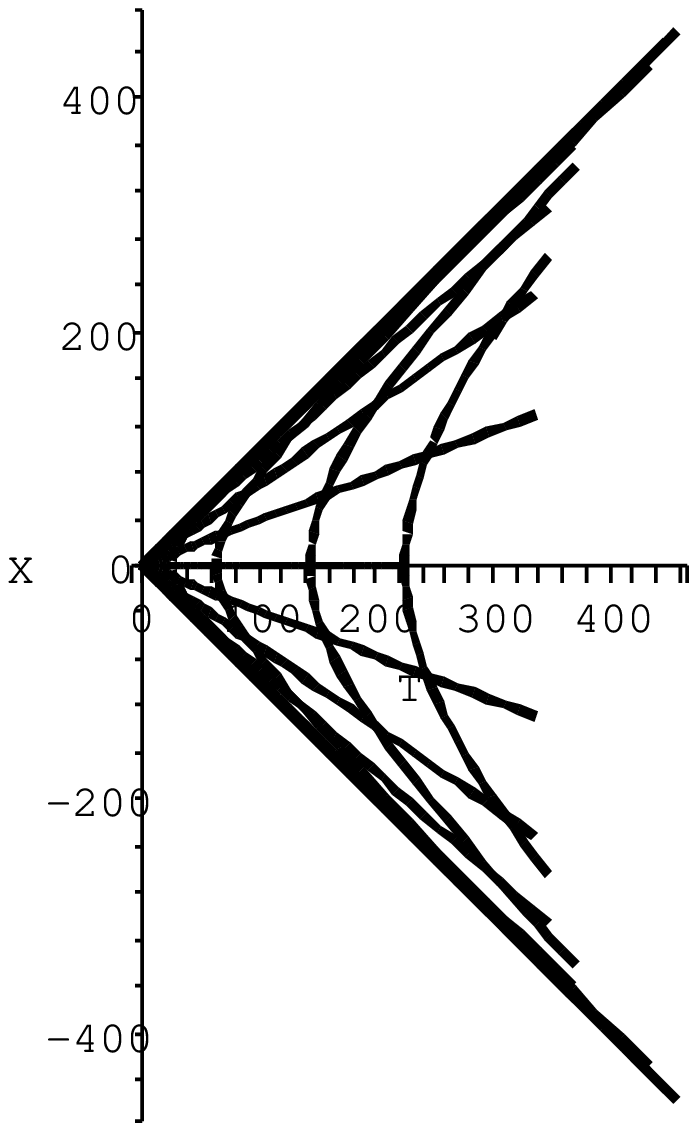}
    \medskip\nopagebreak\\ Fig.~\thefigure. The hyperbolic polar system of coordinates (left)
    and the image of its first quadrant under the mapping $h\mapsto h^2$. }

{\centering\small\refstepcounter{figure}\label{mapindex2}
    \includegraphics[width=.5\textwidth]{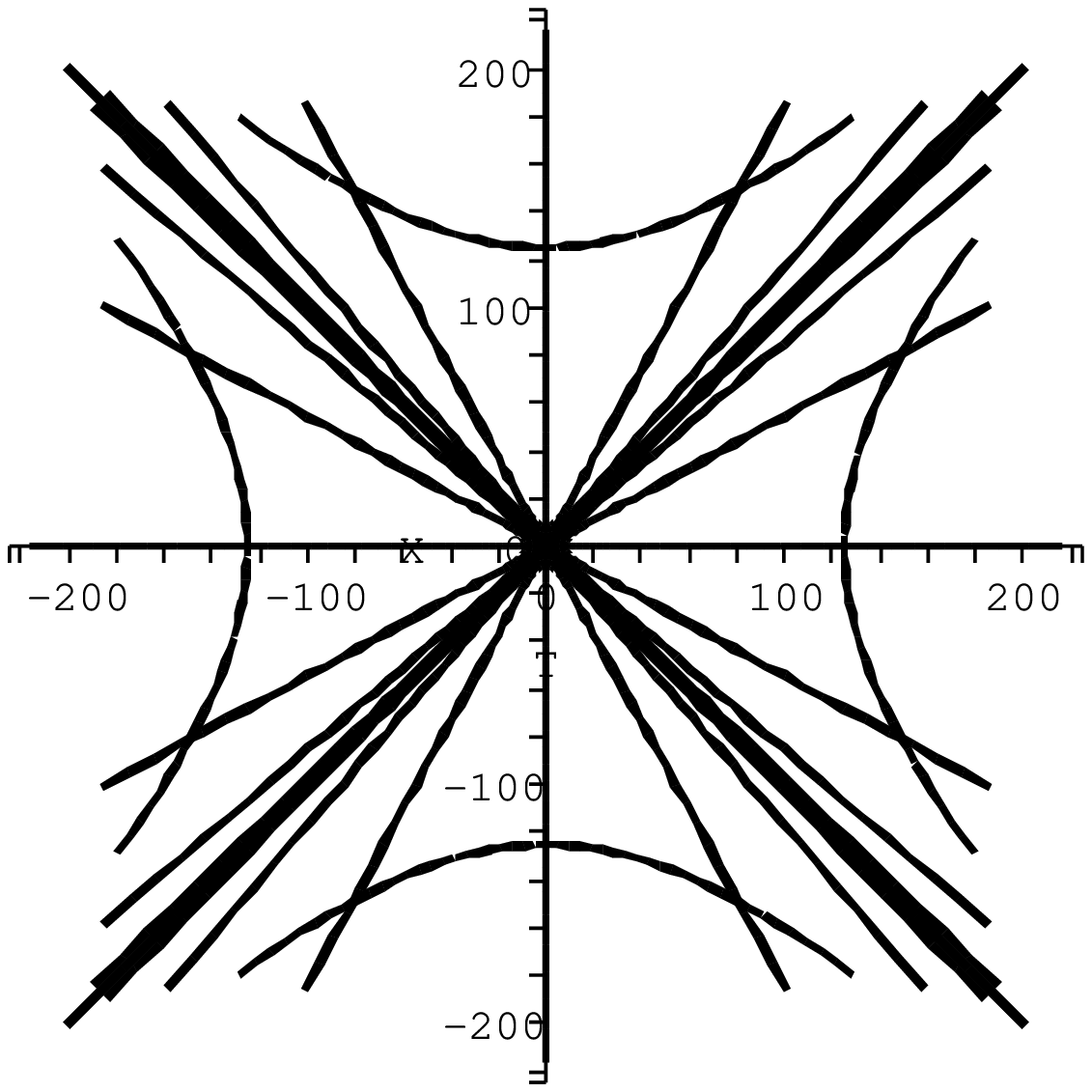}\includegraphics[width=.5\textwidth]{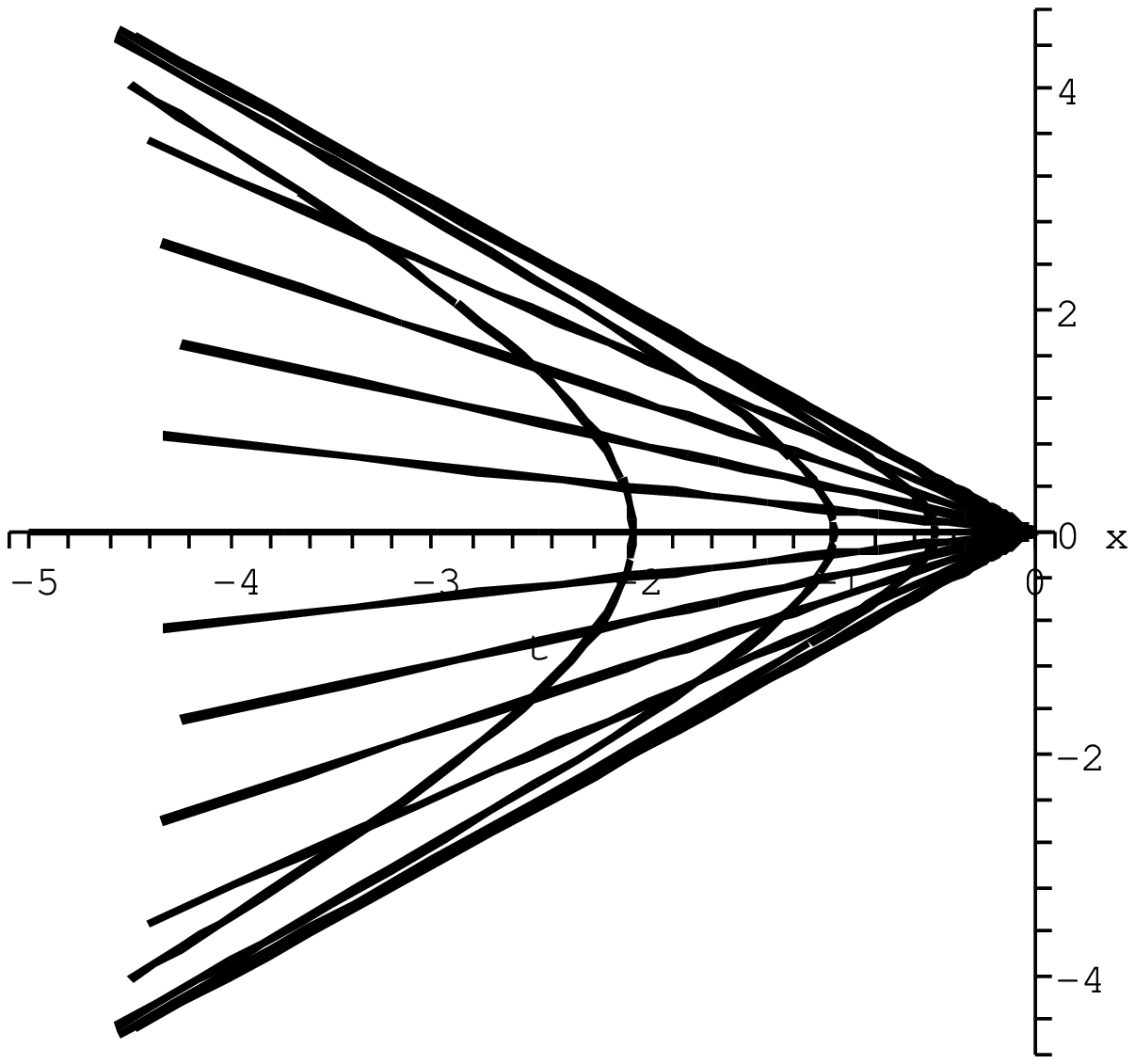}
    \medskip\nopagebreak\\ Fig.~\thefigure. The image of the polar coordinate system
    (Fig.\ref{mapindex1} at left) under the mapping $h\mapsto h^3$ (at left) and the image
    of the first quadrant under the mapping $h\mapsto h^{-1}$ (at right). }\par\medskip

\begin{figure}[htb]\centering\input{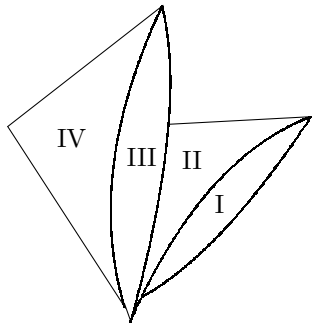}
    \caption{\small The hyperbolic Riemannian surface of the 4-valued mapping $h\mapsto h^{1/2k}$,
    $k\in\Z$}.\label{riemind}
\end{figure}

From the properties of power functions it is easy to derive the properties of roots of different
    orders and of the rational powers $h\mapsto h^{1/n}$ and $h\mapsto h^{m/n}$.
Each root $\sqrt[n]{h}$ of even order is defined in quadrant I. Such a root is a 4-valued
    function.
Each leaf of the hyperbolic Riemannian surface of this function represents a unit copy of the
    first quadrant I, as shown on Fig.\ref{quaddm}. On each leaf, the mapping is bijective.
All the leaves can be glued together into a Riemannian surface, which represents $\R^2$,
    with the point $(0;0)$ belonging to all the leaves and being a hyperbolic analogue of
    the branching point. On the Riemannian surface the roots of even orders can be illustrated
    by means of a sheet of paper, folded as shown on Fig.\ref{riemind}.

The roots of odd order are 1-valued in each of the 4 quadrants.
%
%
\subsection{The exponential of double variable $w=e^h$}\label{exppp}

The relations $e^h=e^{t+jx}=e^te^{jx}$ naturally lead to the global structure of the
    exponential mapping, which is represented in Fig.\ref{exp}.

{\centering\small\refstepcounter{figure}\label{exp}
    \includegraphics[width=.4\textwidth]{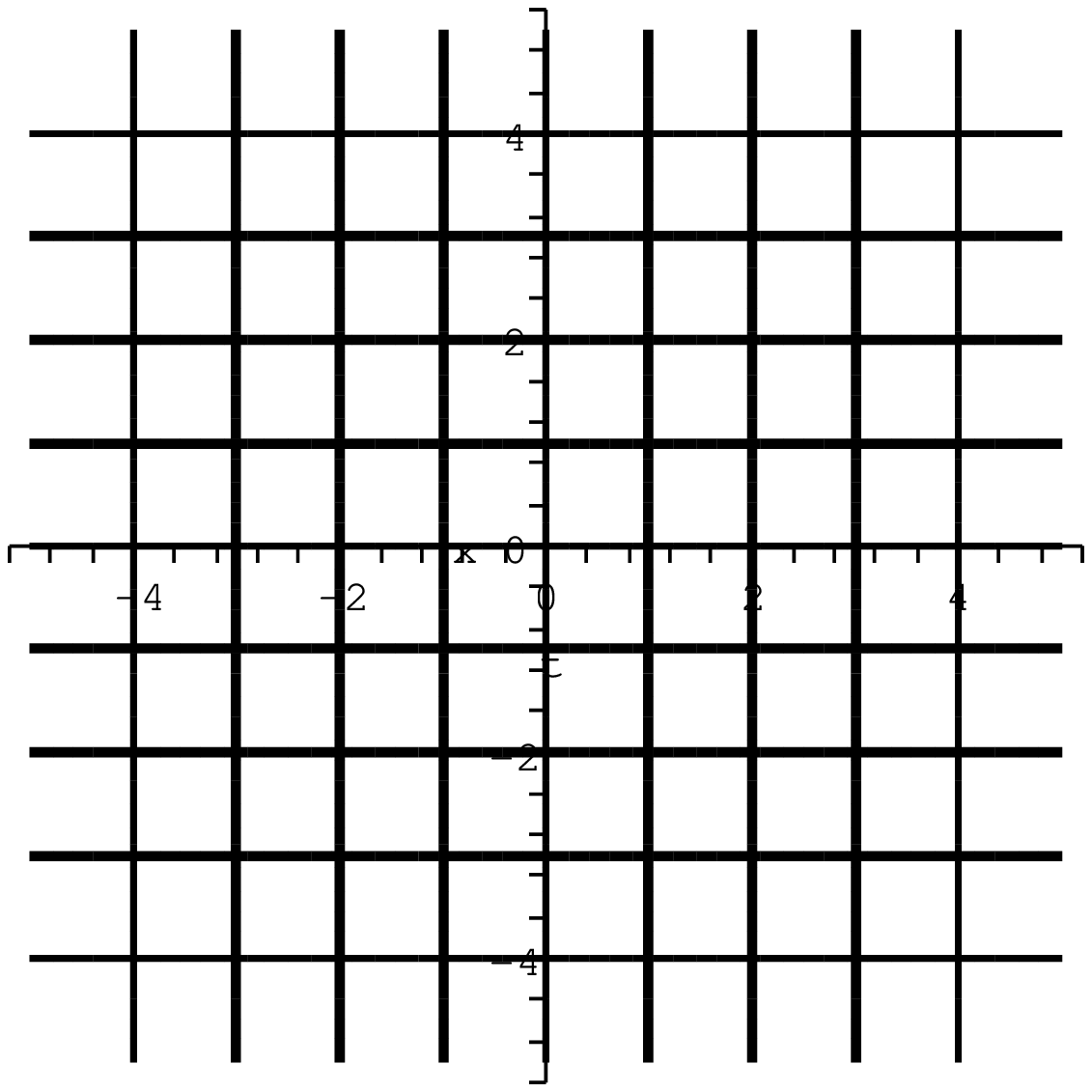}\includegraphics[width=.4\textwidth]{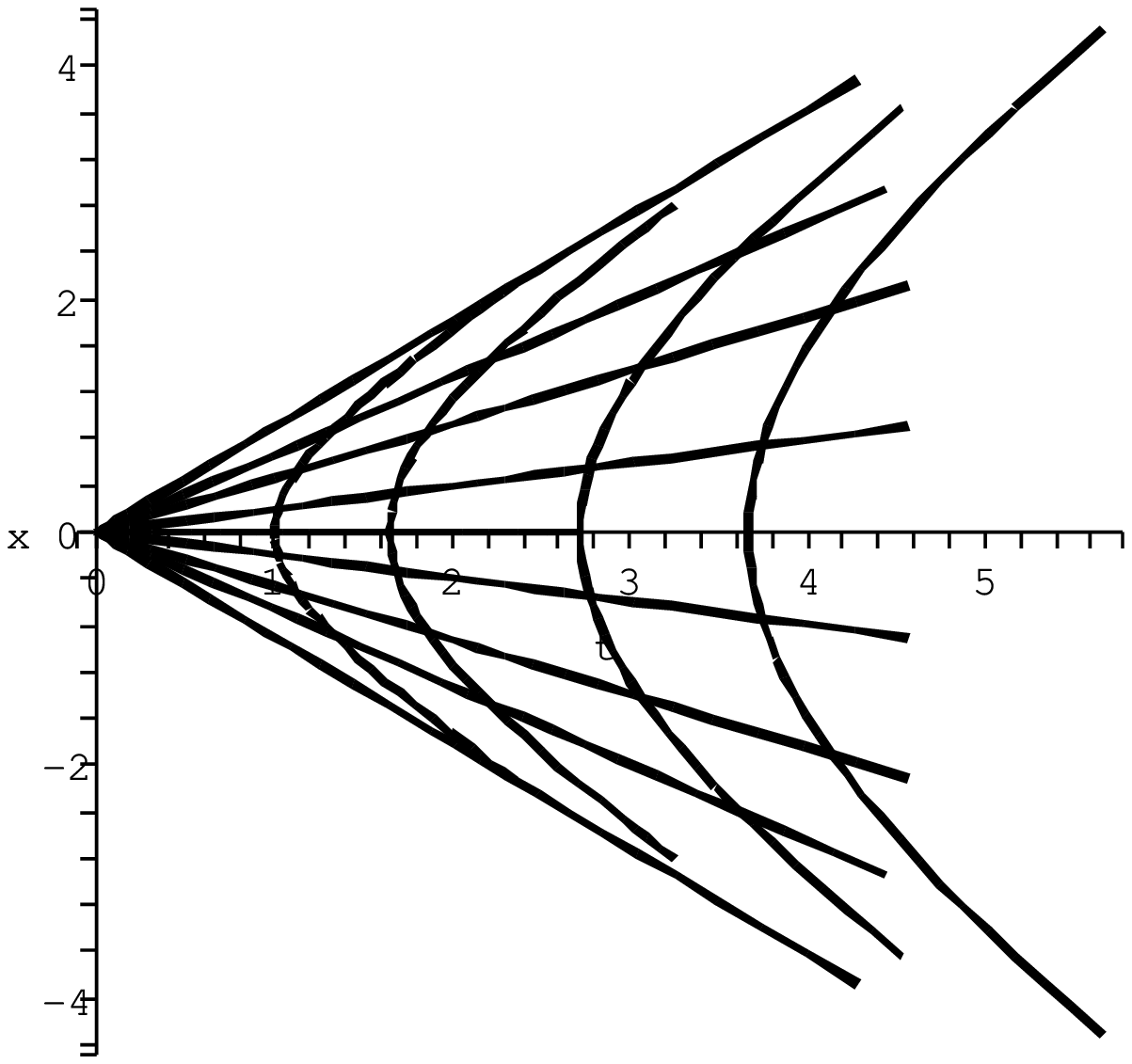}
    \medskip\nopagebreak\\ Fig.~\thefigure. The global structure of the mapping $h\mapsto e^h$}.\par\medskip

The rectangular pseudo-orthogonal net on the plane with variable $h$ is mapped by the
    exponential into the pseudo-orthogonal net, which consists of the rays and hyperbolas
    from the first quadrant and the vertex at the point $h=0$. The mapping $h\mapsto e^h$ is bijective.
Obviously, the inverse mapping $\ln h=\ln\varrho+j\psi$ is defined inside the first quadrant.
    On its border (i.e., on the cone $\text{Con}^{\uparrow}(0)$) the polar coordinate system is
    incorrect and we need supplementary investigation regarding the behavior of the mapping
    $h\mapsto e^h$, which we shall not address here.
%
%
\subsubsection{The hyperbolic Zhukowskiy function: $h\mapsto (h+h^{-1})/2$}

\begin{figure}[htb]\centering\input{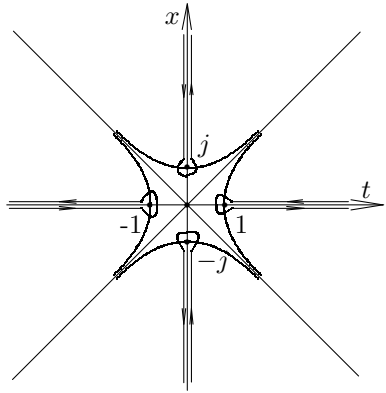}
    \caption{\small The action of the Zhukowskiy hyperbolic function}.\label{zhuk}
\end{figure}

We shall examine, at last, the hyperbolic version of the Zhukowskiy function:
\[Z(h)\equiv\frac{1}{2}\left(h+\frac{1}{h}\right).\]
This transformation maps the unit hyperbolic circle $\text{HS}^1(0)$ into pieces of the coordinate axes,
    enveloping them twice (see Fig.\ref{zhuk}).

Like in the case of complex variable, the Zhukowskiy mapping has two leaves:
    the outer part of the unit circle and its inner side, which it maps onto the double plane.
    At points $\pm1,\pm j$ the conformality of functions $Z(h)$ is broken, since at these points
    we have $Z'(h)=0$. The shape of Zhukowskiy hyperbolic function is shown in Fig.\ref{zhh}.

{\centering\small\refstepcounter{figure}\label{zhh}
    \includegraphics[width=.5\textwidth]{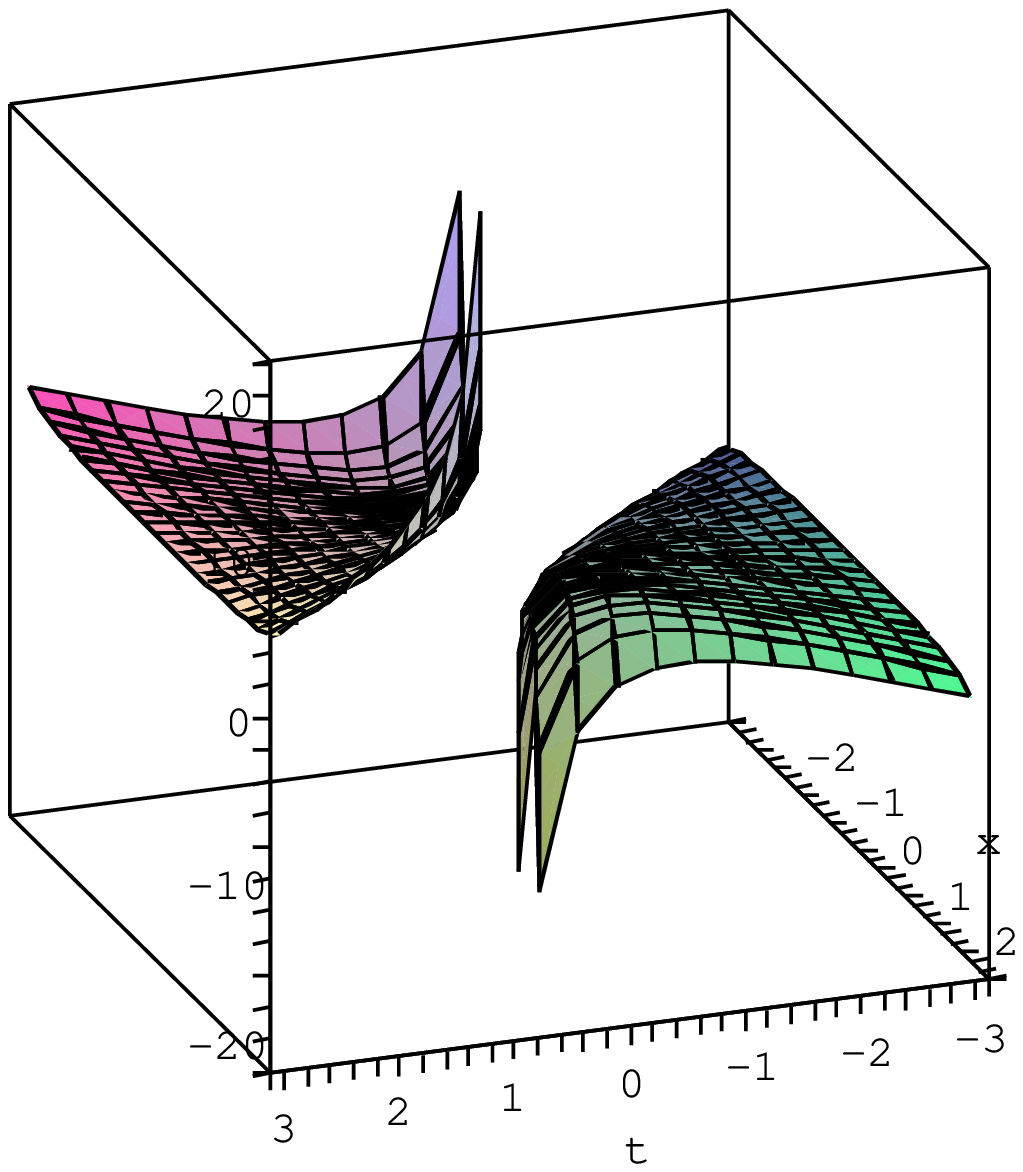}\includegraphics[width=.5\textwidth]{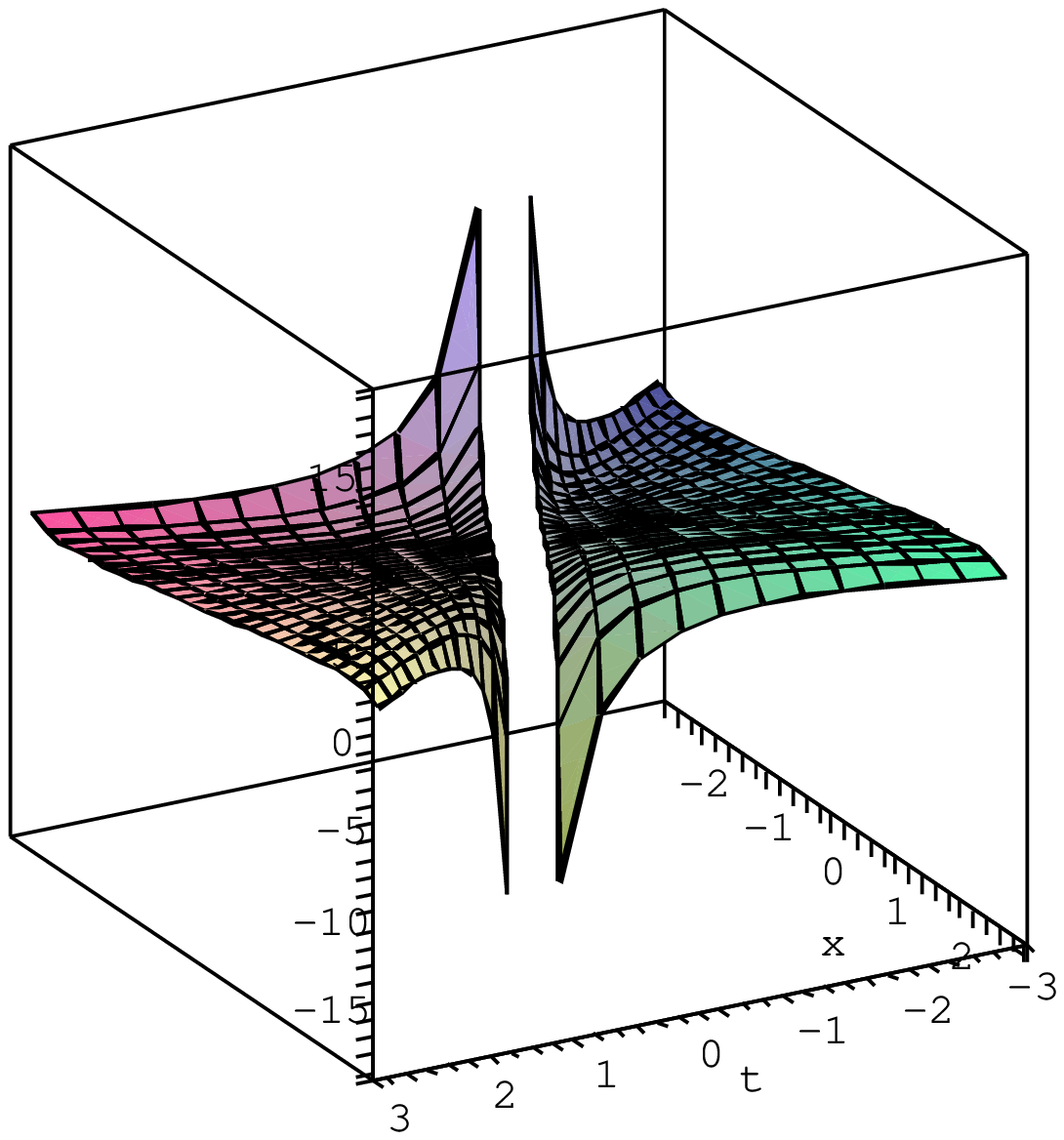}
    \medskip\nopagebreak\\ Fig.~\thefigure. The structure of the mapping $h\mapsto (h+h^{-1})/2$.
    (real and imaginary parts)}.\par\bigskip

\begin{figure}[htb]\centering\input{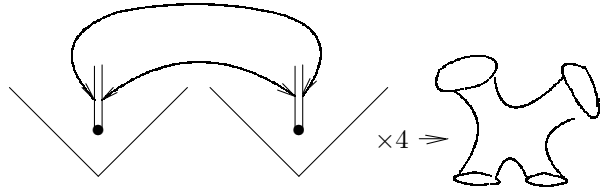}
    \caption{\small The Riemannian surface of the inverse Zhukowskiy function $Z^{-1}$.
    At left is shown the rule of gluing the two quadrants along the cut given by the coordinate axis.
    Four copies of such pieces are glued together in the plane which is topologically
    equivalent to the one represented at the right side}.\label{invzh}
\end{figure}

The Riemannian surface of the double function $Z^{-1}(h)$ is represented in Fig.\ref{invzh}.

%
%
\subsection{$h$-holomorphic functions of double variable: physical interpretation}\label{physh}

We shall be regarding further any $h$-holomorphic function $F(h)=U+jV$ as a complex
    $h$-potential of some 2-dimensional vector field ({\it $h$-pole}) on the plane of double variable.
We shall associate the real part $U$ of this function with the potential of the field,
    ({\it the $h$-potential function}), and the imaginary part $V$ with {\it the force function}
    of this field.
In other words, like in the complex case, the lines $U=\text{const}$ are equi-potential lines
    of the $h$-field, and the lines $V=\text{const}$ coincide with the strength lines of this field
These families of lines are mutually orthogonal (formula (\ref{ortoh})), and each of the
    functions $U$ and $V$ satisfies the wave equation (\ref{harmh}) due to the hyperbolic
    Cauchy-Riemann conditions.

We define the strength $\mathcal{E}$ of the $h$-field by the formula:
\begin{equation}\label{voltageh}\mathcal{E}=\mathcal{E}_t+j\mathcal{E}_x
    =-\overline{\frac{dF}{dh}}=-\frac{d\bar F}{d\bar h}=-U_{,t}+jU_{,x},\end{equation}
which can be regarded as a double form of representing the gradient vector field of the function
    $U$ relative to the pseudo-Euclidean metric.
We obtain the formula (\ref{voltageh}) by taking into consideration the relations
    (\ref{diffh}) and the Cauchy-Riemann conditions (\ref{crh}).

From relation $\mathcal{E}=\mathcal{E}(\bar z)$ (the anti-holomorphicity of the strength),
    which results from the definition (\ref{voltageh}), by considering (\ref{diffh}),
    we get the following identity:
\begin{equation}\label{potsolh1}\frac{\partial \mathcal{E}}{\partial h
    }=\frac{1}{2}[\mathcal{E}_{t,t}+\mathcal{E}_{x,x}+j(\mathcal{E}_{t,x}+\mathcal{E}_{x,t})]=0,\end{equation}
which is equivalent to the two identities:
\begin{equation}\label{potsolh2}\text{divh}\, \mathcal{E}\equiv \mathcal{E}_{t,t}+\mathcal{E}_{x,x}=0;\quad
    \text{roth}\, \mathcal{E}\equiv \mathcal{E}_{t,x}+\mathcal{E}_{x,t}=0,\end{equation}
which accordingly express the {\it solenoidality} and the {\it $h$-potentiality} of the
    electrostatic field.
\footnote{We note that divergence of the vector field is identically defined in the complex and the
    hyperbolic cases, unlike the curl operation on vector fields, which ion the hyperbolic case
    includes a symmetric combination of partial derivatives }.
We note that the condition of $h$-potentiality results from the commutativity of the partial derivatives
    of a smooth scalar function, and the condition of solenoidality is equivalent to the wave equation
    $\Box U=0$, which is automatically satisfied, when the potential $U$ is the real part of some
    $h$-holomorphic function.

Like in the complex case, we shall examine now the integral
\begin{equation}\label{potcirch1}\Omega[\mathcal{E},\gamma]=\int\limits_{\gamma}\mathcal{E}\, d\bar h
    =\int\limits_{\gamma}\mathcal{E}_{t}\,dt-\mathcal{E}_{x}\, dx+j\int\limits_{\gamma}\mathcal{E}_{x}\,dt-
    \mathcal{E}_{t}\,dx=\Upsilon[\mathcal{E},\gamma]-j\Xi[\mathcal{E},\gamma]\end{equation}
along some path $\gamma$. Its real part $\Upsilon[\mathcal{E},\gamma]$ will be called
    {\it the circulation of the field $\mathcal{E}$ along the path $\gamma$}, and the quantity
    $\Xi[\mathcal{E},\gamma]$, which is opposed to the imaginary part, will be called
    {\it the current of the field $\mathcal{E}$ through the curve $\gamma$}.
Taking into consideration (\ref{voltageh}) and the hyperbolic Cauchy-Riemann conditions, we get for
    these quantities the following expression of the complex potential in terms of the
    variations of the components:
\begin{equation}\label{cpoth}\Upsilon[\mathcal{E},\gamma]=-\delta_\gamma U;\quad \Xi[\mathcal{E},
    \gamma]=-\delta_\gamma V,\end{equation}
which can be regarded as the illustrating physical meaning of the components of the
    complex $h$-potential $F(h)$.
%
%
\subsection{The field of the hyperbolic point-like source}\label{physh1}

\begin{figure}[htb]\centering\input{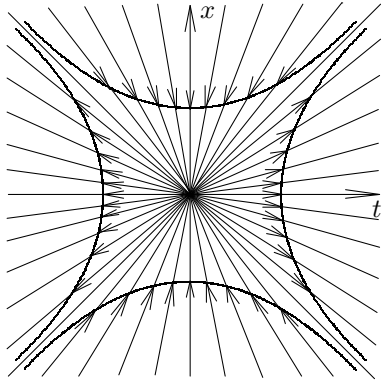}
    \caption{\small The sketch portrait of the force lines of the hyperbolic point source.
    The field has constant absolute value on hyperbolic circles (Euclidean hyperbolas).
    On the cone $\text{Con}_0$ the field is not defined, and in the neighboring quadrants
    it changes its character (positive or negative source)}.\label{hqull}
\end{figure}

We shall examine the $h$-potential of the form
\begin{equation}\label{hqulll}F(h)=-q\ln h,\end{equation}
which is clearly a hyperbolic generalization of the Coulomb potential (\ref{qul2}).
    The strength of the corresponding field is computed by the formula (\ref{voltageh}) and has the form:
\begin{equation}\label{hqul1}\mathcal{E}=\frac{q}{\bar h}
    =\frac{qh}{|h|^2}=q\left(\frac{t}{t^2-x^2}+j\frac{x}{t^2-x^2}\right).\end{equation}
A peculiar new feature is the difference between the definition domains of the formulas (\ref{hqulll})
    and (\ref{hqul1}): the first provides the $h$-potential only in the first quadrant,
    while the second is well defined in each of the four quadrants of the double plane.
The formal explanation os this feature relies on the fact, that the logarithm of a double
    number taken from the quadrants 2,3 or 4 can be represented in the form $\ln h+\ln\epsilon$,
    where $h$ is some double number from the first quadrant,  $\epsilon$ is one of the sign factors,
    defined in (\ref{regih}).
Here, $\ln\epsilon$ is some algebraic constant%
\footnote{We note that this does not belong to the algebra $\mathcal{H}_2$.}, which "disappears"
    by differentiation from the finite expression (\ref{hqul1}).
The force lines of the hyperbolic point source are radial lines with $\psi=\text{const}$, and the
     equi-potential lines are the hyperbolas $\varrho=\text{const}$.
The portrait of the force lines in each of the four quadrants are represented in Fig.\ref{hqull}.

\par\bigskip

From formulas (\ref{cpoth}) we get for the hyperbolic circulation and flow the expression:
\begin{equation}\label{potcylh1}\Upsilon[\mathcal{E},\mathcal{C}]=0;\quad
    \Xi[\mathcal{E},\mathcal{C}]=0\end{equation}
for the contour $\mathcal{C}$, which is represented in Fig.\ref{contt} or is homotopic to it,
    and the expression
\begin{equation}\label{potcylh2}\Upsilon[\mathcal{E},\mathcal{C}]=0;\quad
    \Xi[\mathcal{E},\mathcal{C}]=\ell_Hq\end{equation}
for the contour, which is represented in Fig.\ref{cont3} or is homotopic to it. The formulas
    (\ref{potcylh2}) express the hyperbolic potentiality and the hyperbolic
    Gauss theorem for the field $\mathcal{E}$.
%
%
\section{The $h$-dual interpretation}\label{physh2}

\begin{figure}[htb]\centering\input{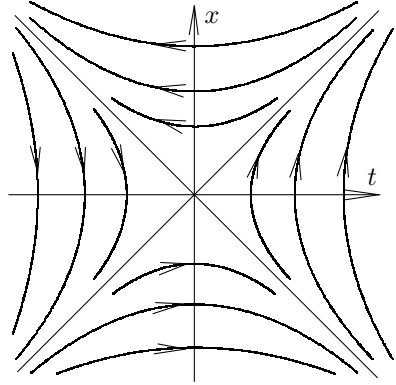}
    \caption{\small Force lines of the point vortex. The orientation of the lines is identical
    for all the four quadrants (anti-clockwise)}.\label{hcurl}\end{figure}

We obtain the dual interpretation of the point-like hyperbolic source by passing from the potential
    $F(h)$ from (\ref{hqulll}) to the potential $jF(h)$.
For the new dual field $\mathcal{B}$ we get the expression:
\begin{equation}\label{dualh}\mathcal{B}=j\frac{d\bar F}{d\bar h}=-\frac{qj}{\bar h}
    =-q\frac{x+jt}{t^2-x^2}.\end{equation}
The field $\mathcal{B}$ is a hyperbolic analogue of the point-like vortex.
    Its field lines are the hyperbolas shown on Fig.\ref{hcurl}.

\par\bigskip

Like in the complex case, the vector field $\mathcal{B}$, is orthogonal to the field $\mathcal{E}$
    relative to the pseudo-Euclidean metric and is obtained by "rotating"\, the field $\mathcal{E}$
    with the hyperbolic angle "{}$-\ell_H+\psi$,{}"\, where $\psi$ is the hyperbolic angle
    of the field $\mathcal{E}$.

From formulas (\ref{cpoth}) we get for the hyperbolic circulation and flow of the field
    $\mathcal{B}$ the expression:
\begin{equation}\label{potcylh3}\Upsilon[\mathcal{B},\mathcal{C}]=0;\quad
    \Xi[\mathcal{B},\mathcal{C}]=0\end{equation}
over the contour $\mathcal{C}$, which is represented in Fig.\ref{contt}, or is homotopic
    to it and the expression
\begin{equation}\label{potcylhh2}\Upsilon[\mathcal{B},\mathcal{C}]=\ell_Hq;\quad
    \Xi[\mathcal{B},\mathcal{C}]=0\end{equation}
for the contour, which is represented in Fig.\ref{cont3}, or is homotopic to it.
    The formulas (\ref{potcylhh2}) express the hyperbolic solenoidality and the hyperbolic
    analogue of the law of total fields, for the field $\mathcal{E}$.
%
%
\subsection{The $h$-vortex source}\label{physh3}

By analogy to the complex case, we can unify the two previous situations into a single one,
    by considering the concept of {\it hyperbolic vortex source} with complex
    charge $\mathcal{Q}=q-jm$. The potential gets the form:
\begin{equation}\label{vihh}F(z)=-\mathcal{Q}\ln h=-q\ln\varrho+m\psi-j(-m\ln\varrho+q\psi).\end{equation}
Such a potential can be more naturally interpreted in the framework of
    the dual-symmetric hyperbolic field theory, in which the hyperbolic
    electric and magnetic charges and currents co-exist on "equal terms".
The equation for the force lines of such a field is obtained from (\ref{vihh})
    by making equal to a constant the imaginary part:
\begin{equation}\label{vihh1}m\ln\varrho+q\psi=C \Leftrightarrow
    (t^2-x^2)e^{-(2q/m)\text{Arth}(x/t)}=C,\end{equation}
or, after several simple calculations:
\begin{equation}\label{vihh2}(t+x)^{1-\alpha}(t-x)^{1+\alpha}=\text{const},\end{equation}
where $\alpha=q/m$. The portrait of force lines for $\alpha=-2$ is shown in Fig.\ref{vihrrr}

{\centering\small\refstepcounter{figure}\label{vihrrr}
    \includegraphics[width=.5\textwidth]{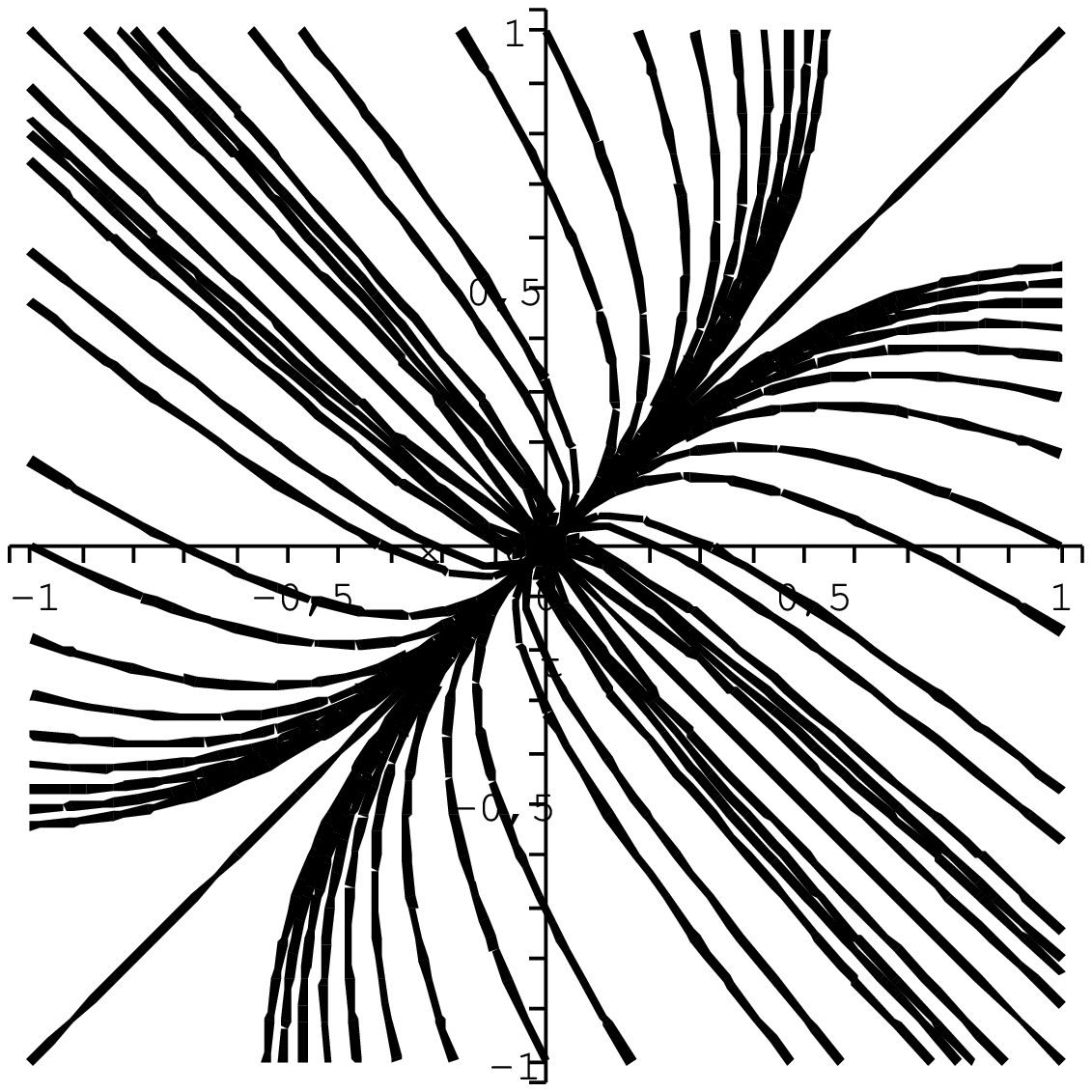}
    \medskip\nopagebreak\par Fig.~\thefigure. Force lines of the point-like vortex source for $q/m=-2$.
    The lines emerge from the center in the second and fourth quadrants, and converge to
    the center in the first and third quadrants}.\par\bigskip

By formulas (\ref{cpoth}) we get for the hyperbolic circulation and flow of the dual-symmetric
    field $\mathcal{E}$ the expression:
\begin{equation}\label{potcylh6}\Upsilon[\mathcal{B},\mathcal{C}]=0;\quad
    \Xi[\mathcal{B},\mathcal{C}]=0\end{equation}
for the contour $\mathcal{C}$, which is represented in Fig.\ref{contt} or is homotopic to it,
    and of the expression
\begin{equation}\label{potcylh7}\Upsilon[\mathcal{B},\mathcal{C}]=-\ell_Hm;\quad
    \Xi[\mathcal{B},\mathcal{C}]=\ell_Hq\end{equation}
for the contour, which is represented in Fig.\ref{cont3}, or is one homotopic to it.
    The formulas (\ref{potcylh7}) express the hyperbolic Gauss Theorem and the hyperbolic
    analogue of the law of complete flow for the dual-symmetric field $\mathcal{E}$.
%
%
\subsection{The hyperbolic cylinder displaced in constant field}\label{physh4}

We shall further examine the hyperbolic analog of a conductive cylinder displaced in
    a constant field $\mathcal{E}_0$. This problem is described by the potential
\begin{equation}\label{cylhh1}F(h)=-2\mathcal{E}_0RZ(h/R)=-\mathcal{E}_0(h+R^2/h),\end{equation}
where $R$ is the constant hyperbolic radius of the cylinder.
    The strength of the field in the neighborhood of such a cylinder is given by the formula:
\begin{equation}\label{cylhh2}\mathcal{E}=\mathcal{E}_0-\frac{\mathcal{E}_0R^2}{\bar h^2}
    =\mathcal{E}_0-\mathcal{E}_0R^2\left(\frac{t^2+x^2}{(t^2-x^2)^2}-j\frac{2tx}{(t^2-x^2)^2}\right).\end{equation}
The strength lines of the field $\mathcal{E}$, which can be obtained from the force function from
    (\ref{cylhh2}), is shown in Fig.\ref{voltzh}.

{\centering\small\refstepcounter{figure}\label{voltzh}
    \includegraphics[width=.5\textwidth]{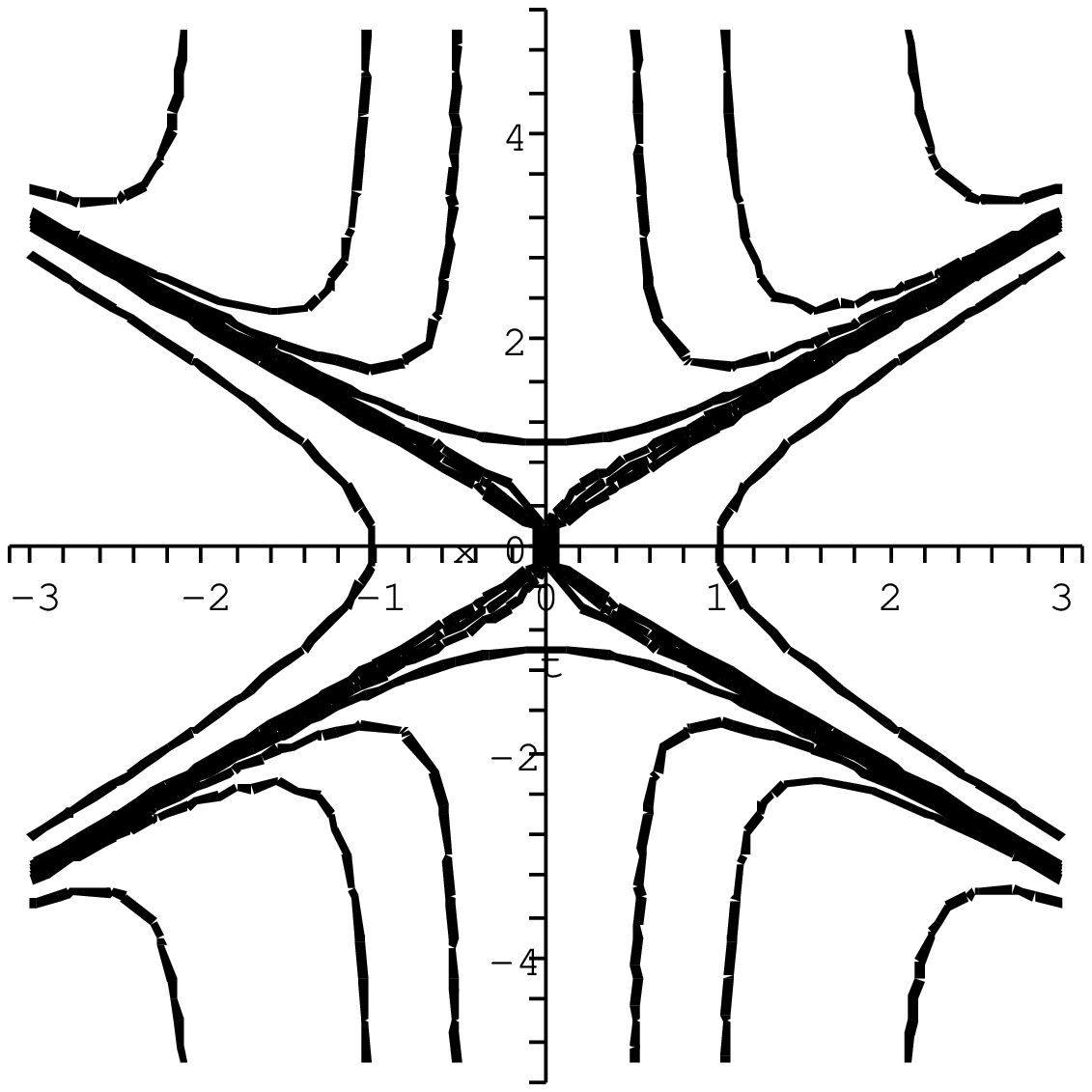}
    \medskip\nopagebreak\par Fig.~\thefigure. Force lines in the neighborhood of the
    hyperbolic cylinder (two side hyperbolas at left and right) of radius $R=1$,
    displaced in a homogeneous force field $\mathcal{E}_0=1$}.
%
%
\subsection{The $h$-multi-fields}\label{physh5}

By analogy to the complex case, we define by the induction formula
\begin{equation}\label{diph1}F_n(h)=\frac{\mathcal{Q}^{(n)}}{\mathcal{Q}^{(n-1)}}\frac{dF_{n-1}}{dh}
    =(-1)^{n+1}\frac{Q^{(n)}}{h^n}=-(-\epsilon)^{n}\epsilon_Q\frac{|\mathcal{Q}^{(n)}|e^{-j(n\psi-
    \delta_n)}}{\varrho^n},\end{equation}
the potential of the point-like hyperbolic $2(n-1)$-multi-field with power $\mathcal{Q}^{(n)}$.
    Here we have $|\mathcal{Q}^{(n)}|=\sqrt{|(\mathcal{Q}^{(n)}_e)^2-(\mathcal{Q}^{(n)}_m)^2|}$,
    $\delta_n=\text{Arth}{\mathcal{Q}^{(n)}_m/\mathcal{Q}^{(n)}_e}$, $\epsilon$ and $\epsilon_Q$
    are sign factors for $h$ and of the complex charge $\mathcal{Q}$, accordingly.
The equation for force lines has in polar coordinates the form
\begin{equation}\label{diph2}\varrho=C\sqrt[n]{\sinh(n\psi-\delta_n)}\end{equation}
The field lines for $n=2,3$ are shown in Fig.\ref{diph3}.

{\centering\small\refstepcounter{figure}\label{diph3}
    \includegraphics[width=.5\textwidth,clip]{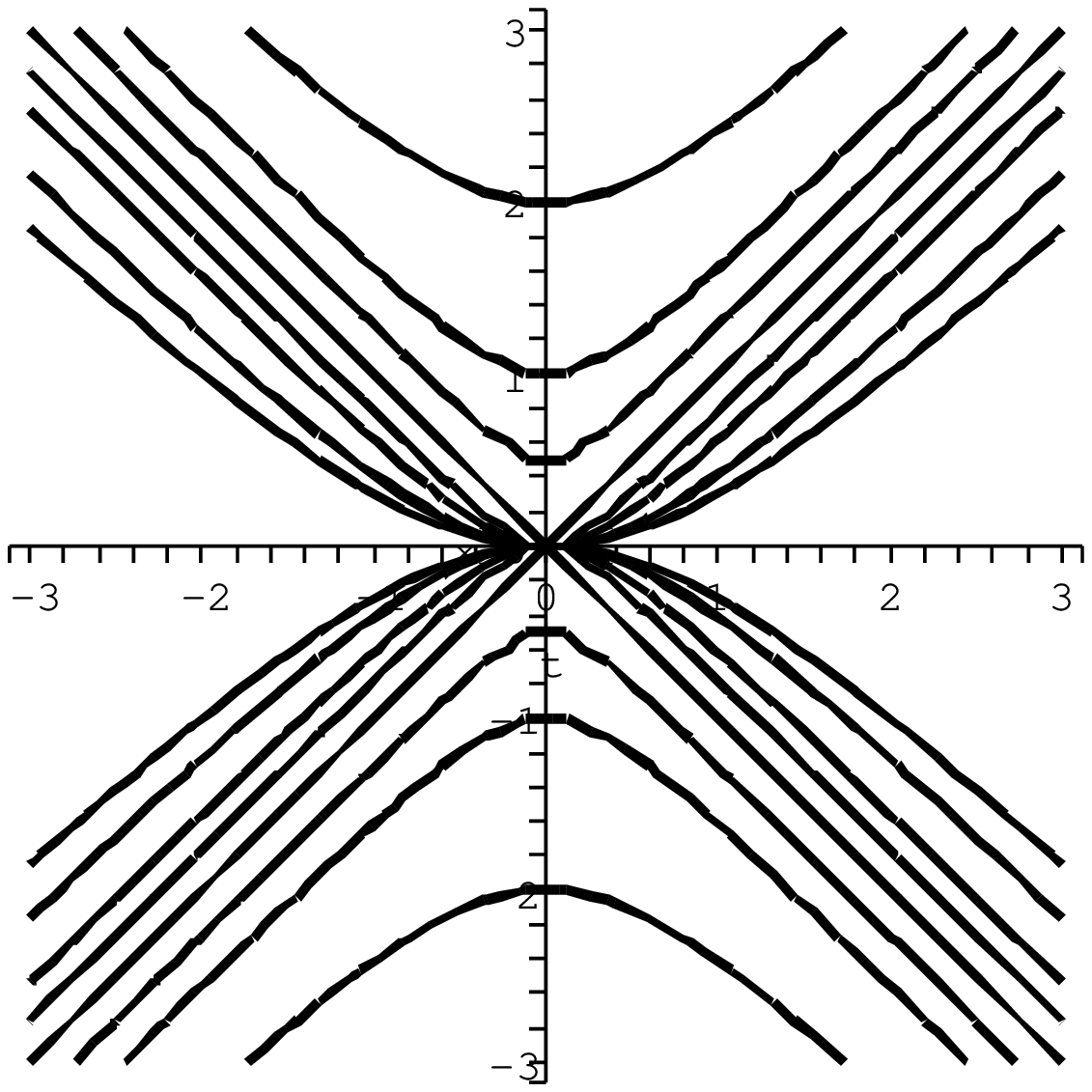}\includegraphics[width=.5\textwidth,clip]{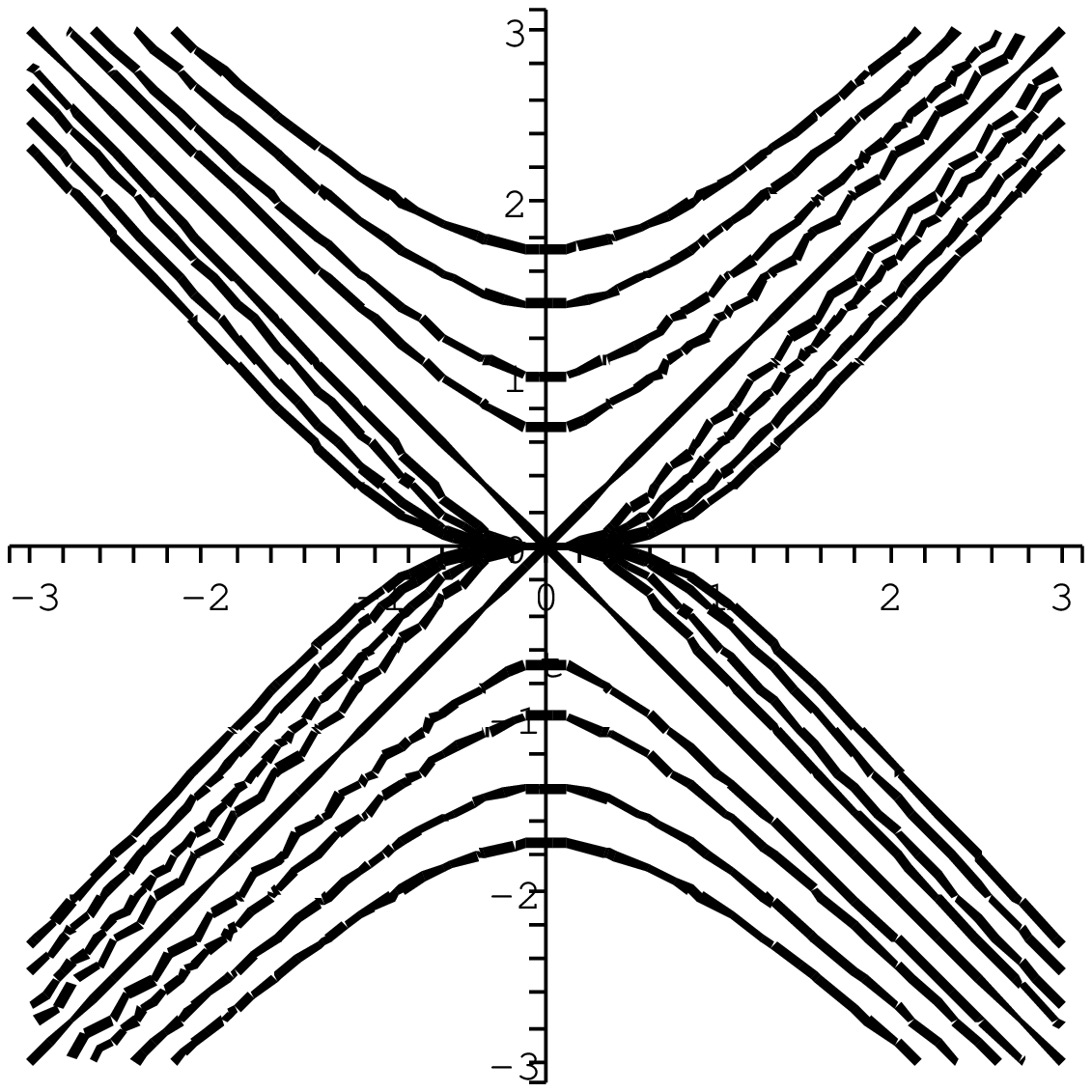}
    \nopagebreak\par Fig.~\thefigure. Force lines of the hyperbolic $h$-field in the neighborhood
    of the point-like hyperbolic dipole and quadrupole with unit power, oriented along the real axis.\par}
%
%
\subsection{The $h$-field in isotropic basis}\label{isobase}

As noted before, a remarkable feature of the double plane is the occurrence of the structure
    of real cone $\text{Con}(h)$ for each point $h\in\mathcal{H}_2$.
Geometrically, the points of the cone $\text{Con}(h)$ are displaced at zero hyperbolic distance
    from the point $h$. Algebraically we have the product $(h_1-h)(h_2-h)=0$, for $h_1$ and $h_2$
    lying in different components of the cone $\text{Con}(h)$, i.e., the structure of the cone
    is tightly related to the existence of zero divisors in the algebra of double numbers.
The occurrence of the cone structure allows us to reformulate the geometry and the algebra
    of the double plane in terms of the so called isotropic coordinates, in which the geometric
    and the algebraic properties of the double plane have a simpler form.
The isotropic basis of the algebra $\mathcal{H}_2$ is defined by the relations:
\begin{equation}\label{iso}e_1=\frac{1+j}{2};\quad e_2=\frac{1-j}{2}.\end{equation}
From the definition (\ref{iso}) it follows the multiplication table of the algebra $\mathcal{H}_2$
    in the isotropic basis:
\begin{equation}\label{iso1}e_i\cdot e_i=e_i;\quad e_i\cdot e_j=0\quad (i\neq j).\end{equation}
The double number $h=t+jx$ in the isotropic basis is expressed as follows:
\begin{equation}\label{iso2}h=\xi_1e_1+\xi_2e_2,\end{equation}
where $\xi_1=t+x$, $\xi_2=t-x$ are the advanced and retarded combination of the Cartesian coordinates
    of the double number.
The product of double numbers $h_1=\xi_1e_1+\xi_2e_2$ and $h_2=\eta_1e_1+\eta_2e_2$ gets a simple form:
\begin{equation}\label{iso3}h_1\cdot h_2=\xi_1\eta_1e_1+\xi_2\eta_2e_2.\end{equation}
In particular, any integer positive power of the double number $h^n$ in isotropic basis gets the form of a combination of powers of components:
\begin{equation}\label{iso4}h^n=\xi_1^ne_1+\xi_2^ne_2.\end{equation}
It is easy to verify, that the operation of complex conjugation in isotropic basis reduces to
    the interchange of components:
\begin{equation}\label{iso5}\bar h=\overline{\xi_1e_1+\xi_2e_2}=\xi_2e_1+\xi_1e_2.\end{equation}
From (\ref{iso3}) it is easy to note, that the operation of division is defined to any double number,
    whose all the isotropic coordinates are nonzero, i.e., when $\xi_1\xi_2\neq0$. In this case we have
\begin{equation}\label{iso6}\frac{h_2}{h_1}=\frac{\eta_1e_1+\eta_2e_2}{\xi_1e_1+\xi_2e_2}
    =\frac{\eta_1}{\xi_1}e_1+\frac{\eta_2}{\xi_2}e_2.\end{equation}
The double numbers which have one vanishing isotropic coordinate lie on the cone $\text{Con}(0)$,
    and have zero norm.

Due to the formulas
\begin{equation}\label{iso7}t=\frac{\xi_1+\xi_2}{2};\quad x=\frac{\xi_1-\xi_2}{2},\end{equation}
any mapping $\mathcal{H}_2\to\mathcal{H}_2$ can be locally represented by a pair of
    functions $F_1(\xi_1,\xi_2)$ and $F_2(\xi_1,\xi_2)$:
\begin{equation}\label{iso8}F(h)=F_1(\xi_1,\xi_2)e_1+F_2(\xi_1,\xi_2)e_2.\end{equation}
We shall show that in the case of an analytic mapping, i.e., one which can be provided by a convergent
    Taylor series in a neighborhood (or, for simplicity, by a Maclaurin series), the functions $F_1$
    and $F_2$ satisfy the relation:
\begin{equation}\label{iso9}F_{1}(\xi_1,\xi_2)=F_2(\xi_2,\xi_1).\end{equation}
First, from (\ref{iso4}), we have:
\begin{equation}\label{iso10}F(h)=\sum\limits_{k=0}^\infty c_kh^k=\sum\limits_{k=0}^\infty
    c_k(\xi_1^ke_1+\xi_2^ke_2)=F(\xi_1)e_1+F(\xi_2)e_2\end{equation}
for any analytic function of variable $h$. For an analytic function of variable $\bar h$ we similarly have:
\begin{equation}\label{iso11}F(\bar h)=\sum\limits_{k=0}^\infty c_k\bar h^k=\sum\limits_{k=0}^\infty
    c_k(\xi_2^ke_1+\xi_1^ke_2)=F(\xi_2)e_1+F(\xi_1)e_2.\end{equation}
Further, for an analytic function $F(h,\bar h)$, due to (\ref{iso10})-(\ref{iso11}) we
    obtain the chain of equalities:
\begin{equation}\label{iso12}F(h,\bar h)=\sum\limits_{k,l=0}^\infty c_{kl}h^k\bar h^
    l=\sum\limits_{k,l=0}^\infty c_{kl}(\xi_1^ke_1+\xi_2^ke_2)(\xi_2^le_1+\xi_1^le_2)
    =\sum\limits_{k,l=0}^\infty c_{kl}(\xi_1^k\xi_2^le_1+\xi_2^k\xi_1^le_2)=\end{equation}
\[F(\xi_1,\xi_2)e_1+F(\xi_2,\xi_1)e_2,\]
whence (\ref{iso9}) follows. For the basic operators of differentiation we easily
    obtain the following relations:
\begin{equation}\label{iso13}\frac{\partial}{\partial h}
    =e_1\frac{\partial}{\partial\xi_1}+e_2\frac{\partial}{\partial\xi_2};\quad\frac{\partial}{\partial\bar h}
    =e_1\frac{\partial}{\partial\xi_2}+e_2\frac{\partial}{\partial\xi_1}.\end{equation}
By applying the operator $\partial/\partial \bar h$ to the formula (\ref{iso12}),
    we infer the condition of $h$-holomorphicity in isotropic basis:
\begin{equation}\label{iso14}\frac{\partial F}{\partial\xi_2}(\xi_1,\xi_2)=0;\quad
    \frac{\partial F}{\partial\xi_1}(\xi_2,\xi_1)=0,\end{equation}
which essentially leads to formula (\ref{iso10}).

We shall examine as an example the point source of an $h$-field in isotropic coordinate system.
    Its potential (\ref{hqulll}) in isotropic coordinate system gets the form:
\begin{equation}\label{hquli}F(h)=-q\ln h=-q(\ln \xi_1\, e_1+\ln\xi_2\, e_2).\end{equation}
The strength of the $h$-field is
\begin{equation}\label{hquli1}\mathcal{E}=-\frac{\partial \overline{F(h)}}{\partial \bar h}
    =q\left(\frac{\partial}{\partial \xi_2}e_1+\frac{\partial}{\partial\xi_1}e_2\right)(\ln\xi_2\,
    e_1+\ln\xi_1\, e_2)=q\left(\frac{1}{\xi_2}e_1+\frac{1}{\xi_1}e_2\right).\end{equation}
The equations for its strength lines can be obtained by setting the characteristic equation:
\begin{equation}\label{iso15}\xi_2d\xi_1=\xi_1d\xi_2,\end{equation}
whose first integral $\xi_1/\xi_2=C=\text{const}$ is a family of straight lines which pass
    through the origin. For a point-like vortex, the strength  is
\begin{equation}\label{iso16}\mathcal{B}=j\mathcal{E}=(e_1-e_2)\mathcal{E}
    =q(\frac{1}{\xi_2}e_1-\frac{1}{\xi_1}e_2)\end{equation}
and the equation of characteristics
\begin{equation}\label{iso17}\xi_2d\xi_1=-\xi_1d\xi_2\end{equation}
has the first integral $\xi_1\xi_2=C=\text{const}$, a family of equilateral hyperbolas.
%
%
\section{Conformal $h$-fields}\label{confo}

We shall examine some generalization of holomorphic mappings of the double plane,
    which in isotropic basis are given by the relation:
\begin{equation}\label{cf}\mathcal{F}=\mathcal{F}_1(\xi_1)e_1+\mathcal{F}_2(\xi_2)e_2,\end{equation}
where $\mathcal{F}_1,\mathcal{F}_2$ is an arbitrary smooth function of real variable or, in components:
\begin{equation}\label{cf0}\xi_1'=\mathcal{F}_1(\xi_1);\quad \xi_2'=\mathcal{F}_2(\xi_2).\end{equation}
Due to the relations (\ref{iso10})-(\ref{iso12}), such a mapping cannot be regarded as an analytic function
    of one of the variables $h,\bar h$ or even of them both. The characteristic property of the
    mapping of the form (\ref{cf}) becomes clear, if we note that the metric (\ref{quadh}),
    which is induced by the algebra $\mathcal{H}_2$, in isotropic coordinates gets the form:
\begin{equation}\label{cf1}\Theta=\text{Re}(dh\otimes d\bar h)
    =d\xi_1\otimes d\xi_2+d\xi_2\otimes d\xi_1.\end{equation}
It is obvious, that under the transformation (\ref{cf0}), the metric(\ref{cf1}) transforms
    according to the rule:
\begin{equation}\label{cf2}\Theta'=\mathcal{F}_1'\mathcal{F}_2'\Theta,\end{equation}
i.e., conformally. In other words, {\it the transformations of the form (\ref{cf})
    describe the general class of conformal transformations of the pseudo-Euclidean metric,
    which preserve the flatness of this metric}.
We shall call such a class of transformations {\it conformal-isotropic}.
    The holomorphic (or anti-holomorphic) functions represent only a subset of this class.

The occurrence of the class of conformal-isotropic transformations are a characteristic consequence
    of the algebra of double numbers structure, and more precisely, of the presence of zero divisors
    and of both the cone structure and isotropic coordinates.
For these reasons, there exists no analogue for conformal-isotropic transformations on the
    plane of complex variable.
Another non-analytic generalization of the class of holomorphic transformations of the
    double plane could have been the transformations of the form:
\begin{equation}\label{cf3}\mathcal{F}=\mathcal{F}(\xi_1,\xi_2)e_1+\mathcal{F}(\xi_2,\xi_1)e_2,\end{equation}
where $\mathcal{F}$ is a smooth function of the pair of real variables, and where we note that
    $\mathcal{F}(\xi_1,\xi_2)\neq\mathcal{F}(\xi_2,\xi_1)$.
Due to the identity (\ref{iso12}), such a function cannot be regarded as a function of the
    pair of variables $\mathcal{F}(h,\bar h)$, which could be represented in the form
    of a convergent Taylor series relative to the variables $h$ and $\bar h$.
%
%
\section{Generalization to the multi-dimensional case}

For physical applications, it is compulsory to extend the concept of $h$-field to the
    case of commutative-associative algebras of higher dimensions.
We shall illustrate the idea of such an extension for the example of $P_n$ -- the algebra of poly-numbers.
    The existence of an isotropic basis annihilates some formal differences between the algebras
    $P_n$ for different $n$, and hence for convenience we shall make our reasonings using the
    algebra and geometry of poly-numbers $P_3$.
A substantial part of these properties can be trivially extended to general poly-numbers $P_n$.
%
%
\subsection{Algebra and operations}

The associative-commutative algebra $P_3$ over the field $\R$ (algebra of 3-numbers)
    generalizes well the well known algebra of double numbers on the plane.
    Its general element has the form
\begin{equation}\label{poly}A=A_1e_1+A_2e_2+A_3e_3,\end{equation}
where $\{e_i\}$ is a special set of generators of the algebra, which satisfy the relations:
\begin{equation}\label{alge}e_ie_j=\delta_{ij}e_i\quad \text{(without summation!)}\end{equation}
From the relations (\ref{alge}), there follow the simple rules of multiplying and dividing the poly-numbers:
\[AB=A_1B_1e_1+A_2B_2e_2+A_3B_3e_3;\quad \frac{A}{B}=\frac{A_1}{B_1}e_1+\frac{A_2}{B_2}e_2+\frac{A_3}{B_3}e_3,\]
where the division is defined only on the so called non-degenerate elements, for which all $B_i\neq0$.
    The role of the unity of the algebra $P_3$ is played by the element $I=e_1+e_2+e_3$.
%
%
\subsection{Complex conjugation and (pseudo)norm}

We define in $P_3$ two operations of complex conjugation:
\[A^{\dag}=(A_1e_1+A_2e_2+A_3e_3)^{\dag}\equiv A_3e_1+A_1e_2+A_2e_3,\]
\[A^{\ddag}=(A_1e_1+A_2e_2+A_3e_3)^{\ddag}\equiv A_2e_1+A_3e_2+A_1e_3\]
and examine the 3-number $AA^{\dag}A^{\ddag}$. A simple calculation shows that this is real
    and equal to $A_1A_2A_3I$. Hence, by analogy to the module from complex numbers, we can
    introduce in $P_3$ a (pseudo)norm by means of the formula:
\begin{equation}\label{norm}\|A\|\equiv (AA^{\dag}A^{\ddag})^{1/3}=(A_1A_2A_3)^{1/3}.\end{equation}
For non-degenerate 3-numbers, the norm practically has all the properties of the usual norm,
    and in particular, for such 3-numbers the following equality holds true:
\begin{equation}\label{propn}\|AB\|=\|A\|\cdot\|B\|.\end{equation}
%
%
\subsection{Zero divisors and the group of inner automorphisms}

Unlike the field of complex numbers and the field of quaternions, the algebra $P_3$ has zero-divisors, i.e.,
    there exist non-zero elements $N$, which satisfy the condition : $\|NA\|=0$
    for any $A\in P_3$.
Such elements are called {\it degenerate} and are characterized by the fact that their
    representation (\ref{poly}) contains zero coefficients.
We note that the set of degenerate elements is closed  relative to the multiplication from $P_3$.
    We shall denote this subset by $P_3^{\circ}$.

The operation of multiplication on non-degenerate elements from $P_3$ is related to the group of
    inner automorphisms $\text{Aut}(P_3)$, which is isomorphic to the subgroup (by multiplication)
    of non-degenerate elements:
\begin{equation}\label{defaut}\text{Aut}(P_3)\sim P_3\setminus P_3^{\circ},\quad
    \text{Aut}(P_3)\ni\sigma:\ A\to\sigma(A)\equiv \sigma A.\end{equation}
In this group we point out the subgroup $D_2\subset P_3$ of the isometries group $\text{Iso}(P_3)$,
    whose elements preserve the norm (\ref{norm}).
Having in view the definition (\ref{defaut}) and the properties (\ref{propn}),
    the elements of this subgroup are characterized by the condition:
    $\|\sigma\|=1$ or, in components: $\sigma_1\sigma_2\sigma_3=1$.
The group $D_2$ is 2-parametric and abelian, and the group $\text{Iso}(P_3)$ is 5-parametric and non-abelian.
    Except the subgroup $D_2$, the latter one includes the subgroup of translations,
    which is isomorphic to $\R^3$.
%
%
\subsection{Series and exponential representation}

In the space $P_3$ (and in any $P_n$) we can define powers of elements of any order,
    and further, analytic functions of poly-number variable.
As an example, the function $e^A$ can be defined by the standard exponential series:
\begin{equation}\label{exppl}e^A\equiv I+A+\frac{A^2}{2!}+\dots=e^{A_1}e_1+e^{A_2}e_2+e^{A_3}e_3.\end{equation}
We define now the exponential representation of poly-numbers, by the formula:
\begin{equation}\label{exprep1}A=\|A\|e^{B},\end{equation}
where $B$ is some 3-number, which is invariant under the action of the group $D_2$, which
    preserves the norm $\|A\|$.
The components of this number in some special basis are called {\it exponential angles}.
    There exist two exponential angles since, as we shall see, the space of numbers $B$
    for a given number $A$ with fixed norm $\|A\|$, is 2-dimensional.
For computing the explicit form of exponential angles, we perform the following chain
    of identical transformations:
\[A=A_1e_1+A_2e_2+A_3e_3=(A_1A_2A_3)^{1/3}\times\]
\[\left(\frac{A_1^{2/3}}{(A_2A_3)^{1/3}}e_1+\frac{A_2^{2/3}}{(A_1A_3)^{1/3}}e_2+
    \frac{A_3^{2/3}}{(A_1A_2)^{1/3}}e_3\right)=\]
\[|A|(e^{\ln(A_1^{2/3}/(A_2A_3)^{1/3})}e_1+e^{\ln(A_2^{2/3}/(A_1A_3)^{1/3})}e_2+
    e^{\ln(A_3^{2/3}/(A_1A_2)^{1/3})}e_3)=\]
\begin{equation}\label{expug}|A|(e^{\chi_1}e_1+e^{\chi_2}e_2+e^{\chi_3}e_3)
    =|A|e^{\chi_1e_1+\chi_2e_2+\chi_3e_3},\end{equation}
where the quantities
\begin{equation}\label{appare}\chi_1\equiv\ln\left[\frac{A_1^{2/3}}{(A_2A_3)^{1/3}}\right];\
    \chi_2\equiv\ln\left[\frac{A_2^{2/3}}{(A_1A_3)^{1/3}}\right];\
    \chi_3\equiv\ln\left[\frac{A_3^{2/3}}{(A_1A_2)^{1/3}}\right]\end{equation}
are exactly the exponential angles. Due to the relations
\begin{equation}\label{relat}\chi_1+\chi_2+\chi_3=0,\end{equation}
which by formulas (\ref{appare}), are identically satisfied, there exist only two
    independent angles and the representation (\ref{expug}) can be written in the
    following equivalent forms:
\[A=|A|e^{-\chi_2E_3+\chi_3E_2}=|A|e^{\chi_1E_3-\chi_3E_1}=|A|e^{-\chi_1E_2+\chi_2E_1},\]
where $E_1=e_2-e_3$,  $E_2=e_3-e_1$, $E_3=e_1-e_2$ are combinations of basic vectors,
    which are generators of the group $D_2$.

The operations of complex conjugation act on the exponential angles as follows:
\[\dag:\quad \chi_1\to\chi_3;\  \chi_2\to\chi_1;\ \chi_3\to\chi_2;\quad
    \ddag:\  \chi_1\to\chi_2;\  \chi_2\to\chi_3;\ \chi_3\to\chi_1\]
and validate the formula (\ref{norm}) written in exponential representation.
%
%
\subsection{Generalization of exponential angles}

In the previous subsection the formulas (\ref{appare}) assume that all $A_i>0$.
    In other words, formula (\ref{exprep1}) holds true only for the "positive octant".
It is easy to define the exponential representation in other octants as well,
    by slightly generalizing (\ref{exprep1}).
Namely, we define the exponential representation for poly-numbers from the interior
    of an arbitrary octant by the formula
\begin{equation}\label{exprep2}A=I_{(j)}\|A'\|e^{B'},\end{equation}
where $I_{(j)}$ $(j=1,\dots,8)$ is the unit vector in the direction of the space bisector
    (in Euclidean sense) corresponding to the coordinate octant,
    the number $A'=A/I_{(j)}$ lies in the positive octant, and $B'$ is obtained from $B$
    using the formulas of the previous  subsection with the replacement $A\to A/I_{(j)}$.
Such a definition removes the artificial limitation on the definition domain of exponential angles
    and more adequately reflects their meaning as quantities, computed from the corresponding
    directions $I_{(j)}$. We note that in our notations we consider $I_1\equiv I$.
In the following, unless this is specified, all the considerations will be performed in the
    positive octant.

The generalization of the formulas of the exponential representation to poly-numbers $P_n$
    is elementary. For this, we have the formula (\ref{exprep2}), in which the index $j$
    can take $2^n$ values according to the number of domains (multi-dimensional octants)
    which are split by the coordinate planes. In the positive octant one has
\[\|A\|=[\prod\limits_{i=1}^n A_i]^{1/n};\quad \chi_k=\ln\left[\frac{A_k}{\|A\|}\right]\]
or take place similar formulas for the corresponding primed quantities for any of the $2^n$
    multi-dimensional octants.
%
%
\subsection{The scalar 3-product (poly-product)}

Using the operation $\dag$ and $\ddag$, we can define the real number $\langle A,B,C\rangle$,
    called the {\it scalar 3-product of the elements $A,B,C$}, which is built for any
    three vectors from $P_3$ by the rule:
\begin{equation}\label{3lin}{}^{(3)}G(A,B,C)\equiv\langle A,B,C\rangle\equiv\sum\limits_{X,Y,Z
    =S(ABC)}XY^{\dag}Z^{\ddag}
    =\text{perm}\left(\begin{array}{ccc}A_1&A_2&A_3\\B_1&B_2&B_3\\C_1&C_2&C_3\end{array}\right),\end{equation}
where $S(ABC)$ is the set of permutations of the elements $A,B,C$, and $\text{perm}\,(M)$ is the permanent
    of the matrix $M$, which copies the structure of its determinant, but all the 6 terms are
    taken with sign "plus".

For poly-numbers $P_n$ there takes place an analogous definition of the $n$-product:
\[{}^{(n)}G(X_1,\dots X_n)\equiv\text{perm}\langle X_1,\dots,X_n\rangle,\]
where ${X_i}$ is a set of $n$ $n$-numbers.
%
%
\subsection{The Berwald-Moor space}

If we detach from the algebra, and from the very beginning we examine the vector space
    endowed with the 3-scalar product, which in special coordinates has the form (\ref{3lin}),
    we are led to {\it the Finsler 3-dimensional Berwald-Moor space} (BM), which will be denoted by
    $\mathcal{H}_3$.
Unlike $P_3$, we do not assume inside this the existence of any multiplicative algebra.
    We can say, that the relation between $P_3$ and $\mathcal{H}_3$ is similar to the relation
    between the complex plane $\C$ and the Euclidean plane $\R^2$.
%
%
\subsection{Construction of tangency}

\begin{figure}[htb]\centering \unitlength=0.50mm \special{em:linewidth 0.4pt}
    \linethickness{0.4pt} \footnotesize \unitlength=0.70mm \special{em:linewidth 0.4pt}
    \linethickness{0.4pt}\input{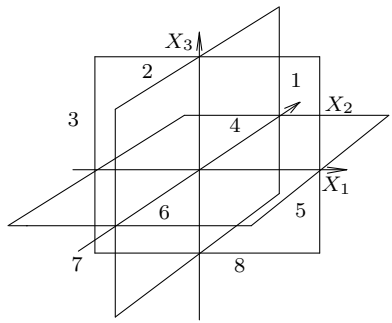}
    \caption{\small Isotropic coordinate planes and octants in $\mathcal{H}_3$}.\label{coord}
\end{figure}

The vectors $\mathcal{H}_3$, whose norm is equal to zero are called in the Berwald-Moor geometry
    {\it isotropic vectors}.
As one can see from the definition (\ref{norm}), each isotropic vector lies in some of the
    three coordinate planes of the isotropic coordinate system.
In particular, the vectors $e_1=(1,0,0)$, $e_2=(0,1,0)$, and $e_3=(0,0,1)$ of the isotropic basis
    of this system are isotropic. Hence, the whole coordinate space $\mathcal{H}_3$ is split by the
    coordinate planes into 8 octants, inside which the norms of the vectors are nonzero and have
    certain signs (see Fig.\ref{coord}).
On the coordinate planes, the metric (\ref{3lin}) becomes
\\
geometrically degenerate, since all the contained vectors have vanishing norm.
    For a good description of the geometric properties of the coordinate planes (which are 2-dimensional
    pseudo-Euclidean spaces) we can use the contact construction \cite{kok2}.
Its essence resides in the transition from the Finsler metric ${}^{(3)}G$ of the form (\ref{3lin})
    to the tangent to it along the vector $e_j$ quadratic metric:
\[{}^{(2)}G_{(j)}\equiv{}^{(3)}G(e_j,\ ,\  ),\]
which acts in the hyperplane of directions $x^j=\text{const}$. As an example, for the case $j=3$ we have:
\[{}^{(2)}G_{(3)}\equiv {}^{(3)}G(e_3,\ ,\ )=dX_1\otimes dX_2+dX_2\otimes dX_1\]
i.e., the Berwald-Moor metric on the planes $X_3=\text{const}$, which is a 2-dimensional Minkowski metric.
%
%
\subsection{The indicatrix}

The metric properties of the space $\mathcal{H}_3$ are illustrated by means of the unit sphere
    $\mathcal{S}^2_{\text{BM}}$ (the indicatrix of $\mathcal{H}_3$), which is defined by the equation:
\begin{equation}\label{ind}|\,\|X\|\,|=|(X_1X_2X_3)^{1/3}|=1,\end{equation}
where $X=(X_1,X_2,X_3)$ is the position vector in $\mathcal{H}_3$.
    The surface $\mathcal{S}^2_{\text{BM}}$ is 8-connected and non-compact.
Its connected components are symmetrically displaced in all the 8 octants and have
    a discrete symmetry relative to any permutation of coordinates.
The sections of these surfaces with the planes $X_i=\text{const}$ are families of hyperbolas
    (in Euclidean sense).
One of the components of the indicatrix in Euclidean representation is shown in Fig.\ref{indica}

{\centering\small\refstepcounter{figure}\label{indica}
    \includegraphics[width=.4\textwidth, height=0.5\textwidth]{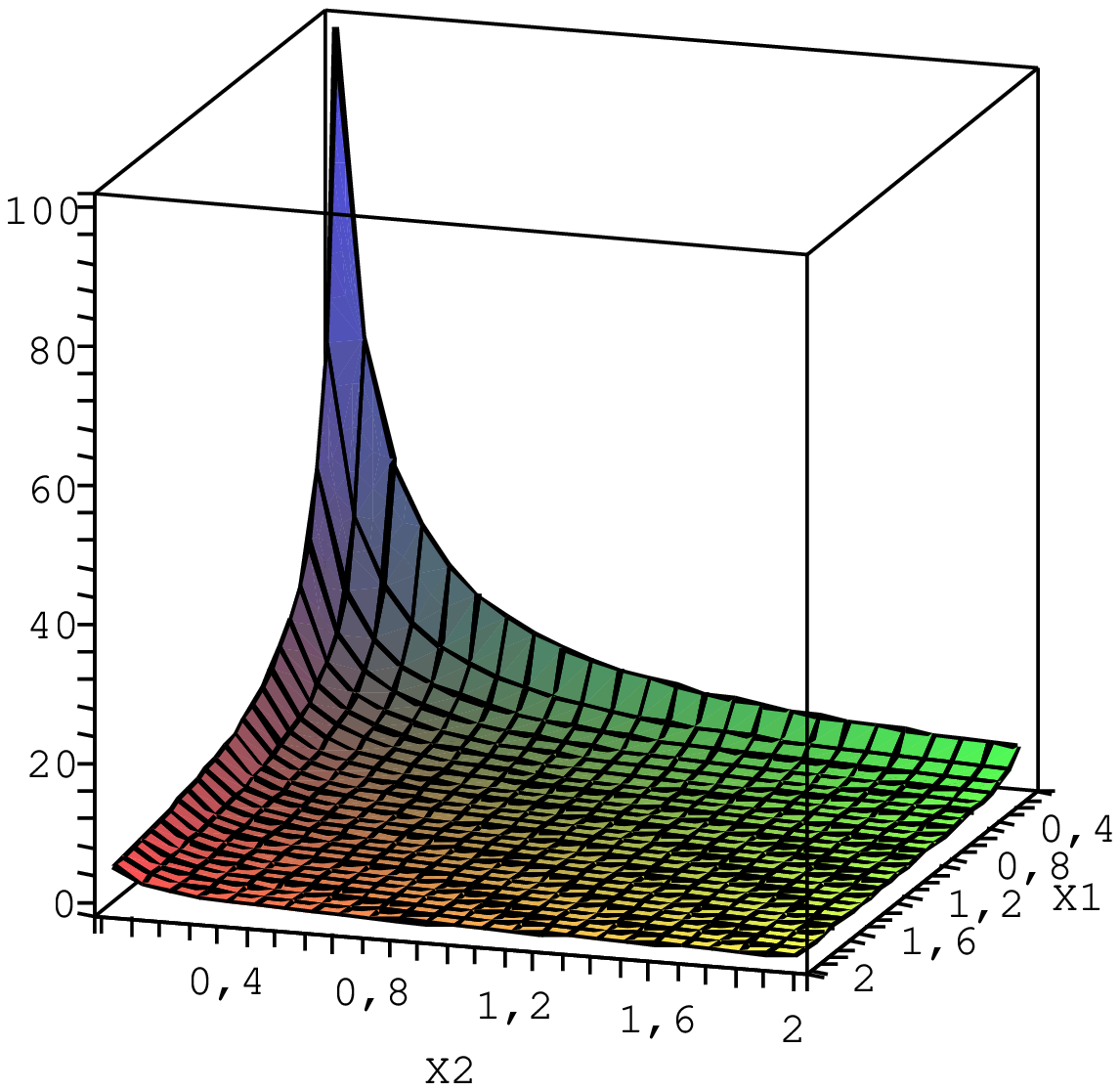}
    \includegraphics[width=.4\textwidth, height=0.5\textwidth]{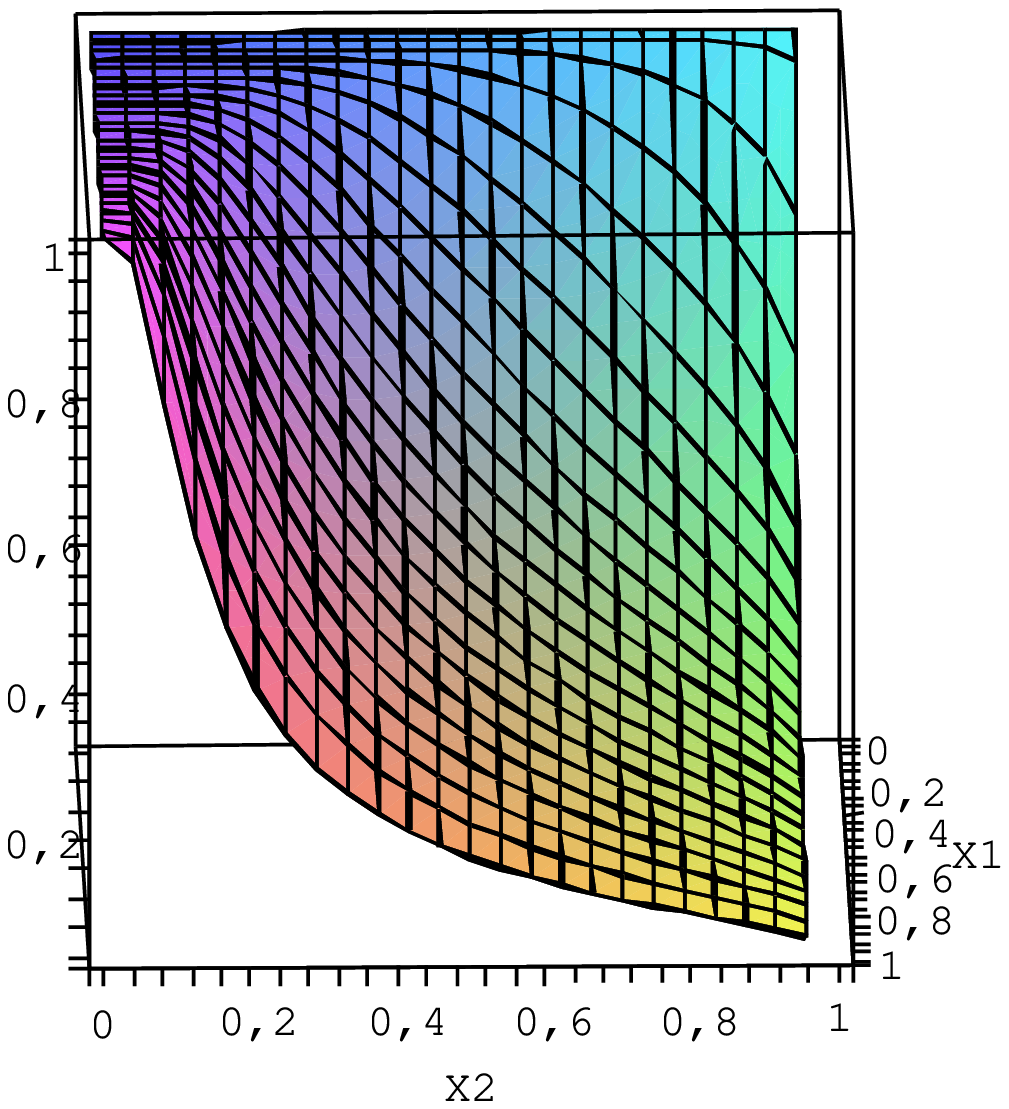}
    \medskip\nopagebreak\par Fig.~\thefigure. The component of the indicatrix $\mathcal{S}^2_{\text{BM}}$,
    which lies in the first octant. On the right drawing this component is compactified within the unit cube
    by means of the mappings $X_i\mapsto \tanh(X_i\ln3/2)$. The coefficient in the argument of
    the hyperbolic tangent is chosen so, that the point $(1,1,1)$ transforms into the point
    $(1/2,1/2,1/2)$ of the unit cube}.\par\medskip
%
%
\subsection{Geometric interpretation of the group $\text{Iso}(\mathcal{H}_3)$.}

As already specified before, the isometry group $\text{Iso}(\mathcal{H}_3)$ of the metric (\ref{3lin})
    consists of a 3-parametric abelian subgroup of translations with elements $T_A:\ X\to X+A$
    and the 2-parametric abelian subgroup of unimodular dilatations ${D}_2$ described by
    $D_{\sigma_1,\sigma_2,\sigma_3}:\ (X_1,X_2,X_3)\to(\sigma_1X_1,\sigma_2X_2,\sigma_3X_3)$, where
\begin{equation}\label{normir}\sigma_1\sigma_2\sigma_3=1.\end{equation}
From algebraic point of view, on $\mathcal{H}_3$ the group ${D}_2$ is exactly the described above
    subgroup of $\text{Iso}(P_3)$ multiplied with elements of norm one.
We note once more, that the group $\text{Iso}(\mathcal{H}_3)$ is non-abelian and has the
    structure of semi-direct product: $\text{Iso}(\mathcal{H}_3)=\mathcal{R}^3\rtimes{D}_2$.
The group ${D}_2$ plays the role of rotations in the space $\mathcal{H}_3$ and generalizes
    the hyperbolic rotations of the pseudo-Euclidean plane.
In particular, the action of the group ${D}_2$ on the indicatrix, is transitive:
    $\mathcal{S}^2_{\text{BM}}\stackrel{{D}_2}{\to}\mathcal{S}^2_{\text{BM}}$.

For Berwald-Moor spaces $\mathcal{H}_n$ of higher dimension, we similarly have
    $\text{Iso}(\mathcal{H}_n)=\R^n\rtimes D_{n-1}$.
%
%
\subsection{$h$-analytic fields in the space $\mathcal{H}_3$.}

Like in the previously examined 2-dimensional cases, the general smooth mapping
    $\mathcal{H}_3\to\mathcal{H}_3$ can be represented in the form $F=F(h,h^\dag,h^\ddag)$.
    We shall call a mapping $F$:
\begin{enumerate}
\item $h$-holomorphic, if
\begin{equation}\label{hh1}\frac{\partial F}{\partial h^\dag}=\frac{\partial F}{\partial h^\ddag}=0;\end{equation}
\item $h^\dag$-holomorphic, if
\begin{equation}\label{hh2}\frac{\partial F}{\partial h}=\frac{\partial F}{\partial h^\ddag}=0;\end{equation}
\item $h^\ddag$-holomorphic, if
\begin{equation}\label{hh3}\frac{\partial F}{\partial h}=\frac{\partial F}{\partial h^\dag}=0;\end{equation}
\item $hh^\dag$-holomorphic, if
\begin{equation}\label{hh4}\frac{\partial F}{\partial h^\ddag}=0;\end{equation}
\item $hh^\ddag$-holomorphic, if
\begin{equation}\label{hh5}\frac{\partial F}{\partial h^\dag}=0;\end{equation}
\item $h^\dag h^\ddag$-holomorphic, if
\begin{equation}\label{hh6}\frac{\partial F}{\partial h}=0.\end{equation}
\end{enumerate}

As it can be seen from the definitions from above, the increase in the dimension of the
    commutative-associative algebra, the number of types of holomorphicity increases too.
It is clear, that the first definition is the analogue of the usual complex holomorphicity, or
    $h$-holomorphicity on the double plane.
The second and the third definitions are the analogs of the complex or double anti-holomorphicity
    and the last three definitions have no analogs in two dimensions and and represent "weakened"\,
    variants of holomorphicity.
In this paper, we focus on the $h$-holomorphicity, its properties and consequences.

We start with an isotropic basis in $\mathcal{H}_3$,, relative to which the $h$-holomorphic
    function has the following representation:
\begin{equation}\label{hh7}F(h)=F(\xi_1)e_1+F(\xi_2)e_2+F(\xi_3)e_3.\end{equation}
The differentiation operators in terms of $h,h^\dag,h^\ddag$ have the form:
\begin{equation}\label{hh8}\frac{\partial}{\partial h}=e_1\frac{\partial}{\partial\xi_1}+
    e_2\frac{\partial}{\partial\xi_2}+e_3\frac{\partial}{\partial\xi_3};\
    \frac{\partial}{\partial h^\dag}=e_1\frac{\partial}{\partial\xi_3}+
    e_2\frac{\partial}{\partial\xi_1}+e_3\frac{\partial}{\partial\xi_2};\
    \frac{\partial}{\partial h^\ddag}=e_1\frac{\partial}{\partial\xi_2}+
    e_2\frac{\partial}{\partial\xi_3}+e_3\frac{\partial}{\partial\xi_1}.\end{equation}
Due to (\ref{hh8}), a direct verification in component writing confirms the validity of
    the following equality for a function $F$ of the form (\ref{hh7}):
\begin{equation}\label{hh9}\frac{\partial F}{\partial h^\dag}=\frac{\partial F}{\partial h^\ddag}=0;\quad
    \frac{\partial F}{\partial h}=\frac{\partial F_1}{\partial \xi_1}e_1+
    \frac{\partial F_2}{\partial \xi_2}e_2+\frac{\partial F_3}{\partial \xi_3}e_3,\end{equation}
where here and further $F_i\equiv F(\xi_i)$ is the same function of different isotropic variables.
    The aim of the following study is to write the holomorphicity conditions in some
    non-isotropic basis, i.e., to derive a multi-dimensional analogue of the standard
    Cauchy-Riemann conditions.
Among all the possible non-isotropic bases of the algebra $\mathcal{H}_3$, one which is
    remarkable due to its symmetry is the one which consists of three hyperbolic imaginary units
    $\{j_1,j_2,j_3\}$, and which is related to the isotropic basis by the relations:
\begin{equation}\label{hh10}j_1=e_1-e_2-e_3;\quad j_2=-e_1+e_2-e_3;\quad j_3=-e_1-e_2+e_3,\end{equation}
from which there follow the rules of multiplication for these units:
\begin{equation}\label{hh11}j_i^2=-(j_1+j_2+j_3);\quad j_i\cdot j_k=j_l\quad (j\neq k\neq l).\end{equation}
By writing the number $h$ in the decomposed form $h=x_1j_1+x_2j_2+x_3j_3$, expressing the $j$-basis
    in terms of the $e$-basis by means of formulas (\ref{hh10}), collecting the coefficients of
    $e_i$ and identifying them with $\xi_i$, we get the following relation between the coordinates
    of a 3-number expressed in the $e$-basis and the one for the $j$-basis:
\begin{equation}\label{hh12}\xi_1=x_1-x_2-x_3;\quad \xi_2=-x_1+x_2-x_3;\quad \xi_3=-x_1-x_2+x_3.\end{equation}
We shall further also need the formulas for the inverse transformations for the bases and coordinates:
\begin{equation}\label{hh13}e_1=-\frac{j_2+j_3}{2};\quad e_2=-\frac{j_1+j_3}{2};\quad
    e_3=-\frac{j_1+j_2}{2};\quad\end{equation}
\begin{equation}\label{hh14}x_1=-\frac{\xi_2+\xi_3}{2};\quad x_2=-\frac{\xi_1+\xi_3}{2};\quad
    x_3=-\frac{\xi_1+\xi_2}{2}.\end{equation}
It is easy to verify that the action of complex conjugation reduces to the rules:
\begin{equation}\label{hh15}j_1^\dag=j_2;\   j_2^\dag=j_3;\ j_3^\dag=j_1;\ j_1^\ddag=j_3;\
    j_2^\ddag=j_1;\ j_3^\ddag=j_2.\end{equation}
As a further step we write the differentiation operator in terms of a 3-number expressed in the $j$-basis.
    To this aim, we perform in formulas (\ref{hh8}) the substitutions $e_i$ taken from (\ref{hh13}),
    and the differentiation operators, by means of the formulas:
\begin{equation}\label{hh16}\frac{\partial}{\partial\xi_i}=-\frac{1}{2}
    \left(\frac{\partial}{\partial x_k}+\frac{\partial}{\partial x_l}\right),\end{equation}
where $i,k,l$ take all the possible distinct possible values. After these substitutions
    and elementary calculations we get the needed formulas:
\begin{equation}\label{hh17}\frac{\partial}{\partial h}=\frac{1}{4}[j_1(\bar\partial+\partial_1)+
    j_2(\bar\partial+\partial_2)+j_3(\bar\partial+\partial_3)],\end{equation}
where $\partial_i\equiv\partial/\partial x_i$, $\bar\partial\equiv\sum\limits_i\partial_i$
    and we consider also another pair of formulas, which result from (\ref{hh17}) by means of
    applying the operations $\dag$ and $\ddag$, and by the use of the formulas (\ref{hh15}):
\begin{equation}\label{hh18}\frac{\partial}{\partial h^\dag}=\frac{1}{4}[j_2(\bar\partial+\partial_1)+
    j_3(\bar\partial+\partial_2)+j_1(\bar\partial+\partial_3)],\end{equation}
\begin{equation}\label{hh19}\frac{\partial}{\partial h^\ddag}=\frac{1}{4}[j_3(\bar\partial+\partial_1)+
    j_1(\bar\partial+\partial_2)+j_2(\bar\partial+\partial_3)].\end{equation}
We shall examine now some $h$-holomorphic function, which in the $j$-basis has the representation
\begin{equation}\label{hh20}F(h)=U_1j_1+U_2j_2+U_3j_3.\end{equation}
By acting on this by the operators (\ref{hh18}) and (\ref{hh19}), and writing the result in components
    by considering the rules (\ref{hh11}), we get the following pair of systems of equations
    with partial derivatives, which have to be satisfied by any $h$-holomorphic function:
\begin{equation}\label{hh21}\left(\begin{array}{ccc}-(\bar\partial+\partial_3)&\partial_{2-1}&\partial_{1-2}\\
    \partial_{2-3}&-(\bar\partial+\partial_1)& \partial_{3-2}\\
    \partial_{1-3}&\partial_{3-1}&-(\bar\partial+\partial_2)\end{array}\right)\,
    \left(\begin{array}{c}U_1\\U_2\\U_3\end{array}\right)=0;\quad\end{equation}
\begin{equation}\label{hh22}\left(\begin{array}{ccc}-(\bar\partial+\partial_2)&\partial_{1-3}&\partial_{3-1}\\
    \partial_{1-2}&-(\bar\partial+\partial_3)& \partial_{2-1}\\
    \partial_{3-2}&\partial_{2-3}&-(\bar\partial+\partial_1)\end{array}\right)\,
    \left(\begin{array}{c}U_1\\U_2\\U_3\end{array}\right)=0.\end{equation}
For shortening the writing, the system is reduced to the matrix form, in which the multiplication
    by the "operator matrix"\, with the column matrix of components is performed by the usual rules
    of matrix multiplication. We have $\partial_{i-j}\equiv\partial_i-\partial_j$.
Due to the invariance property of the $h$-holomorphicity relative to the choice of the basis
    of the algebra, we can state that the general solution of the system (\ref{hh21})-(\ref{hh22})
    can be written by means of the representation of $U_i$ in terms of $F_i$, expressed in $x$-coordinates:
\[U_1=F(x_2-x_1-x_3)+F(x_3-x_1-x_2);\quad U_2=F(x_1-x_2-x_3)+F(x_3-x_1-x_2);\]
\begin{equation}\label{hh23}U_3=F(x_1-x_2-x_3)+F(x_2-x_1-x_3).\end{equation}
This fact can be verified by the straightforward substitution of (\ref{hh23}) into (\ref{hh21})-(\ref{hh22}).
    The combinations of coordinates in the arguments of the function $F$ are higher analogs
    of the prior and future arguments on the double plane.

The operator of third order
\begin{equation}\label{oper}\Delta^{(3)}\equiv\frac{\partial}{\partial h}\,\frac{\partial}{\partial h^\dag}\,
    \frac{\partial}{\partial h^\ddag}=(e_1+e_2+e_3)\frac{\partial}{\partial \xi_1}
    \frac{\partial}{\partial \xi_2}\frac{\partial}{\partial \xi_3}\end{equation}
is proportional to the algebraic unit, and hence we have
\begin{equation}\label{oper2}\Delta^{(3)}F=(\Delta^{(3)}U_1)j_1+(\Delta^{(3)}U_2)j_2+(\Delta^{(3)}U_3)j_3\end{equation}
for each smooth function $F(h,h^\dag,h^{\ddag})$. If the function $F$ is $h$-holomorphic,
    then due to the fact that the operator $\Delta^{(3)}$ contains differentiation by $h^\dag$ and
        $h^\ddag$, it holds true the relation $\Delta^{(3)}F\equiv0$, equivalent on its three components with:
\begin{equation}\label{oper3}\Delta^{(3)}U_i\equiv0\quad (i=1,2,3).\end{equation}
The relations (\ref{oper3}) represent the 3-dimensional analogue of the
    harmonicity conditions or of the hyperbolic harmonicity, which are identically satisfied by
    holomorphic functions of complex or accordingly, of double variable.
In $x$-coordinates, the operator $\Delta^{(3)}$ has the form:
\begin{equation}\label{oper4}\Delta^{(3)}=-\frac{1}{8}(\partial_1+\partial_2)(\partial_2+
    \partial_3)(\partial_3+\partial_1)=-\frac{1}{8}(2\partial_{123}+\partial_{112}+
    \partial_{221}+\partial_{113}+\partial_{331}+\partial_{223}+\partial_{332}),\end{equation}
where $\partial_{ijk}\equiv\partial_i\partial_j\partial_k$.
%
%
\section{The physical interpretation of an $h$-field}

The physical interpretation of the results regarding the plane of double variable will be consistently
    depend on the point of view on the  status of the space $\mathcal{H}_2$.
There exist at least three such viewpoints.
\begin{enumerate}
\item The plane $\mathcal{H}_2$ is a 2-dimensional section of the 4-dimensional Minkowski Space-Time.
    The plane (i.e., reducible to 2 dimensions) problems of the field theory in Minkowski space
    can be reformulated in terms of $h$-holomorphic functions, similar to the way in which plane
    elliptic problems of the 3-dimensional Euclidean space admit an effective formulation in terms
    of complex variable.
\item The plane $\mathcal{H}_2$ is a "toy"\, 2-dimensional Space-Time, on which the laws of physics
    can be formulated in terms of algebra of double numbers.
\item The plane $\mathcal{H}_2$ is the 2-dimensional projection of the hierarchy of spaces
    $\mathcal{H}_n$. The multi-dimensional generalizations of the double plane bring to
    the investigated circle metrics with non-quadratic intervals and equations of higher order fields,
    and the corresponding fields are considered to be fundamental.
\end{enumerate}
We shall separately examine each point viewpoint.
%
%
\subsection{2-dimensional problems of the field theory in the 4-dimensional Minkowski Space-Time}

By analogy to the applications of conformal transformations on the standard (elliptic)
    complex plane, the $h$-analytic functions can be used for solving problems of field theory,
    which are related to the 2-order wave equation: $\Box_2\varphi=0$.
The lack of acknowledgement which the hyperbolic conformal mapping receive is firstly related
    by the non-traditional way in which are posed the initial-boundary problems, which can be solved
    by the method of hyperbolic conformal transformations.
Indeed, the use of conformal transformations in the plane of complex variable for solving
    problems of elliptic type, which are related to the Laplace equation, relies on the
    circumstances which were formerly described: the analytic mapping $w=f(z)$, whose real part
    represents the solution of some boundary problem, maps the boundary from outside the sources
    of the field into the straight line $\text{Re}\, w=\text{const}$.
This requirement reflects the condition of having a constant potential on the boundary of the domain,
    in which we solve the Laplace equation and guarantees the uniqueness of the solution up to
    an arbitrary choice of the value of potential on the boundary.

In initial-boundary problems of hyperbolic type we also use another setting of the problem:
    usually, one examines the Space-Time domain as a multiple-boundary cylinder
    $D^3\times \R_+$ (or a topologically equivalent to it alias)%
\footnote{In our case, $D^3$ is the 3-dimensional ball, $\R_+=[0,\infty)$}.
    and there are provided the initial conditions on the initial surface $D^3\times\{0\}$
    (the initial values for the field and its derivatives by time) and the initial-boundary conditions
    on the flank surface $\partial D^3\times \R_+$ (the values for the and its derivatives
    by space coordinate).
If the problem is well-posed, then these initial boundary condition data lead a unique solution
    with good properties at each moment of time $t>0$. In the 2-dimensional Space-Time, the
    boundary of the domain represents a time-like rectangle $I\times \R_+$ or a topologically equivalent
    to it figure.
The use of conformal transformations -- represented by $h$-analytic functions, assumes the
    transformation of the elliptic problem on the plane to the hyperbolic problem on the plane.
In other words, a $h$-analytic function $w=f(h)$ represents the solution of some initial boundary
    problem of hyperbolic type -- namely, the one for which this function transforms the 1-dimensional
    border of the domain outside the sources into the line $\text{Re}\, w=\text{const}$.
It is obvious that such a setting of the problem differs from a standard one, since the initial
    boundary conditions change to {\it providing the form of the surface (lines) of constant
    potential}. This surface has a Space-Time character.
In principle, it can be obtained by means of measuring the wave field $\varphi$ in different points
    of the space at different moments of time using the existent (sometimes, sufficiently large)
    number of devices.
The points-events of Space-Time for which the readings of the devices provide $\varphi=\text{const}$
    form the claimed surface.
According to the considerations from above, the form of this surface uniquely determines the solution
    of the wave equation.
Still, usually such a setting of the problem is not used in practice, since the data are "distributed"\,
    in space and time.
As an example, we examine the problem of determining the wave field, which has the constant value
    $\varphi_0$ on the hyperbolic circle $t^2-x^2=R^2$. According to the results of section \ref{exppp},
    the appropriate solution has the form: $\varphi=\varphi_0+\ln[(t^2-x^2)/R^2]=\text{Re}
    (\ln h+\varphi_0)$. Its 3-dimensional graph and the subsequent time slices are shown in Fig.\ref{3ddd}.

{\centering\small\refstepcounter{figure}\label{3ddd}
    \includegraphics[width=.4\textwidth]{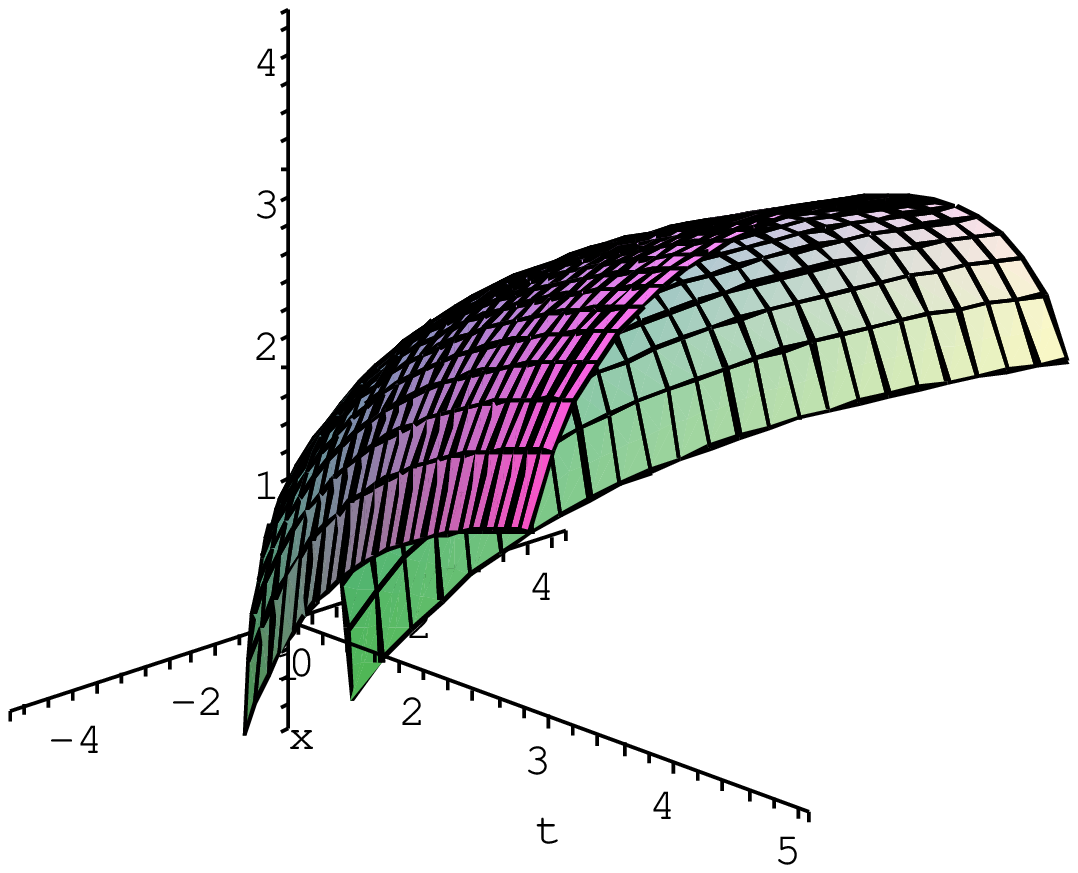}\includegraphics[width=.4\textwidth]{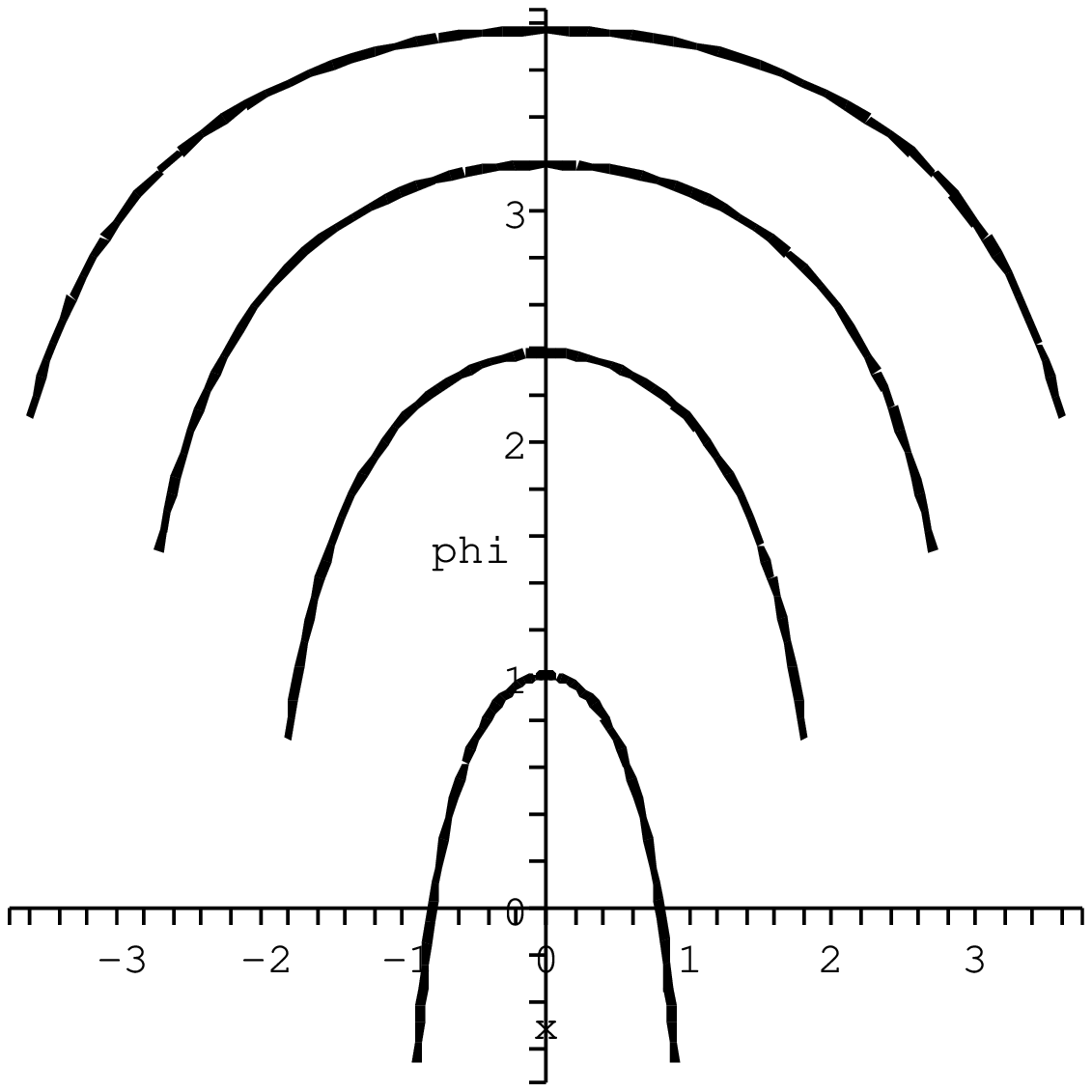}
    \medskip\nopagebreak\\ Fig.~\thefigure. The wave solution, which is constant on the
    hyperbolic circle $(R=1,\,\varphi_0=1)$. In the right picture there are represented
    the subsequent time slices of the surface, which is shown in the left picture, for $t=1,2,3,4$}.
%
%
\subsection{The 2-dimensional Minkowski universe}\label{stoooo}

In this subsection we show, that the algebra of double numbers is the natural sufficient framework
    for presenting the facts of the 2-dimensional Special Relativity Theory and 2-dimensional
    electrodynamics. Regardless of some apparent artificiality of such an approach, it proves in
    many situations to be useful and meaningful.
In particular, it dims the algebraic aspect of the pseudo-Euclidean geometry and the geometric
    aspects of the algebra of double numbers, and represents a natural foundation for further generalizations.
%
%
\subsubsection{The 2-dimensional Space-Time and related vector operations}

We shall identify the elements of $\mathcal{H}_2$ with the point-like events of the 2-dimensional
    Minkowski Space-Time $\mathcal{M}_2$.
Hence, to each element $h\in\mathcal{H}_2$ we associate a 2-dimensional position vector $h=t+jx$.
    The elements of $\mathcal{H}_2$, regarded as elements of the algebra, generate a 2-dimensional
    real linear space. We shall examine a couple of elements $h_1=t_1+jx_1$ and $h_2=t_2+jx_2$, and
    the complex-valued semi-linear form which is computed at these elements:
\begin{equation}\label{2form}h_1\bar h_2=t_1t_2-x_1x_2+j(t_2x_1-t_1x_2).\end{equation}
It is obvious, that such a form, from one side, is completely defined by means of the algebra
    $\mathcal{H}_2$, and from the other side, it defines both a real symmetric $h_1\star h_2$ and
    a skew-symmetric $h_1\times h_2$ forms by means of the formulas:
\begin{equation}\label{scal2}h_1\star h_2\equiv\text{Re}(h_1\bar h_2);\quad h_1\times h_2\equiv-
    \text{Im}(h_1\bar h_2).\end{equation}
The symmetric product%
\footnote{We still let the usual dot $\cdot$ denote the operation of multiplication in the algebra
    $\mathcal{H}_2$.}
is the 2-dimensional variant of the Minkowski pseudo-Euclidean metric, and the skew-symmetric form is the
    2-dimensional variant of the cross product, which becomes here a (pseudo)scalar and
    is essential for the geometry of oriented volumes (i.e., areas) of the space$\mathcal{M}_2$.
%
%
\subsubsection{The algebra of isometries}

The isometries groups $\text{Iso}^\star$ and $\text{Iso}^\times$ of bilinear forms (\ref{scal2})
    are well known%
\footnote{Here we ignore the translations. Their inclusion leads to the Poincar\'e groups $P_2$
    and to the affine-unimodular group $SAff(2,\R)$}: the first represents the 2-dimensional Lorentz
    group $\text{Lor}_2$, while the second represents the group of unimodular transformations $SL(2,\R)$.
    We shall examine an arbitrary (non-degenerate) inner automorphism of the algebra $\mathcal{H}_2$,
    generated by the products $h\mapsto h'=\alpha\cdot h$, where
    $\alpha=\alpha_1+j\alpha_2\in\mathcal{H}_2$ and we write it in matrix form:
\begin{equation}\label{matr1}\left(\begin{array}{c}t'\\x'\end{array}\right)
    =\left(\begin{array}{cc}\alpha_1&\alpha_2\\\alpha_2&\alpha_1\end{array}\right)
    \left(\begin{array}{c}t\\x\end{array}\right).\end{equation}
By straightforward verification we check the validity of the formulas:
\[(\alpha h_1)\star(\alpha h_2)=(\alpha_1^2-\alpha_2^2)h_1\star h_2;\quad
    (\alpha h_1)\times(\alpha h_2)=(\alpha_1^2-\alpha_2^2)h_1\times h_2.\]
By means of (\ref{matr1}), we infer the important equality:
\[\text{Aut}^{\text{int}}(\mathcal{H}_2)\supset\text{Iso}^\star
    =\text{Iso}^\times\cap\text{Aut}^{\text{int}}(\mathcal{H}_2),\]
where $\text{Aut}^{\text{int}}(\mathcal{H}_2)$ is the group of inner automorphisms of the algebra
    $\mathcal{H}_2$, generated by the multiplications with non-degenerate elements.

We note that the occurrence of the metric, associated to the skew product, allows us to
    regard the elements of $\mathcal{H}_2$ as real spinors.

The matrix of the transformation from (\ref{matr1}), which describes the inner
    automorphisms of multiplications in the algebra $\mathcal{H}_2$,
    exhibits a symmetry under its transposition relative to both the main and the second diagonals.
    We shall call such a matrix {\it absolutely symmetric}.
From the triviality of the equality $\alpha_1\cdot\alpha_2=\alpha_3\in\mathcal{H}_2$,
    which expresses the algebraic closedness of $\mathcal{H}_2$, it follows that
    {\it the absolutely symmetric matrices generate a group relative to matrix multiplication}.
This group represents a direct product $\R\times\text{Lor}_2$ and, besides the Lorentz
    transformations, it contains the homogeneous dilations: $h\mapsto \lambda h$,
    $\lambda\in \R$.
For the sake of brevity, we shall further denote this group $\text{Cn}_2$ and
    call it {\it the homogeneous conformal group on $\mathcal{H}_2$}.
It is obvious that the geometric invariant objects of this group are the cones
    $\text{Con}(h)$, $h\in\mathcal{H}_2$.

We shall examine now the discrete transformations of $\mathcal{H}_2$: $h=t+jx\mapsto\sigma_ih$
    of the following independent types:
\begin{equation}\label{dis}\sigma_th\equiv-t+jx;\quad \sigma_xh\equiv t-jx;\quad \sigma_I\equiv x+jt.\end{equation}
It is evident that, in algebraic terms, these operations can be written as :
\[\sigma_xh=\bar h;\quad \sigma_th=-\bar h;\quad \sigma_I=j\cdot h,\]
whence we can see that only the operation $\sigma_I$ allows the representation (\ref{matr1}).
%
%
\subsubsection{The co-algebra $\mathcal{H}_2^\ast$}

We shall examine the algebra $\mathcal{H}_2^\ast$, which is dual to $\mathcal{H}_2$,
    and whose elements are the linear functionals (1-forms, co-vectors) over $\mathcal{H}_2$.
We introduce the notation $\omega(h)$ for the value of the 1-form $\omega\in\mathcal{H}_2^\ast$
    on the element $h\in\mathcal{H}_2$ (this is a real number).
By choosing a basis $\{1_\ast,j_\ast\}$ in the algebra $\mathcal{H}_2^\ast$, such that this is
    dual to the basis $\{1,j\}$ in $\mathcal{H}_2$, we get the system of relations:%
\footnote{Now, strictly speaking, we should distinguish between the unity of the algebra
    $1_H\in\mathcal{H}_2$ and the real unit $1\in\R$. In the brackets of the left hand sides
    of the relations (\ref{dual}) are present exactly the units of the algebra $\mathcal{H}_2$,
    and in the right sides lie the real units from $\R$. In the case when does lead to any confusion,
    we keep the notation $1$ for both units}.:
\begin{equation}\label{dual}1_\ast(1)=1;\quad 1_\ast(j)=0;\quad j_\ast(1)=0;\quad j_\ast(j)=1.\end{equation}
Then the value of the arbitrary 1-form $\omega=T 1_\ast+X j_\ast$ on the element $h=t+jx$ will be equal to:
\begin{equation}\label{contr}\omega(h)=T t+X x.\end{equation}
Having at our disposal two nondegenerate metrics, associated with the operations $\star$
    and $\times$, we can introduce two mappings of conjugation $\mathcal{H}_2\to\mathcal{H}_2^\ast$
    by the formulas:
\begin{equation}\label{conj}h\mapsto h^\star\in\mathcal{H}_2^\ast:\ h^\star(q)\equiv h\star q;\quad
    \text{i}\quad h\mapsto h^\times\in\mathcal{H}_2^\ast:\ h^\times(q)\equiv h\times q.\end{equation}
We shall call the first conjugation {\it vectorial}, and the second -- {\it spinorial}.
    In components, considering (\ref{dual}) and (\ref{conj}), for an arbitrary $h=t+jx$ we shall have:
\begin{equation}\label{coord1}\text{Re}(h^\star)=t;\quad \text{Im}(h^\star)=-x;\quad
    \text{Re}(h^\times)=-x;\quad\text{Im}(h^\star)=t.\end{equation}
The formulas (\ref{coord1}) corresponds to the well known rules of "moving the indices"\, by
    means of the pseudo-Euclidean and of the spinorial metrics in index notations and
    establish well known isomorphisms between the linear metric spaces endowed with
    non-degenerate metrics.

Like $\mathcal{H}_2$, the co-algebra $\mathcal{H}_2^\ast$ induces a couple of operations
    of scalar product: ${\star}$ and ${\times}$ accordingly, by the rules:%
\footnote{These should have been denoted by $\stackrel{\ast}{\star}$ and $\stackrel{\ast}{\times}$, since
    these are defined among the elements of $\mathcal{H}_2^\ast$, whose nature is different
    from the one of $\mathcal{H}_2$. It will further result from the context, on which elements
    act these operations, and for avoiding complex notations we use the same symbol}.
\begin{equation}\label{scal2d}\omega_1{\star} \omega_2\equiv\text{Re}(\omega_1\bar \omega_2);\quad
    \omega_1{\times}\omega_2\equiv-\text{Im}(\omega_1\bar \omega_2),\end{equation}
which will be further called {\it co-scalar} and {\it co-skew} or {\it co-spinorial}.

The algebras $\mathcal{H}_2$ and $\mathcal{H}_2^\ast$ are mutually conjugate, i.e.,
    $(\mathcal{H}_2^\ast)^\ast=\mathcal{H}_2$.
This means, that the elements $\mathcal{H}_2$ can be regarded as 1-forms relative to the elements
    of $\mathcal{H}_2^\ast$.
We can elementarily check the following identities:
\[h_1^\star{\star}h_2^\star=h_1\star h_2;\quad h_1^\times{\times}h_2^\times=h_1\times
    h_2;\quad h_1^\times{\star}h_2^\times=-h_1\star h_2;\quad h_1^\star{\times}h_2^\star=-h_1\times h_2;\quad\]
\[h_1^\star{\star}h_2^\times=-h_1\times h_2;\quad h_1^\star{\times}h_2^\times=h_1\star h_2;\quad
    h_1^\times{\times}h_2^\star=-h_1\star h_2;\quad h_1^\times{\star}h_2^\star=h_1\times h_2.\]
Symbolically, these rules can be condensed, if we define {\it the conjugation table}
    by means of the pair of elements $\{(\star),(\times)\}$ relative to the pair of
    operations $\{\star,\times\}$:
\begin{equation}\label{bigr} \star:\quad \begin{tabular}{c|c|c}$2\setminus 1$&$(\star)$&$(\times)$\\ \hline
    $\star$& $\star$& $\times$\\$\times$&$-\times$&$-\star$\end{tabular}\qquad \times:\quad
\begin{tabular}{c|c|c}$2\setminus 1$&$(\star)$&$(\times)$\\ \hline $\star$& $-\times$&$ -\star$\\
    $\times$&$\star$&$\times$\end{tabular}\end{equation}
%
%
\subsubsection{Reference systems in $\mathcal{M}_2$}

From the position of the standard Special Relativity Theory, the elements of the introduced
    co-algebra $\mathcal{H}_2^\ast$ represent different reference systems.
More exactly, we define {\it the class $\mathcal{IR}$ of inertial reference systems in
    $\mathcal{M}_2$ as the set of elements of the projective subalgebra $P\mathcal{H}_2^\ast$
    by multiplication.}
Its elements, at their turn, can be regarded as points on the unit hyperbolic circle
    $|\omega\bar \omega|=1$ in $\mathcal{H}_2^\ast$.
This circle has in $\mathcal{H}_2^\ast$ 4 connected components: on two of them, we have
    $\omega\bar\omega=+1$, and on the other two, $\omega\bar\omega=-1$.
We shall call the subclass of the reference systems from the first two components
    {\it causal} (it corresponds to reference systems with sub-luminal speeds)
    and we denote it by $\mathcal{IR}_+$, and the class of reference systems from the other two components
    {\it a-causal} (it corresponds to reference systems with sub-luminal speeds)
    and we denote it by $\mathcal{IR}_-$.
In each of these subclasses there we distinguish two more connected components: in the
    positive and in the negative semiplanes $\text{Im}\omega\gtrless0$ for $\mathcal{IR}_+$ and
    $\text{Re}\omega\gtrless0$ for $\mathcal{IR}_-$.
We shall them respectively by $\mathcal{IR}^\uparrow$ and $\mathcal{IR}^\downarrow$, and
    we shall call them {\it positive} and {\it negative} components.
Hence, the complete class $\mathcal{IR}$ of all the inertial reference systems allows the following splitting:
\begin{equation}\label{decfr}\mathcal{IR}=\mathcal{IR}_+^\uparrow\bigcup\mathcal{IR}_+^\downarrow
    \bigcup\mathcal{IR}_-^\uparrow\bigcup\mathcal{IR}_-^\downarrow.\end{equation}
It is easy to verify that the components of this splitting result from the class
    $\mathcal{IR}_+^\uparrow$ of positive causal reference systems by means of the
    discrete operation (\ref{dis}):
\begin{equation}\label{divid}\mathcal{IR}_+^\downarrow=\sigma_t\mathcal{IR}_+^\uparrow;\quad
    \mathcal{IR}_-^\uparrow=\sigma_I\mathcal{IR}_+^\uparrow\quad
    \mathcal{IR}_-^\downarrow=\sigma_x\sigma_I \mathcal{IR}_+^\uparrow.\end{equation}
It should be noted that, in view of the formal equivalence of all the
    coordinate quadrants on the planes $\mathcal{H}_2$ or $\mathcal{H}_2^\ast$, the splitting
    (\ref{divid}) looks -- to a certain extent -- conditional.

We shall examine now a "normal"\, (i.e., sub-luminal and future-oriented) reference system,
    chosen from the component $\mathcal{IR}_+^\uparrow$.
To this, there corresponds some element $\tau\in\mathcal{H}_2^\ast$, which in the chosen by us basis
    has the form: $\tau=T 1_\ast+X j_\ast$, and its components $T$ and $X$ satisfy the conditions:
\begin{equation}\label{cond}T>0;\quad T^2-X^2=1.\end{equation}
The first condition expresses the positiveness of the orientation of the
    time reference, the second -- the universal (constant) "unitary proper time increase".
Formally, the second one relates to  the unitary norming of the co-vector $\tau$.
    The conditions (\ref{cond}) are automatically satisfied by the parametrization:
\begin{equation}\label{param}T=\cosh\psi;\quad X=\sinh\psi,\end{equation}
where the parameter $\psi$ has the geometric meaning of hyperbolic angle in the hyperbolic
    polar coordinate system, and the physical meaning of the renowned speed parameter
    ($\tanh\psi=v$, where $v$ is the spatial speed of the reference system).
Inside the components (in terms of parametrized speed $v$) the 1-form $\tau$ gets the form:
\begin{equation}\label{param1}\tau=\frac{1}{\sqrt{1-v^2}}1_\ast-\frac{v}{\sqrt{1-v^2}}j_\ast.\end{equation}

Now, for the given reference system $\tau\in\mathcal{H}_2^\ast$ and for any element-event
    $h\in\mathcal{H}_2$, we can define its {\it time component $h_T^\tau$ relative to the
    reference system $\tau$} by means of the simple formula:
\begin{equation}\label{T}h_T^\tau\equiv \tau(h).\end{equation}
For arbitrary $h=t+jx$, using the formulas (\ref{contr}), (\ref{param}) and (\ref{param1}),
    the definition (\ref{T}) gives the temporal part of the Lorentz transformation:
\[h_T^\tau\equiv \frac{t-vx}{\sqrt{1-v^2}}.\]
For the vector $\Delta h=\Delta t+j\cdot0$ which characterized the time interval in
    the rest system of some clocks, we get the formula of relativistic stretching
    of time intervals:
\[h_T^\tau(\Delta h)=\Delta t'=\frac{\Delta t}{\sqrt{1-v^2}}.\]

In order to provide the definition of spatial projections of events, we need to define
    the unitary 1-form $s$ from the component $\mathcal{IR}_-^\uparrow$, which is
    orthogonal to $\tau$, i.e., which satisfies the relation: $s\star \tau=0$.
Using the formulas (\ref{scal2d}) and (\ref{param1}) it is easy to find its coordinate form:
\begin{equation}\label{param2}s=-\frac{v}{\sqrt{1-v^2}}1_\ast+\frac{1}{\sqrt{1-v^2}}j_\ast.\end{equation}
Now, using the formulas (\ref{contr}) and (\ref{param2}), we can provide for {\it the
    space projection of an arbitrary event $h\in\mathcal{H}_2$ relative to the reference
    system $\tau$}, the following definition:
\begin{equation}\label{s}h_X^\tau\equiv s(h)=\frac{x-vt}{\sqrt{1-v^2}},\end{equation}
which gives in fact the space part of the Lorentz transformation.

We shall examine the identity linear operator $\hat I\equiv 1\otimes1_\ast+j\otimes j_\ast$.
    By direct verification using the formulas (\ref{param1}) and (\ref{param2}),
    we can check the rightness of the following decomposition of this operator:
\begin{equation}\label{unit}\hat I=\tau^\star\otimes \tau-s^\star\otimes s.\end{equation}
Acting by this operator on the vector-events or 1-forms, we get their
    decomposition into spatial and temporal components:
\[h=h^\tau_T \tau^\star+h^\tau_X s^\star;\quad \omega=\omega^\tau_T\tau+\omega^\tau_X s,\]
where $h\in \mathcal{H}_2$, $\omega\in\mathcal{H}_2^\ast$ and
\[h^\tau_T\equiv \tau(h);\quad h_X^\tau\equiv s(h);\quad \omega_T^\tau\equiv
    \omega(\tau^\star);\quad \omega^\tau_X\equiv\omega(s^\star).\]
Similarly, by replacing the expansion of the unit operator in the tensor bundle
    $T^{(r,s)}(\mathcal{H}_2)$:
\[\hat I^{\otimes (r+s)}=(\tau^\star\otimes \tau-s^\star\otimes s)^{\otimes(r+s)}\]
we can split any tensor on $\mathcal{H}_2$ into spatial-temporal components.
    As an example, the metric tensors $g^\star$ and $g^\times$, associated with
    the symmetric and skew products, respectively have the following representations:
\[g^\star=\tau\otimes\tau-s\otimes s;\quad g^\times= \tau\wedge s,\]
which essentially have the meaning of the 2-dimensional (diadic) analogue of the tetradic description
    of the quantities in Special Relativity Theory and General Relativity Theory \cite{vlad}.
We note that the 2-dimensional character resides in the occurrence of the bijective correspondence
    between the spatial and the temporal elements of the diad $\{\tau,s\}$.

We shall examine a couple of elements $\tau_1$ and $\tau_2$ from $\mathcal{H}_2^\ast$,
    which is parametrized by the speeds $v_1$ and $v_2$ by formula (\ref{param1}).
It is easy to check that their derivatives in the double algebra $\mathcal{H}_2^\ast$
    are defined by the element
\[\tau=\tau_1\cdot\tau_2=\frac{1}{\sqrt{1-v^2}}1_\ast-\frac{v}{\sqrt{1-v^2}}j_\ast,\]
where
\begin{equation}\label{relsum}v=\frac{v_1+v_2}{1-v_1v_2}.\end{equation}
In other words, a consecutive change of the reference system is described by the
    multiplication of the algebra of double numbers with the corresponding elements from
    $\mathcal{H}_2^\ast$.
This multiplication automatically induces a relativistic law of velocity addition.
    An interesting consequence of this fact is related to the algebraic interpretation
    of active and passive transformations: {\it the multiplication with of the normed to unity elements
    of the co-algebra $\mathcal{H}_2^\ast$ describe the passive Lorentz transformations
    (the change of point of view for the same events), while the
    multiplication with of the normed to unity elements of the algebra $\mathcal{H}_2$
    describe the active Lorentz transformations (the transition to other events, which
    we observe from the same viewpoint)}.
From here, in particular, there follow the identities:
\[\tau_v(\alpha_v\cdot h)=t;\quad s_v(\alpha_v\cdot h)=x\]
for each element $h=t+jx$ and elements $\alpha_v\in P\mathcal{H}_2$,
    $\tau_v,s_v\in P\mathcal{H}_2^\ast$, parametrized by the same parameter $v$.
These identities are the mathematical expression of the following assertion:
    {\it the event which transforms by an active boost into a new event, does not change
    its Space-Time projections in the reference system which corresponds to this boost}.

All the constructions from above allow a localization i.e., a transition to
    differential-geometric objects (tangent to the vectors and to differential 1-forms).
For this it suffices to allow the dependence of the parameter $v$ on $t$ and $x$,
    and to regard all the constructions in the tangent and cotangent spaces
    $T_{(t,x)}\mathcal{H}_2$ and $T_{(t,x)}^\ast\mathcal{H}_2=T_{(t,x)}\mathcal{H}_2^\ast$.
Such a transition permits to examine extended deformed reference systems and even
    to take into account gravitation.
%
%
\subsubsection{Dynamics of Special Relativity Theory in representation of double numbers}

A smooth curve $\Gamma$ on $\mathcal{H}_2$ can be described by means of the parametric dependence
    $h(w)=t(w)+jx(w)$, where $w$ is a real parameter, and $t(w),x(w)$ are smooth functions
    which depend on it.
The velocity vector field on this curve has the form:
\[\dot h=\frac{dh}{dw}=\dot t+j\dot x,\]
and the physical speed $v$ is defined by the relation:
\begin{equation}\label{velf}v=\frac{\text{Im}\dot h}{\text{Re}\dot h}=\frac{\dot x}{\dot t}\end{equation}
and does not depend on parametrization. We shall call the curve {\it causal and future-oriented},
    if for any reference system we have $\tau\in \mathcal{IR}_+^\uparrow$ $\tau(\dot h)>0$ for all $w$.
Every such curve allows a natural parametrization
\begin{equation}\label{nat}s=\int\limits_{w_0}^w\sqrt{\dot h\star \dot h}\,
    dw=\int\limits_{t_0}^t\sqrt{1-v^2}\, dt\end{equation}
in which we have $\dot h\star \dot h=1$. The vector $\dot h$ in the natural parametrization will
    be called {\it 2-velocity} $V$ on the curve $\Gamma$.
It is easy to show, using relations (\ref{velf}) and (\ref{nat}), that it can be represented by:
\begin{equation}\label{2vel}V\equiv \frac{dh}{ds}
    =\frac{1}{\sqrt{1-v^2(s)}}+j\frac{v(s)}{\sqrt{1-v^2(s)}}.\end{equation}

Causal and oriented curves can be interpreted as world lines os massive point-like particles,
    which are displaced in a 2-dimensional Space-Time. For each such a particle we can define
    a 2-moment:
\[P\equiv mV=\frac{m}{\sqrt{1-v^2(s)}}+j\frac{mv(s)}{\sqrt{1-v^2(s)}},\]
where $m$ is the rest mass of the particle, which determines the value of the invariant
\begin{equation}\label{inv1}P\star P=P\cdot\bar P=m^2.\end{equation}
We define the 2-force $F_0$ in the reference system, which travels with the particle as an
    element of $\mathcal{H}_2$ of the form:
\begin{equation}\label{2f}F_0=jf,\end{equation}
where $f$ is the 1-dimensional force scalar, measured by means of a dynamometer in the
    moving system. The 2-velocity vector $V_0$, in this system, has the form: $V_0=1$ and,
    obviously we have
\begin{equation}\label{orto}F_0\star V_0=0.\end{equation}
In a laboratory reference system, the 2-velocity is described by the element $V$,
    which is related to $V_0$ via the active boost $\alpha_{v}$: $\alpha_{v}\cdot V_0=V$.
A similar relation links the elements $F$ and $F_0$: $\alpha_{v}\cdot F_0=F$, whence by means of the
    algebra $\mathcal{H}_2$, we get:
\begin{equation}\label{2f2}F=\alpha_{v}\cdot F_0=\frac{vf}{\sqrt{1-v^2}}+j\frac{f}{\sqrt{1-v^2}}.\end{equation}
Due to both the Lorentz-invariance of the operation $\star$ and to (\ref{orto}),
    we have the following general orthogonality relation:
\begin{equation}\label{orto1}F\star V=0,\end{equation}
which reflects the established in Special Relativity Theory circumstance of spatial space-likeness
    of the forces which act onto a point.
This occurrence, in fact, is embedded in the structure of the structure of the relativistic equations from dynamics:
\begin{equation}\label{din2}\dot P=F\end{equation}
where we take into account the relation $\dot P\star P=0$, which follows from (\ref{inv1}).
    In components, the equations (\ref{din2}) get the form
\begin{equation}\label{dyn21}\frac{d}{ds}m(1-v^2)^{-1/2}+j\frac{d}{ds}mv(1-v^2)^{-1/2}=vf+jf.\end{equation}
The real part of this equality describes {\em the law of conservation of relativistic energy}
    $E=\text{Re}P=m(1-v^2)^{-1/2}$ while considering the power $vf$ of the active force,
    a consequence of the imaginary part of (\ref{dyn21}) which expresses the second relativistic Newton law.
%
%
\subsubsection{Particles in "electromagnetic field"\, on the double plane}

A 2-dimensional analogue of the potential of the "electromagnetic field"\, in 2-dimensional
    Space-Time of Special Relativity Theory, is the 1-form $\mathcal{A}=A_01_\ast+A_1j_\ast$.
Its exterior differential (curl) has the form
\[d\mathcal{A}\equiv \nabla\times\mathcal{A}=\frac{\partial A_1}{\partial t}-
    \frac{\partial A_0}{\partial x}\equiv\mathcal{E}\]
and is the unique (pseudo)scalar which describes the strength of the electromagnetic field on
    the double plane.
Here is introduced the complex operator $\nabla\equiv\partial_t+j\partial_x$.
    The pseudo-Euclidean square of the tensor $\mathcal{E}1_\ast\times j_\ast$
    defines the first (and in our case the unique) invariant:
\[I=F_{ik}F^{ik}=-2F_{01}^2=-2\mathcal{E}^2,\]
whose sign is negative. This means that in the 2-dimensional Space-Time, the electromagnetic
    field exists only in its "electric occurrence".

The motion of the charges which are distributed in $\mathcal{H}_2$ is described by the
    2-vector of current density
\begin{equation}\label{curr}J=\rho_0V=\rho+j\rho v,\end{equation}
where $\rho_0$ is the charge density in its rest reference system, $\rho=\rho_0/\sqrt{1-v^2}$ is the
    density of the charge in the laboratory system, $V$ is the 2-velocity field of the charge, and
    $v$ is the field of physical velocity of the charge.
The action for the system "charge + electric field"\, gets the form:
\begin{equation}\label{acttot}S=-\sum\limits_k m_k\int\sqrt{dh_k\star dh_k}-\sum\limits_k
    q_k\mathcal{A}(dh_k) +\frac{j}{8\pi}\int \mathcal{E}^2\, dh\wedge d\bar h.\end{equation}
By varying the actions (\ref{acttot}) by the coordinates of the particles, we get
    the equations of motion for the particles displaced in an exterior field:
\begin{equation}\label{eqmot}\dot V=j\frac{q}{m}\mathcal{E}\cdot V\quad\text{or}\quad
    \frac{dE}{dt}+j\frac{t}{dt}\frac{mv}{\sqrt{1-v^2}}=q\mathcal{E}v+jq\mathcal{E},\end{equation}
which contain as imaginary component the equation of the motion of a charged particle
    under the action of the purely electric Lorentz force.
The variation of the action (\ref{acttot}) by the components of the potential $\mathcal{A}$
    leads to the Maxwell equations:
\begin{equation}\label{maxw2}\bar\nabla \mathcal{E}=-2\pi j J\quad\text{or}\quad
    \frac{\partial\mathcal{E}}{\partial t}-j\frac{\partial \mathcal{E}}{\partial x}
    =-2\pi\rho(v+j).\end{equation}
The imaginary part of these equations corresponds to the 1-dimensional version of the Gauss theorem,
    and the real one -- to the law of the "total field", which simply expresses the canceling of
    the total field.
By applying to both sides of the first equation (\ref{maxw2}) the operation $\bar\nabla\times$, we get
    due to the identity $\bar\nabla\times\bar\nabla=0$, the conservation law of the electric charge:
\begin{equation}\label{conserv2}\bar\nabla\times jJ=0\quad\text{ili}\quad \frac{\partial \rho}{\partial
    t}+\frac{\partial \rho v}{\partial x}=0.\end{equation}
%
%
\subsection{The theory of the fundamental $h$-analytic field}

In this section we provide some hints regarding the theoretic-field approach to $h$-holomorphic
    fields, relying on the principle of least action.
While we construct the Lagrangian, we shall emerge from taking into consideration the covariance,
    symmetry and simplicity. We shall examine the formal theory of the field $F(h,\bar h)$
    on the plane of double variable, whose action is given by the expression:
\begin{equation}\label{act1}\mathcal{S}[F,j]=\frac{j}{2}\int\limits_{\mathcal{H}_2}(F_{,h}\star F_{,h}-
    \mu^2 F\star F-F\star J)\, dh\wedge d\bar h,\end{equation}
where $J$ is the hyper-source of the field $F$, $\mu$ is the mass parameter of the field $F$.
    The whole expression (\ref{act1}) is real and is a 2-dimensional relativistic invariant.
    Passing to components $F=U+jV$ and writing the $\star$-product by formulas (\ref{scal2}),
    we infer the action for the pair of mutually "non-interacting"\, massive scalar fields:
\begin{equation}\label{act2}\mathcal{S}[F,\bar F, J,\bar J]=\mathcal{S}[U,J^0]-\mathcal{S}[V,J^1],\end{equation}
where
\begin{equation}\label{act3}\mathcal{S}[\phi,X]\equiv\int\limits_{\R^2}\frac{1}{2}(\phi_{,t}^2-
    \phi_{,x}^2-\mu^2\phi^2-\phi X)\, dt\wedge dx.\end{equation}
The standard variational procedure for the action (\ref{act1}) in terms of $F$ and $\bar F$
    leads to the Klein-Gordon equation:
\begin{equation}\label{act5}(\Box+\mu^2)F=-J\end{equation}
and to the complex-conjugate equation. The componentwise form of the equation (\ref{act5})
    is equivalent to the system of Klein-Gordon equations for the scalar fields $U$ and $V$,
    which can be straightforward derived from (\ref{act2}) by variation in terms of $U$ and $V$.
From the equation (\ref{act5}) and formula (\ref{hlap}) it follows, that in the space outside sources,
    for $\mu=0$, the solution of the field equations is provided by some $h$-analytic or
    $h$-anti-analytic function.

The generalization of the action (\ref{act2}) on $\mathcal{H}_3$ (further to $\mathcal{H}_n$ this is
    straightforward) has the form:
\begin{equation}\label{act6}\mathcal{S}[F,J]=\frac{1}{4}\int\limits_{\mathcal{H}_3}
    \left\{F_{,h}(F_{h})^{\dag}(F_{h})^{\ddag}-\mu^3 FF^\dag F^\ddag-\langle F,J\rangle\right\}\,
    dh\wedge dh^\dag\wedge dh^\ddag,\end{equation}
where $J$ is the tensorial hyper-current ($J$ is here a symmetric tensor with components
    $J^{\alpha\beta}$, $\alpha,\beta=1,2,3$), and the angular brackets represent the scalar product
    (\ref{3lin}) of the Berwald-Moor metric.
The field equations which are obtained from the action (\ref{act6}) are nonlinear differential equations
    of second order.
We shall not present them here, having in view that we shall specially deal with
    them in a forecoming paper.
%
%
\section{Conclusions}

Confining to the common methodological setting, according to which the active force and
    the ultimate criterion for the validity of scientific knowledge is the experiment,
    in the practice of actual scientific research it is widely applied the opposed,
    "neo-Pythagorean methodology"\,, in which Mathematics provides as source of ideas
    and representations of the ambient world.
At the first sight it might appear that such an approach has weak chances for being successful
    due to the immense number of different mathematical constructions, which might
    be proposed as potential basis for such investigations.
Still, the number of candidates can be effectively limited, if one emerges not from random, but from
    most elementary and fundamental mathematical objects.
Such simple objects which should be first considered, are the {\em numbers}.
    Still, there exist sufficiently many different classes of numbers.
    Besides the common ones --- in which normally are included natural, integers, rational
    real and complex numbers, there are well known such ones, as quaternions, $p$-adic numbers,
    Clifford and Grassmann numbers, etc.
The organic interweaving of the properties of these classes with the real world, is a fact.
    A remarkable illustrative example is provided by the complex numbers, whose properties
    not only have a direct relation to the geometry of the 2-dimensional Euclidean space
    (which, in fact, at its time contributed as basic argument in setting the agreement
    among mathematicians to consider the complex algebra as an extension of real numbers),
    but also have beautiful and fruitful links with Physics, especially in the area of complex analysis.
To each analytic function of complex variable, one can associate a concrete 2-dimensional physical field,
    (e.g., the electro- and magnetostatic ones).
The converse statement is true as well: to each combination of sources and vortices,
    which provides an ideal 2-dimensional field in the vacuum, we can always look for
   "its"\, analytic function of complex variable.
It is considered that the 2-dimensional case is singular - when looking for the
    tightest interference of properties of numbers, geometry and physics.
In algebra it is proved an important theorem, whose proof belongs to Frobenius,
    which states that numeric sets with dimension greater or equal to three,
    which inherit absolutely all the properties of real and complex algebras, do not exist.
Together with this, there exist no possibility of relating the geometry of multi-dimensional spaces
    with the "good"\, multi-component algebras.
The quaternions, discovered by Hamilton, do not provide a numeric field, since their multiplication,
    unlike the multiplication of real and complex numbers, is non-commutative.
Moreover, there exist no non-trivial analytic functions over quaternions: the most "complex"
    such functions are the homographic functions.
The present circumstances are tightly related to the Liouville theorem which, if
    applied to the 4-dimensional Euclidean space related to the algebra of quaternions,
    asserts that the conformal group is 15-parametric, while on the complex plane
    and on the real line, the corresponding conformal groups are infinite-dimensional.
Relating this fact to the multitude of analytic functions of real and complex variables,
    one clearly concludes that the absence of diversity among conformal transformations
    leads to a harsh drop in geometric and physical applications for quaternions and
    functions over them, a situation similar to some extent to the amount of applications of the
    corresponding numeric fields.

Confronting with such circumstances, and becoming somehow reluctant after perceiving the
    successful interaction picture between algebra, geometry and physics in the field of
    complex analysis, most of the mathematicians ceased to look for alternative extensions
    for the known list of numbers.
Few enthusiasts continue to deal with this problem, focusing their attention to searching
    such changes in understanding the notion of analytic function, which might have embedded in itself
    both the algebra of quaternions and the theory of analytic functions.
In particular, the corresponding attempts are effective in the framework of the algebra of
    complex quaternions or bi-quaternions -- as they are called sometimes (\cite{1a,2,3,4}),
    which do not generate a numeric field and have a relatively poor group of symmetries.

In this paper, we made an overview of properties of double numbers and of functions defined over them.
    Doing this, a remarkable fact was clarified: the analogy between the double numbers with
    the complex ones is by far deeper, then a simple formal coincidence.
We can say, that all the links between complex numbers and the geometry of Euclidean plane has
    analogies in the form of links between double numbers and the geometry of 2-dimensional Space-Time.
To the operations of addition and multiplication of double numbers there correspond translations, rotations
    and dilations of the pseudo-Euclidean plane.
There make sense the notions of conjugate number, module and argument, algebraic and
    exponential form of the representation, and there hold true analogues for the formulas due to
    Euler, Stokes, Ostrogradski, Cauchy, etc., in a natural way there are introduced the notions of
    derivative independent of direction, and of analyticity of functions (\cite{6,7}).
It is hard to imagine a property which takes place in the plane of complex variable and
    has no analog on the plane of double variable.
As weird as it seems, in spite of all these, double numbers have relatively few practical
    applications and are less popular than the complex ones.

Apparently, the main reason of such a reckless relation of mathematics and physicists towards
    double numbers is hidden in the too evident simplicity of their their constitution,
    which sometimes is close to triviality.
For these numbers it is easy to find a basis (which is called {\em isotropic}, see Section \ref{isobase}),
    relative to which they split into two independent real algebras.
As a result, one gets the impression that the plane of double variable contains nothing,
    except the properties of such a couple.
In this paper we tried to show the evident wrongness of such a viewpoint.
    It is well known that double numbers practically satisfy all the axioms of a numeric field,
    except the fact that it admits zero-divisors, i.e., objects with nonzero components,
    whose module is equal to zero and which do not admit an inverse relative to the multiplication,
    as in the case of the usual zero number.
Often, the mathematics consider that this quality is "bad"\, or at least non-desirable.
    This is why the presented above Frobenius theorem did not deal in principle with algebras
    which have zero-divisors.
Such a limitation, from the position of the proposed approach -- which is based on searching
    numbers which are tightly and naturally related to geometry and physics -- completely lacks justification.
The algebra of double numbers corresponds to the geometry of the 2-dimensional pseudo-Euclidean
    Space-Time (\cite{6}), and the latter one -- as it is well known -- has a fundamental object,
    which does not appear (at least in its real form) in any Euclidean space.
This is the light cone or, in other words, the set of points whose distance (or, as it is
    also called, interval) towards a fixed point is equal to zero.
The points and the vectors of the light cone are the ones to which there correspond the zero-divisors
    of the algebra of double numbers. Namely, the basic impediment -- which the mathematics consider to
    prevent double numbers to be regarded at the same level as the real and complex numbers,
    in fact proves to be an indispensable element both of pseudo-Euclidean geometry and to the related
    to it relativistic physics.
If from the conditions of the Frobenius theorem one removes the requirement of the absence of
    zero divisors of the algebra and has in view that to numbers may correspond not only Euclidean,
    but also pseudo-Euclidean spaces, then the pessimistic conclusion regarding the limitation of the
    realm of fundamental numeric objects by the complex algebra becomes false and there appear serious
    reasons for completing the list of numeric systems, first of all, by including into it the double numbers.

A considerable fact, which supports the need of considering double numbers as fundamental
    objects, is the occurrence over them of an infinitely-parametrized set of holomorphic
    functions (more correctly, $h$-holomorphic, \cite{6}, since on the plane $\mathcal{H}_2$
    of hyperbolic variable, the natural topology is non-Euclidean).
Moreover, the notion of $h$-holomorphic function of double variable is defined in such a way, that
    the diversity of such functions is comparable to the diversity of holomorphic functions
    of complex variable, and any function from one set corresponds to a unique function
    of the other one (\cite{9}).

One of the remarkable properties of the complex plane, which was recently discovered,
    is the construction on this plane, by means of computer software, of the fractal sets
    Julia and Mandelbrot (\cite{10}).
The beauty, harmony and consistence of such objects (\cite{11}) appear as a supplementary
    confirmation of the existent relation between algebra and
    geometry.
There exists the opinion that -- in principle, one cannot construct on the plane of double variable
    fractal or fractal-like objects of similar complexity.
Many researchers which study this problem obtained trivially smooth squares and rectangles (\cite{12,13,14})
    instead of the infinite repetitive fragmentation of borders -- specific to complex fractals.
But, recent research discovers that the real situation is -- in the hyperbolic case, significantly more
    interesting (\cite{15,16,17}).
If instead of limit fractal sets Julia one studies the so called pre-fractals, which differ from
    the previous ones by taking a finite number of iterations, then instead smooth borders of
    rectangles on the plane of double variable appear objects which little differ from the pre-fractals
    on the complex plane.

We shall briefly examine the interesting potential physical interpretations of the fields
    provided by $h$-holomorphic functions. Like on the complex plane, the group of motions of
    the double plane is produced by the addition of double numbers and by the multiplication
    with numbers of the unitary module, with the essential difference, that rotations on
    the double plane are hyperbolic.
The straightforward physical interpretations of such algebraic and geometric properties
    of double numbers are also natural, and are usually related with problems of
    Special Relativity Theory, when one has at least two significant coordinates:
    one spatial, and one temporal (\cite{18}; see, as well, Section \ref{stoooo}).
A much more interesting question is the question regarding the physical realization of the fields,
    which are given by $h$-holomorphic functions, and the observation of physical effects,
    which are related to them.
We stress the fact that we discuss here about looking for relations, which are -- in a some sense --
    analogous to those, which are known since long ago, between holomorphic functions of complex
    variable, conformal transformations of the Euclidean plane, and their corresponding
    physical applications.
Should one expect the occurrence of similar beautiful and significant for physicists relations
    between $h$-holomorphic functions of double variable and some simple enough
    (effectively 2-dimensional) real physical events?.
The Sections \ref{physh}-\ref{physh5} suggest a positive answer to this question.
    These propose to naturally assume that the Space-Time mapping, which is shown in
    Fig.\ref{hqull} can be regarded as a field with point source with the charge $q$,
    which is located at the point-event with coordinates $(0,0)$. The force lines
    of this field -- like for the source on the complex plane, are radial lines, and the level
    lines are concentric circles -- not Euclidean, but pseudo-Euclidean, since
    these are represented by quadratic hyperbolas.
Of course, this source is of a different nature than the one related to the logarithm
    on the complex plane, since the metric on the plane of double variable is completely different
    and we deal not with a spatial vector field, but with a Space-Time field.
Moreover, if on the complex plane the logarithm was losing its analyticity at the only point
    with coordinates $(0,0)$, then on the plane of double variable the logarithm ceases to be an
    $h$-analytic function not only at the point $(0,0)$, but also on the related to it
    isotropic (light-like) cone $\text{Con}(0)$.
Regardless of these differences, we still deal with a source, which in order to be distinguished
    from the complex analogue, will be named {\em hyperbolic}, and the measure of its charge
    $q$ will be called hyper-charge.

The electric vector field $\mathcal{E}$ (\ref{hqul1}), produced by a hyper-charge has
    much in common with the strength of the 2-dimensional electrostatic field $E(x,y)=E_x+iE_y$.
These have the same formulas which relate them to the potentials and the same portraits of
    field and potential lines, if one studies it by means of of analogous geometric objects
    (straight lines and circles).
These fields have as well similar quantitative characteristics, which are expressed by the formulas
    (\ref{formuc}) and (\ref{potcylh1})-(\ref{potcylh2}).
Hence, we have in fact a complete hyperbolic analog of the 2-dimensional potential field of the
    electric strength $E$ given by the unit charge, with the difference that in
    our case the field $\mathcal{E}$ id produced by a single hyper-charge and fills not the
    whole space, but the Space-Time.
Similarly, we can discuss about the analogy between the dual quantities $\mathcal{B}$ and $B$
    in the dual interpretation of electric fields.

The duality and the hyperbolic-elliptic symmetry between the strength of the pair of fields $E$ and
    $B$ on the complex plane, and the strengths $\mathcal{E}$ and $\mathcal{B}$ on the plane
    of double variable, are surprisingly harmonic and complete.
It would have been strange to exist a 2-dimensional realization in the physical world for the first pair,
    but not to have such a realization for the second one.
A simple selection of fundamental interactions known by the today physics shows, that a straightforward
    realization of the examined pair $\mathcal{E}$ and $\mathcal{B}$ as it seems, does not exist in nature.
Hence we may assume, that these fields are related with the fields of the well known interactions
    only indirectly, or intermediately (by means of some of their combinations) or they provide
    a complementary fundamental interaction, whose properties -- while passing to the cases when
    only two dimensions essentially remain (one of them being the time), are the ones provided
    by $h$-analytic functions of double variable.
We shall call for brevity the pair of fields $\mathcal{E}$ and $\mathcal{B}$ a {\em hyperbolic field}.

Hence, based solely on mathematical considerations and deep mathematical analogies with
    the well known electromagnetic field, we are led to interesting hypotheses which obviously
    need further theoretic development and experimental verification.
\begin{enumerate}\item
In fact, the sources of the hyperbolic field are not the elementary particles -- as it happened
    in the case of the electric field, but the point-level events in Space-Time, where the role of
    charges is taken by hyper-charges, whose relation to charges and masses is subject of clarification.
\item The strength of the electric hyperbolic field, generated by hyper-charges, has hyperbolic
    character, i.e., the force lines are not filling the space, but the Space-Time, and the
    equi-potential lines of the field (in the 2-dimensional case) are hyperbolas.
 \item The hyperbolic electric field -- like the usual electric field on the Euclidean plane,
    has its own dual analogue: the hyperbolic magnetic field, whose force lines are pseudo-Euclidean
    circles, i.e. hyperbolas, and the sources are the hyper-vortices.
\item Hyper-charges and hyper-vortices are related not to a particle, but to a point-like
    elementary event.
\item  The sources of the hyperbolic field may be not necessarily positive or negative, but
    real, imaginary and hyper-complex quantities.
\item The strength of both components of the hyperbolic field, which are related to
    a 2-dimensional single vortex-source, proportionally decrease with the length
    of the interval to the point at which this is displaced.
\item The experimental observation of the complementary interaction, related to hyperbolic fields,
    needs, as it seems, special experiments including the measuring of time intervals,
    relative to different physical processes.
We shall insist on describing such experiments in a separate paper.
\item The natural extensions of the hyperbolic field to three and four dimensions should be related
    with the natural extensions of double numbers $\mathcal{H}_2$ to three-numbers $\mathcal{H}_3$
    and to quadri-numbers $\mathcal{H}_4$.
Such algebras generate no more the pseudo-Euclidean Space-Time, but a linear Finslerian space
    endowed with Berwald-Moor metric (\cite{19,20,21,22,23}).
\item In multi-dimensional Finsler spaces with Berwald-Moor metric, besides lengths and
    angles, there exist supplementary natural metric invariants, e.g., the so called tringles
    and poly-angles \cite{24,25}, to which there are related not only the $h$-holomorphic,
    but also more complex functions.
The diversity of such functions is considerably larger, and their properties are much more interesting.
    It might happen, that the study such functions clarifies the unification of the hyperbolic and
    electromagnetic fields in the framework of the notion of multicomponent hypercomplex numbers.
\end{enumerate}
Further research should shed light onto the rightness of the hypotheses which were stated above.
    A detailed examination of properties of $h$-holomorphic functions of double numbers
    relative to their applications to geometry and physics already suggests the necessity
    of studying other commutative-associative algebras with zero-divisors having in view their
    own physical interpretations.
If one takes into consideration that to such multi-component algebras there correspond not the Euclidean
    or pseudo-Euclidean spaces, but their linear Finslerian extensions (\cite{24}), then the partial or
    total replacement of the Minkowski Space-Time by such extensions, may prove to be consistent and fruitful.

{\small\noindent
D.G. Pavlov\\
Research Institute of Hypercomplex Systems in Geometry and Physics,\\
Fryazino, Russia.\\
E-mail: geom2004@mail.ru\\\\
S.S. Kokarev\\
RSEC "Logos", Yaroslavl, Russia,\\
E-mail: logos-center@mail.ru
}
\end{document}